\documentclass[%
twocolumn,
superscriptaddress,
nofootinbib,
 amsmath,amssymb,
 aps, prd
]{revtex4-2}

\usepackage{graphicx}
\usepackage{tensor}
\usepackage{siunitx}
\usepackage{xcolor} % Remove after notes are no longer needed
\usepackage[normalem]{ulem} % Remove after notes are no longer needed
\usepackage{lipsum} % Remove after filler text is no longer needed
\usepackage[pdfborder={0 0 0}, plainpages=false, hyperfigures=true]{hyperref}

\newcommand{\beq}{\begin{equation}}
\newcommand{\eeq}{\end{equation}}
\newcommand{\beqn}{\begin{eqnarray}}
\newcommand{\eeqn}{\end{eqnarray}}

\newcommand{\lo}{\mathrel{\raise.3ex\hbox{$<$}\mkern-14mu
    \lower0.6ex\hbox{$\sim$}}}
\newcommand{\go}{\mathrel{\raise.3ex\hbox{$>$}\mkern-14mu
    \lower0.6ex\hbox{$\sim$}}}

\newcommand{\UNH}{\affiliation{Department of Physics \& Astronomy, University of New Hampshire, 9 Library Way, Durham NH 03824, USA}}
\newcommand{\TAPIR}{\affiliation{TAPIR, Walter Burke Institute for Theoretical Physics, MC 350-17, California Institute of Technology, Pasadena, California 91125, USA}}
\newcommand{\Cornell}{\affiliation{Cornell Center for Astrophysics and Planetary Science, Cornell University, Ithaca, New York, 14853, USA}}
\newcommand{\MPI}{\affiliation{Max Planck Institute for Gravitational Physics (Albert Einstein Institute), Am M{\H u}hlenberg 1, 14476 Potsdam, Germany}}

\begin{document}
\title{Assessing the Relative Importance of Neutrino–Matter Interaction Channels in Post-Merger Remnant of Binary Neutron Stars}
\author{Samantha Rath}\UNH
\author{Francois Foucart}\UNH
\author{Lawrence E. Kidder}\Cornell
\author{Harald P. Pfeiffer}\MPI
\author{Mark A. Scheel}\TAPIR

\begin{abstract}
Binary neutron star (BNS) mergers are amongst the most promising multimessenger sources for the joint detection of gravitational waves and electromagnetic counterparts, while also being potential sites for the production of heavy r-process elements and probes of the equation of state of matter above nuclear saturation density. The interpretation of these observations relies decisively on numerical simulations incorporating general relativity, relativistic hydrodynamics, neutrino radiation transport, magnetic fields, and nuclear physics. In particular, neutrino-matter interactions during and after merger strongly influence the thermodynamic evolution and composition of the remnant and its outflows, thereby affecting kilonova emission and nucleosynthesis yields. However, existing merger simulations remain limited by significant approximations in the treatment of neutrino transport and neutrino interaction rates. In this work, we assess the thermodynamic conditions under which neutrinos decouple from matter and show the effect of charged-current absorption, quasi-elastic scattering on nucleons and nuclei, pair-production processes, and inelastic neutrino-electron scattering for electron neutrinos, electron antineutrinos, and heavy-lepton neutrinos in the different thermodynamical conditions sampled by a simulation using an energy-dependent Monte Carlo neutrino transport.
We first estimate opacities in the post-merger remnant assuming neutrinos in equilibria with the fluid, and find results consistent with previous studies performed on simulations using a gray two-moment scheme. We note, in particular, the very distinct regions in which nucleon-nucleon Bremmstrahlung and electron-positron annihilation are active (high and low density regions, respectively). We then evaluate opacities using the actual distribution function of neutrinos within a Monte Carlo simulation. We show greatly increased pair annihilation rates in cold, low-density regions, especially for heavy-lepton neutrinos. We also show that inelastic scattering on electrons, which has not been included in merger simulations so far, makes important contributions to the thermalization of heavy-lepton neutrinos.
\end{abstract}

\maketitle

\section{Introduction}
Binary neutron star (BNS) mergers are astrophysical fireworks that occur due to loss of orbital energy through gravitational wave emission. They result in a burst of gravitational waves (GW) and the formation of a compact object (remnant) surrounded by a massive, accreting disk. This is also accompanied by ejection of matter into the interstellar medium~\cite{Rosswog:2015nja,Baiotti:2016qnr,Barack:2018yly,Bernuzzi:2020tgt,Dhani:2025axt}. The properties of the disk and the timescale of possible collapse to a black hole (BH) are determined by the initial mass of the NSs fated for collision and their underlying equation of state~\cite{Hotokezaka:2011dh,Bauswein:2013jpa,hotokezaka2013mass}. 
While this collapse itself occurs on the order of a few milliseconds to seconds, the remnant evolution is a dynamically active phase during which out-of-equilibrium weak interactions impact the thermodynamics and composition of the ejecta.This has a major impact on our observables. The neutron rich ejecta synthesizes heavy elements above Fe via rapid neutron capture nucleosynthesis~\cite{Symbalisty:1982ni,Thielemann:2017acv}. The radioactive decay of these heavy elements powers a UV/Optical/NIR/Electromagnetic transient, kilonova, which is observed days to months after the merger~\cite{Li:1998bw,Metzger:2010sy,Tanaka:2016sbx,Metzger:2016pju}. The elements produced by the r-process and the properties of the associated kilonova are both very sensitive to the composition of the outflows~\cite{Barnes:2013wka,Kullmann:2021gvo,Just:2021vzy,Rosswog:2022tus}. In addition, the remnant is postulated to be the likely progenitor of short-gamma ray bursts~\cite{Eichler:1989ve,Janka:1995cq,Nakar:2007yr,Lee:2007js,Meszaros:2014pca}.

The observation of GW170817, the first GW signal detected from merging BNS by the LIGO-VIRGO collaboration across several detectors on Earth heralded a new era in the field of multimessenger astronomy~\cite{LIGOScientific:2017vwq}.  Observations are expected to continue until the infrastructural limits of these detectors are reached. The next generation of advanced detectors, Cosmic Explorer and Einstein Telescope are expected to make a remarkable leap and achieve detection rates of $10^4$ BNS/year, detecting upto redshifts of 3 for BNS~\cite{Grado:2023pyb}. However, the multimessenger signals observed from GW170817 have already provided us key insights into the physics of compact object mergers. The simultaneous detection of GRB170817A and AT2017gfo~\cite{LIGOScientific:2017zic,LIGOScientific:2017ync} provided an observational confirmation of the predictions that BNS produce short-gamma ray bursts and kilonovae, respectively. Current understanding of the intricacies of these observations is limited by our inability to reproduce such conditions in laboratories on Earth as well as several limitations in the modeling of these systems -- from the difficulty of resolving the broad range of length and time scales and thermodynamical regimes important to BNS mergers to the use of approximate physics in the simulations.
Specifically, different models give different predictions for the mass and geometry of the ejecta, and its opacity to photons~\cite{Kasen:2017sxr,Villar:2017wcc,Perego:2017wtu,Wollaeger:2017ahm,Kawaguchi:2018ptg,Miller:2019dpt}. It is crucial to study the opacity of the ejecta which is itself significantly impacted by the amount of lanthanides/actinides produced. The electron fraction, ($Y_e$), determines the outcome of nucleosynthesis and, in particular, the amount of these high-absorption elements present. Neutrinos are produced in copious amounts in the remnant and, given the high density of material, engage in frequent weak interactions which effectively alter the $Y_e$. Neutrino-matter reactions are thus largely responsible for setting the composition of ejecta, hence motivating their importance in BNS mergers. 

Numerical simulations are indispensable tools for deciphering the multimessenger signals of compact binary mergers. Such simulations require capturing a wide range of coupled physical processes, among which neutrino transport plays a crucial role in determining the thermodynamic evolution and composition of the remnant. The multidimensional nature of these systems makes fully consistent modeling particularly challenging, requiring increasingly sophisticated treatments of neutrino radiation. Early relativistic merger simulations incorporated neutrino effects through leakage schemes, which capture local cooling but neglect detailed neutrino transport~\cite{Ruffert:1996by,Rosswog:2003rv,Deaton:2013sla,Neilsen:2014hha}. These approaches were later extended by coupling leakage with moment-based transport schemes that approximately model neutrino propagation and absorption ~\cite{Wanajo:2014wha, Sekiguchi:2015dma}. In particular, these studies demonstrated that neutrino irradiation can significantly alter the electron fraction of the ejecta, thereby influencing r-process nucleosynthesis yields and the resulting abundance distribution. Subsequent developments have led to a hierarchy of transport schemes with increasing physical fidelity. Gray two-moment (M1) schemes, which evolve neutrino energy and momentum densities using approximate analytical closures, are widely used in merger simulations~\cite{Foucart:2015vpa,Radice:2021jtw}. Recent simulations using the moments method also evolve the number density of neutrinos, providing some information about their energy distribution~\cite{Foucart:2016rxm}. However, these moment-based approaches depend sensitively on the choice of closure relations both for calculations of higher-order moments and when information about the energy distribution of neutrinos is required. They can exhibit numerical artifacts, and do not converge to the exact transport solution. In parallel, Monte Carlo (MC) transport methods have emerged as a promising alternative, offering a more direct solution of the neutrino transport equation without relying on closure assumptions. Recent work has demonstrated the feasibility of general relativistic MC transport in merger simulations~\cite{Foucart:2021mcb,Foucart:2022kon}. Hybrid schemes using MC simulations to provide closures to moment-based evolutions are also being investigated, but have not been used in full merger simulations so far~\cite{Izquierdo:2023fub}. These approaches are particularly attractive for capturing spectral and angular features of the neutrino radiation field, which are approximated in gray transport schemes. We are, however, bound by computational costs as well as theoretical uncertainties. In particular, all non-MC merger simulations rely on gray (energy-integrated) transport schemes due to the high cost of spectral treatments, while Monte Carlo methods remain computationally expensive at high accuracy, especially in dense regions of the remnant. In addition, significant uncertainties persist in the modeling of neutrino--matter interactions. These include incomplete treatments of many-body effects, inelastic processes, and the choice of interaction rates, all of which can impact the predicted neutrino opacities. As a result, the thermodynamic evolution of the remnant, the electron fraction of the ejecta, and ultimately the nucleosynthesis yields and kilonova observables remain sensitive to approximations made to these inputs.

Thus, it is sufficiently clear that we require more involved examination of neutrino-matter interactions in compact object mergers. The current study aims to advance that effort. In particular, it remains unclear which reactions dominate in different thermodynamic regimes, and where improved microphysical modeling is most critical. The present work aims to address this by identifying the regions of the $\rho$--$T$ phase space, encountered in merger simulations, in which different reactions contribute significantly to the total opacity. In optically thick regions, neutrinos are strongly coupled to the fluid. If reactions are fast enough for neutrinos to be in equilibrium with the fluid, knowing the exact reactions rates is not that important -- they just need to be fast enough for that equilibrium to be maintained in the simuations. Far from the remnant, neutrinos free-stream with negligible interactions. The transition between these regimes is more complex, as it depends sensitively on the underlying interaction rates. In radiative transport theory, this transition is often characterized using the optical depth,
\begin{equation}
\tau(R) = \int_R^\infty \frac{dr}{\lambda_{\rm tot}(r)},
\end{equation}
with the effective decoupling region typically defined by $\tau \sim 2/3$ in analogy with the photosphere~\cite{Shapiro:1983du}. This defines the concept of a neutrinosphere, which approximately separates trapped and free-streaming neutrinos ~\cite{Janka:2012wk}.

The neutrino luminosity is constant with radius well outside the neutrinosphere as a reflection of conservation of energy. While this region has been well-examined in astrophysical systems like core-collapse supernovae and proto-neutron stars~\cite{Thompson:2000tu,Buras:2005tb,Langanke:2007ua,Rrapaj:2014yba}, there have only been a couple of studies in the context of compact binary mergers. In reality, the mean free path is strongly energy dependent, such that different neutrino energies decouple at different locations. Furthermore, it is useful to distinguish between a transport sphere, defined by the last interaction, and an effective or thermalization sphere, defined by the last significant energy exchange with the medium. These concepts are particularly relevant for heavy-lepton neutrinos, for which scattering can dominate over absorption processes. In the present work, we do not compute optical depths or neutrinosphere locations directly. Instead, we use energy-integrated (gray) opacities, opacities averaged over the simulated neutrino spectrum, and opacities at fixed neutrino energies as a diagnostic to identify regions where the neutrino mean free path becomes comparable to the characteristic length scale of the remnant ($\sim 1\,\mathrm{km}$). These regions correspond to the transition between trapped and free-streaming regimes. The regions close to and just outside of the neutrinosphere are where neutrino interactions are expected to have the greatest impact on the emergent spectra and fluid composition, while energy deposition in low-density regions may also impact the production of short gamma-ray bursts. This motivates a detailed examination of the relative importance of different reactions in these regimes.

A similar approach was adopted by~\cite{Endrizzi:2019trv} for evaluating the optical depth and decoupling surfaces for two different nuclear EOS, DD2 and Sly4, by postprocessing at 1, 10 and 20 ms after simulation. However, we start from an energy-dependent Monte Carlo simulation for post-processing in our case. In addition, the main objective of our present work is to investigate the effect of different reactions to the total thermalization and scattering opacity. This includes considerations of inelastic scattering on electrons and out of equilibrium effects for pair processes which have not been analyzed previously and are generally not taken into account in simulations, but have been shown to be potentially important in simulations of isolated neutron stars~\cite{Cheong:2024cnb}. 

\section{Simulation}
\label{sec:sim}
The analysis presented in this paper has been performed by post-processing snapshots of a merger simulation performed with the SpEC code. In the simulation, the dense matter is described by the SFHo equation of state, and we consider neutron stars with gravitational masses of $1.26\, M_\odot$ and $1.36\, M_\odot$. Their initial separation is 43km. Within the simulation, the remnant collapses to a black hole $6\,\mathrm{ms}$ after merger, with the merger time defined as the point when the maximum density first rises $3\%$ above its initial value (i.e. when the cores of the two neutron stars collide). We analyze individual snapshots at intervals of $1\,\mathrm{ms}$ between the merger and collapse to a black hole. The SpEC simulation includes general relativity, ideal fluid dynamics, and Monte Carlo radiation transport but no magnetic fields. More details on our numerical methods can be found in~\cite{Duez:2008rb,Foucart:2012vn,Foucart:2021mcb}. Neutrino matter interactions include charged current reactions for electron type neutrinos, $e^+e^- \leftrightarrow \nu\bar \nu$ pair annihilation and nucleon-nucleon Bremmstrahlung for heavy-lepton neutrinos, as well as scattering on nucleons, nuclei and alpha particles for all species. All those rates are computed using the NuLib library~\cite{OConnor:2014sgn}; we discuss individual reactions in more detail below. None of the reactions differentiate between $\nu_\mu, \nu_\tau, \bar\nu_\mu, \bar \nu_\tau$. Accordingly, all heavy-lepton neutrinos are grouped into a single species $\nu_x$ in the simulation. We note, in particular, that the simulation includes neither charged current reactions for muons nor pair processes for electron type neutrinos. For $\nu_x$, pair processes are treated very approximately, assuming neutrinos in equilibrium with the fluid when the distribution function of neutrinos or antineutrinos enters the calculation of the reaction rates~\cite{Foucart:2025nub,Alford:2026kwd}.
\begin{figure*}
\centering
\includegraphics[width=0.32\textwidth]{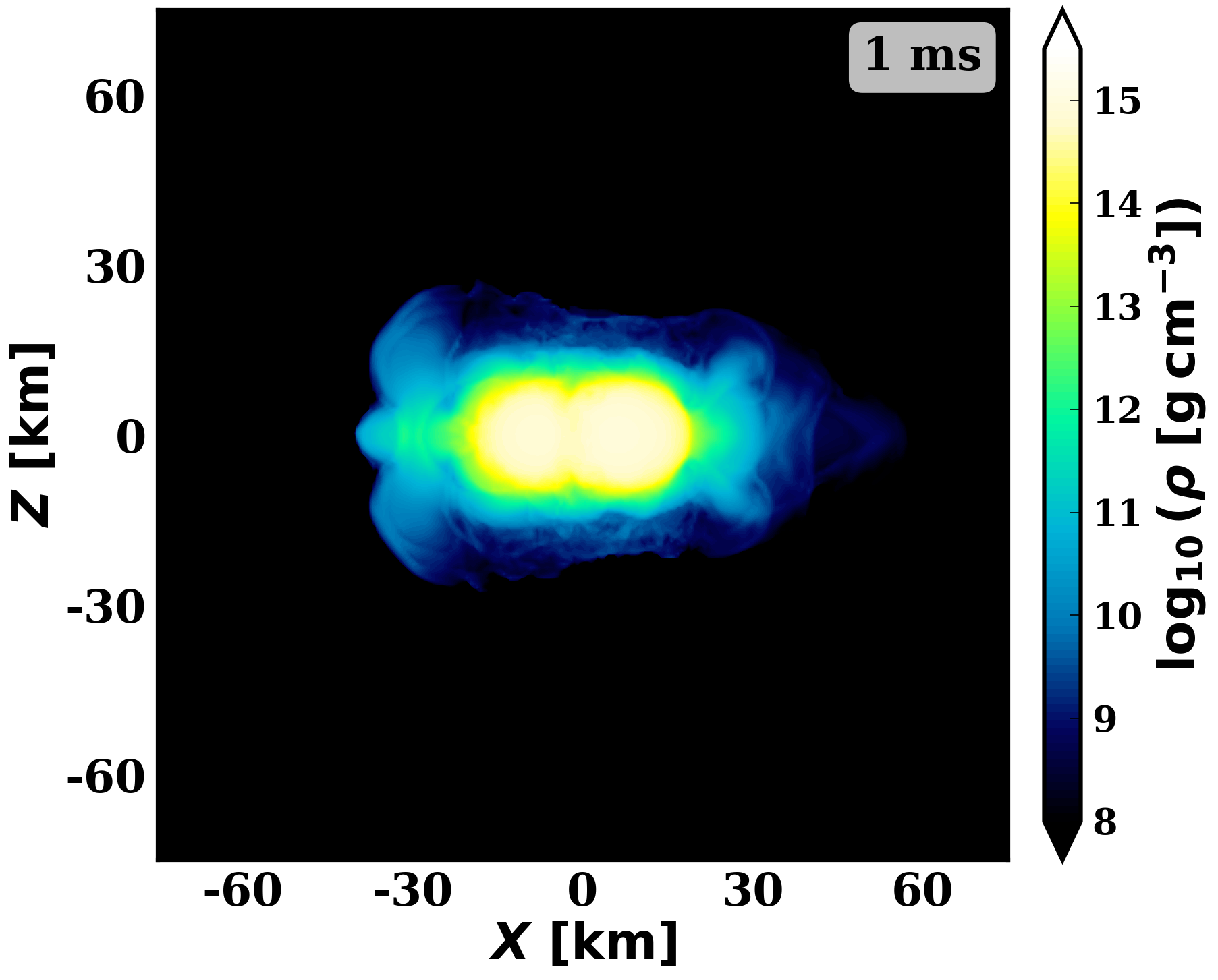}
\includegraphics[width=0.32\textwidth]{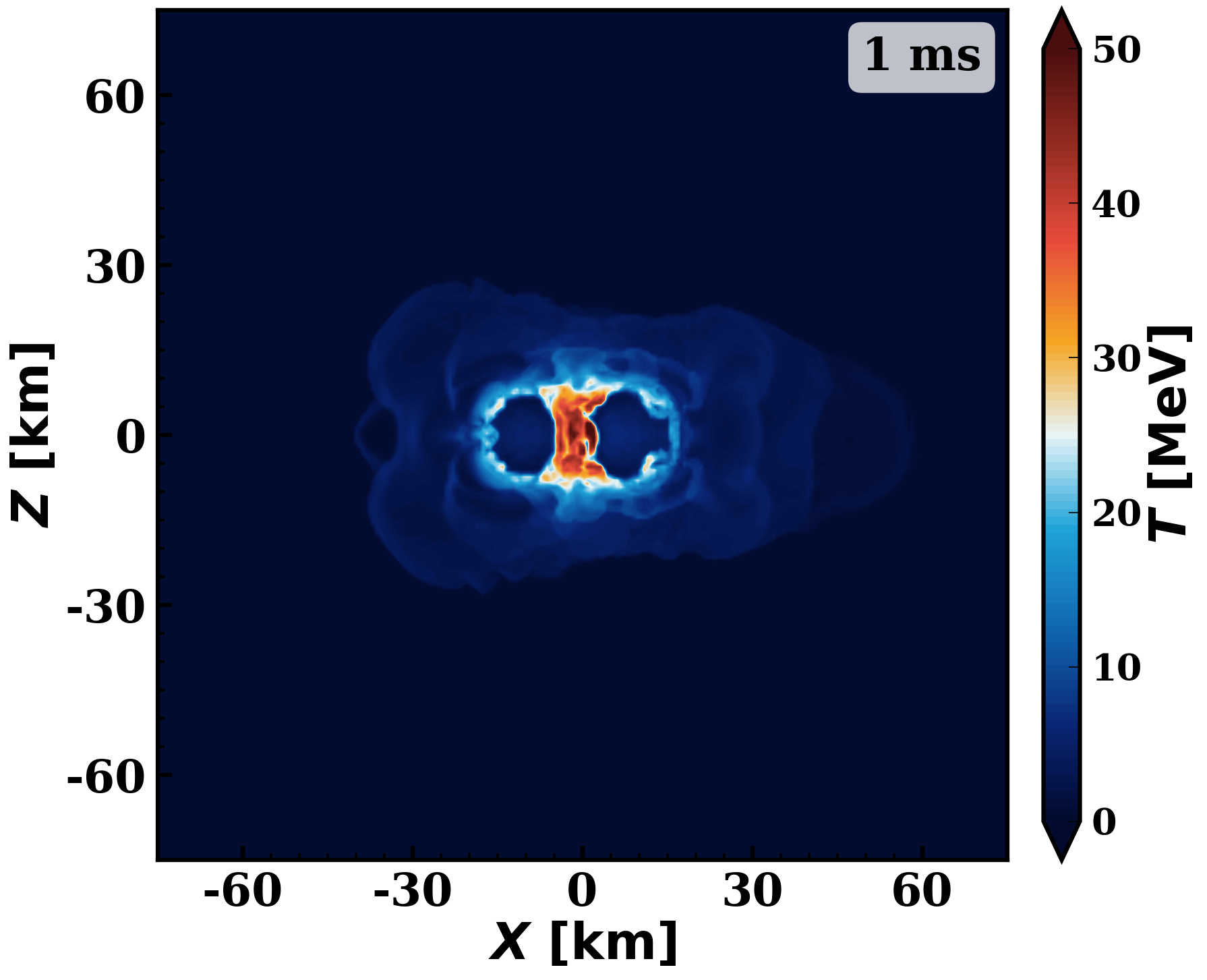}
\includegraphics[width=0.32\textwidth]{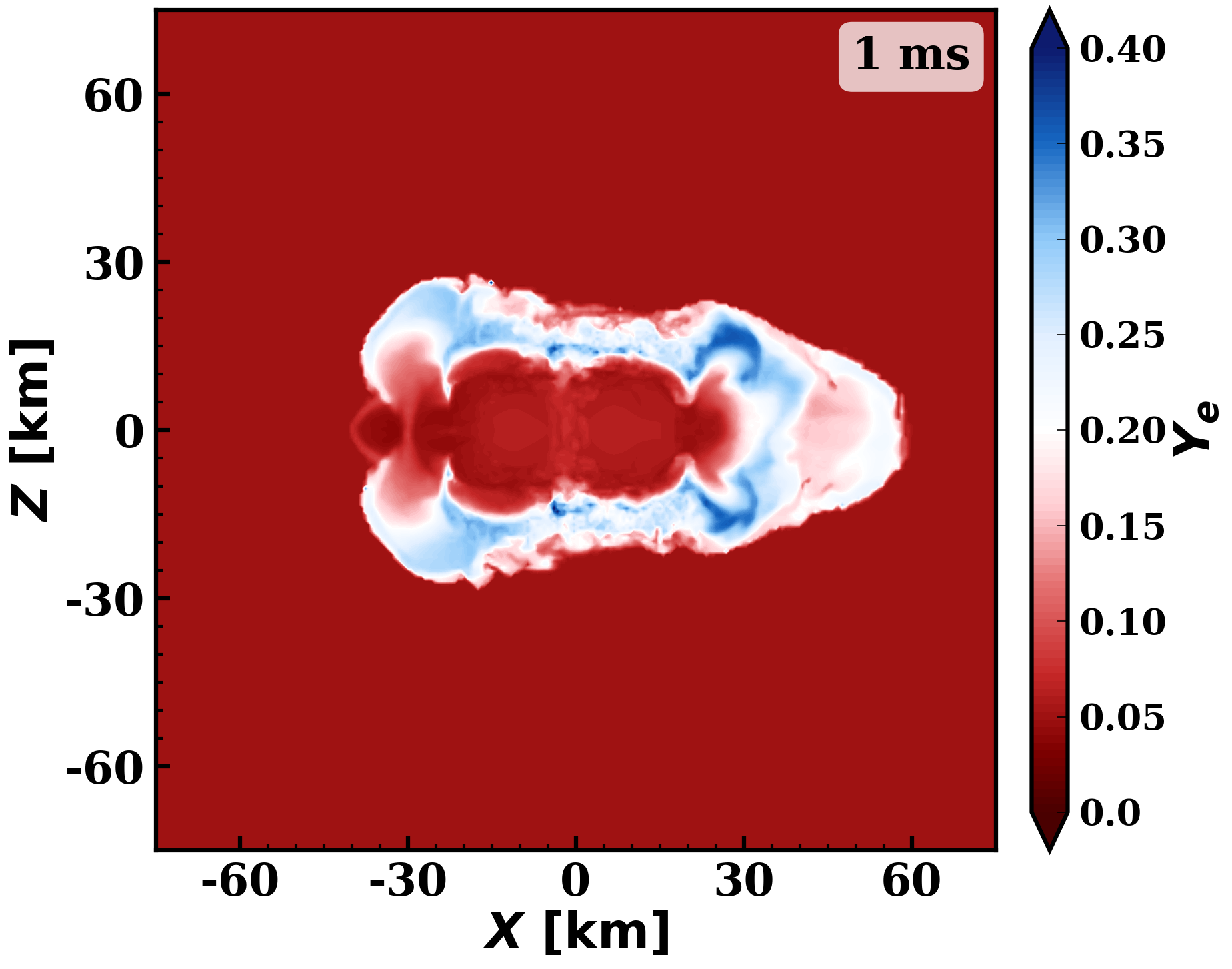}

\vspace{0.3cm}
\includegraphics[width=0.32\textwidth]{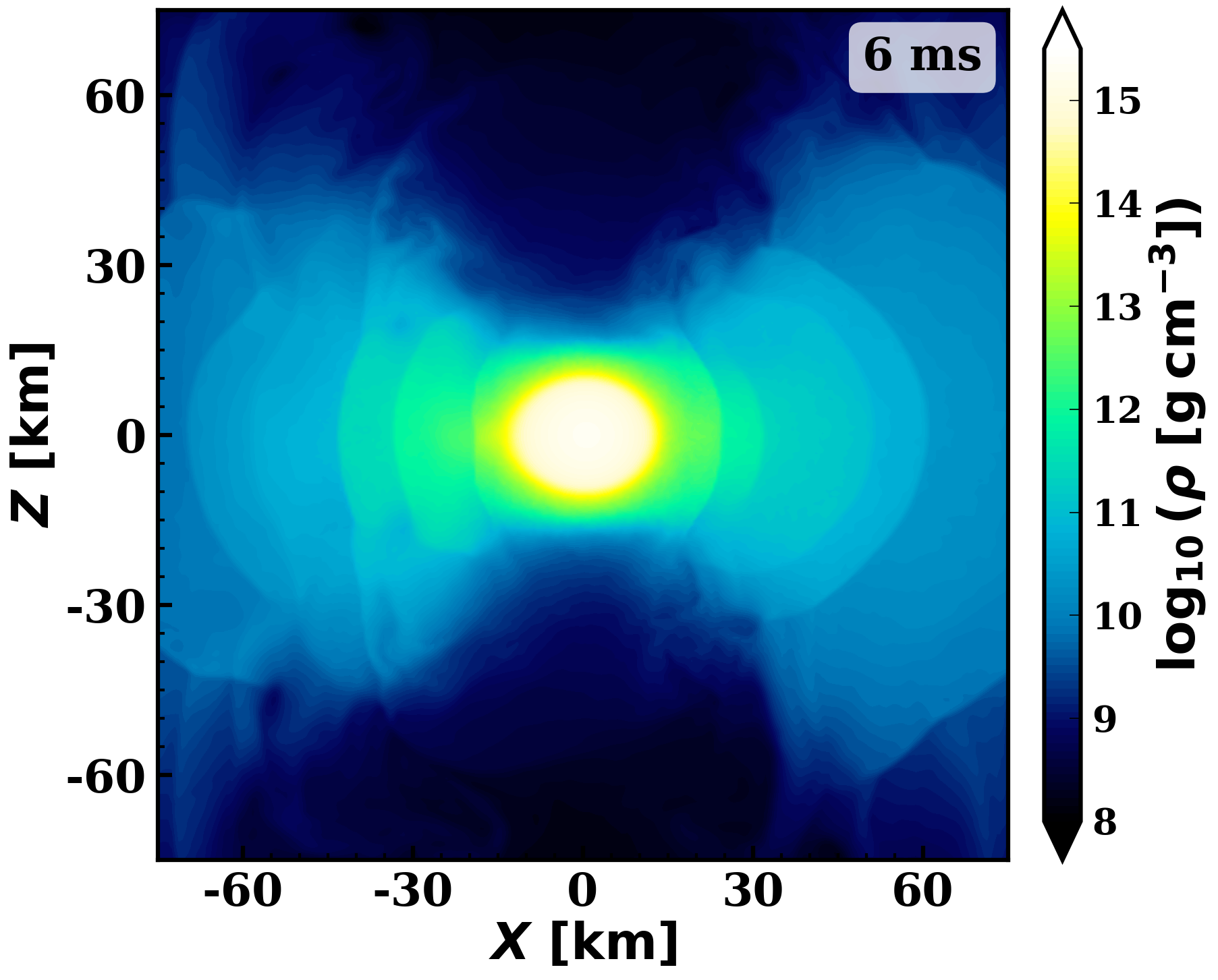}
\includegraphics[width=0.32\textwidth]{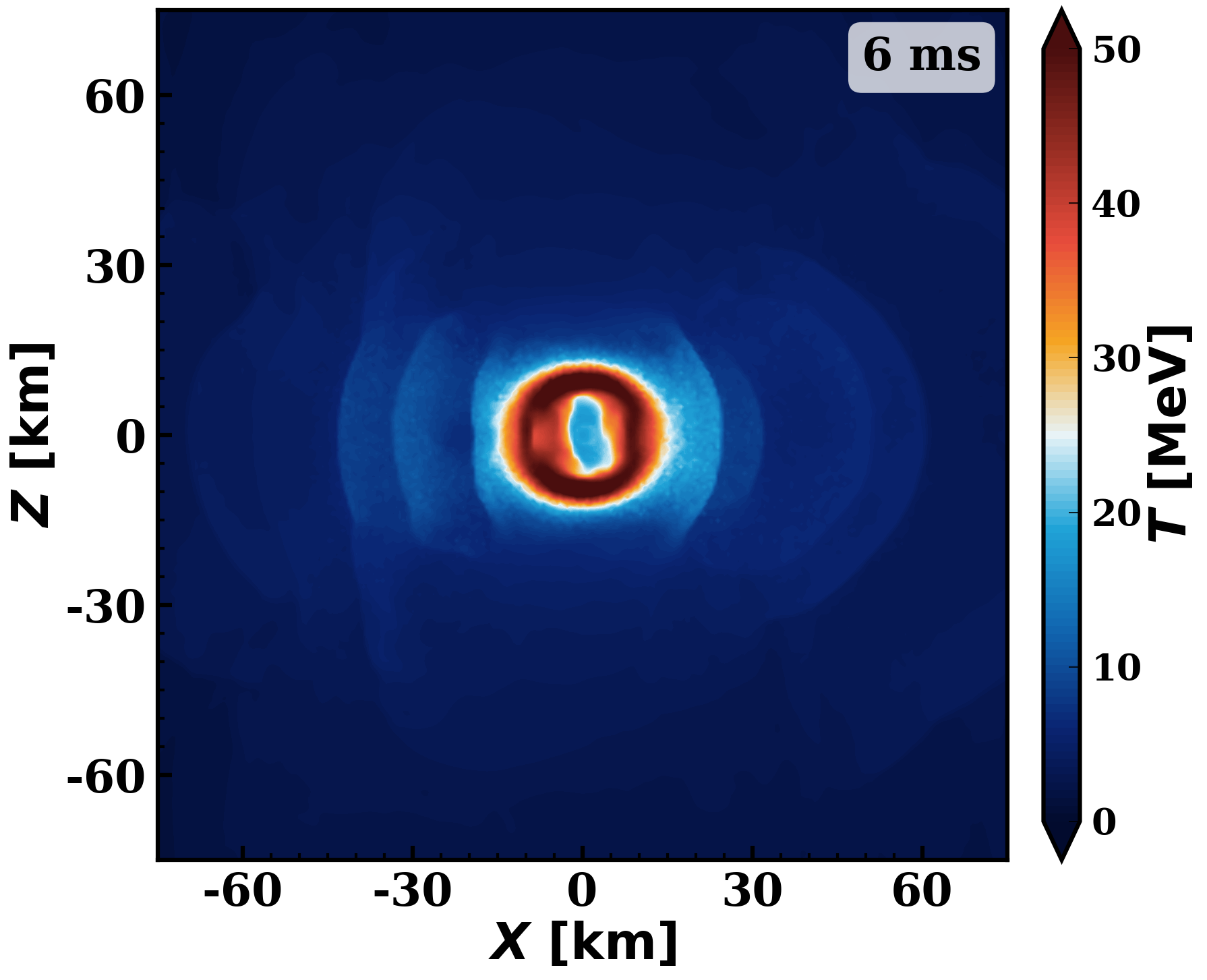}
\includegraphics[width=0.32\textwidth]{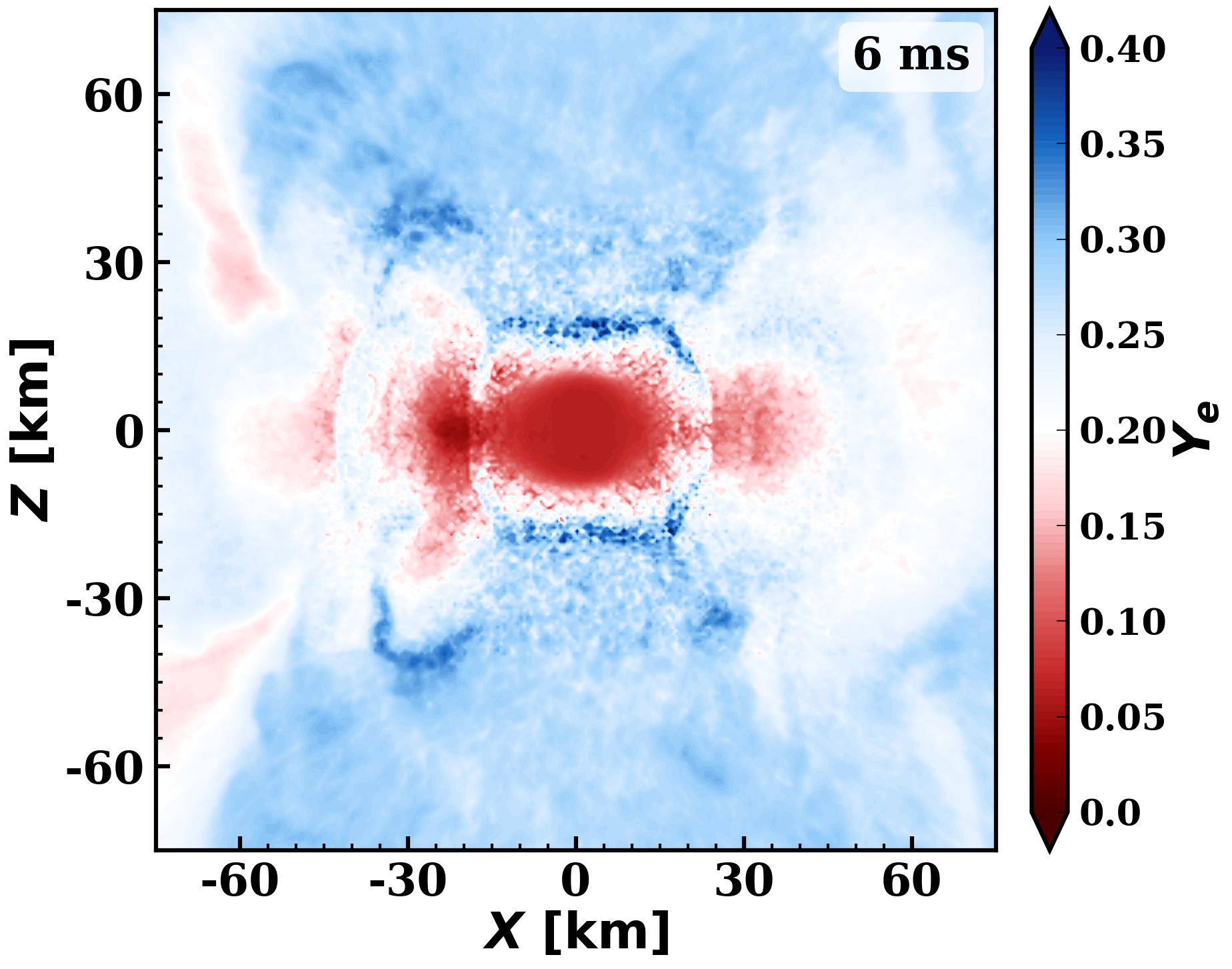}
\caption{
Snapshots of the post-merger remnant in the $x$–$z$ plane at $1\,\mathrm{ms}$ (top row) and $6\,\mathrm{ms}$ (bottom row).
From left to right we show the rest-mass density $\rho$, temperature $T$, and electron fraction $Y_e$.}
\label{fig:xz_snapshots}
\end{figure*}

To build intuition for the physical conditions relevant to neutrino transport, we first present $x–z$ slices in Fig~\ref{fig:xz_snapshots} of the rest-mass density ($\rho$), temperature ($T$), and electron fraction ($Y_e$) at $1\,\mathrm{ms}$ and $6\,\mathrm{ms}$ post-merger. At $1\,\mathrm{ms}$ post-merger, the remnant retains signatures of the binary structure, with two dense cores and strong tidal distortions. Shock heating at the interface between the merging stars produces a localized high-temperature region ($T \sim 50\,\mathrm{MeV}$), marking the site of the most intense energy dissipation. The electron fraction remains low in the densest regions, reflecting the neutron-rich composition of the original stars, while moderately higher-$Y_e$ material appears in the surrounding layers due to early neutrino processing. By $6\,\mathrm{ms}$, the system transitions toward a quasi-equilibrium remnant with a smoother density profile and an extended envelope. While the highest temperature remains confined to the core, there is distribution of heat to the surrounding disk and outflow system. The $Y_e$ distribution correspondingly develops a clear radial gradient, with neutron-rich material confined to the inner remnant and higher-$Y_e$ material in the outer layers, consistent with sustained neutrino irradiation shaping the composition of the outflows.

\begin{figure*}[t]
    \centering
    \includegraphics[width=\textwidth]{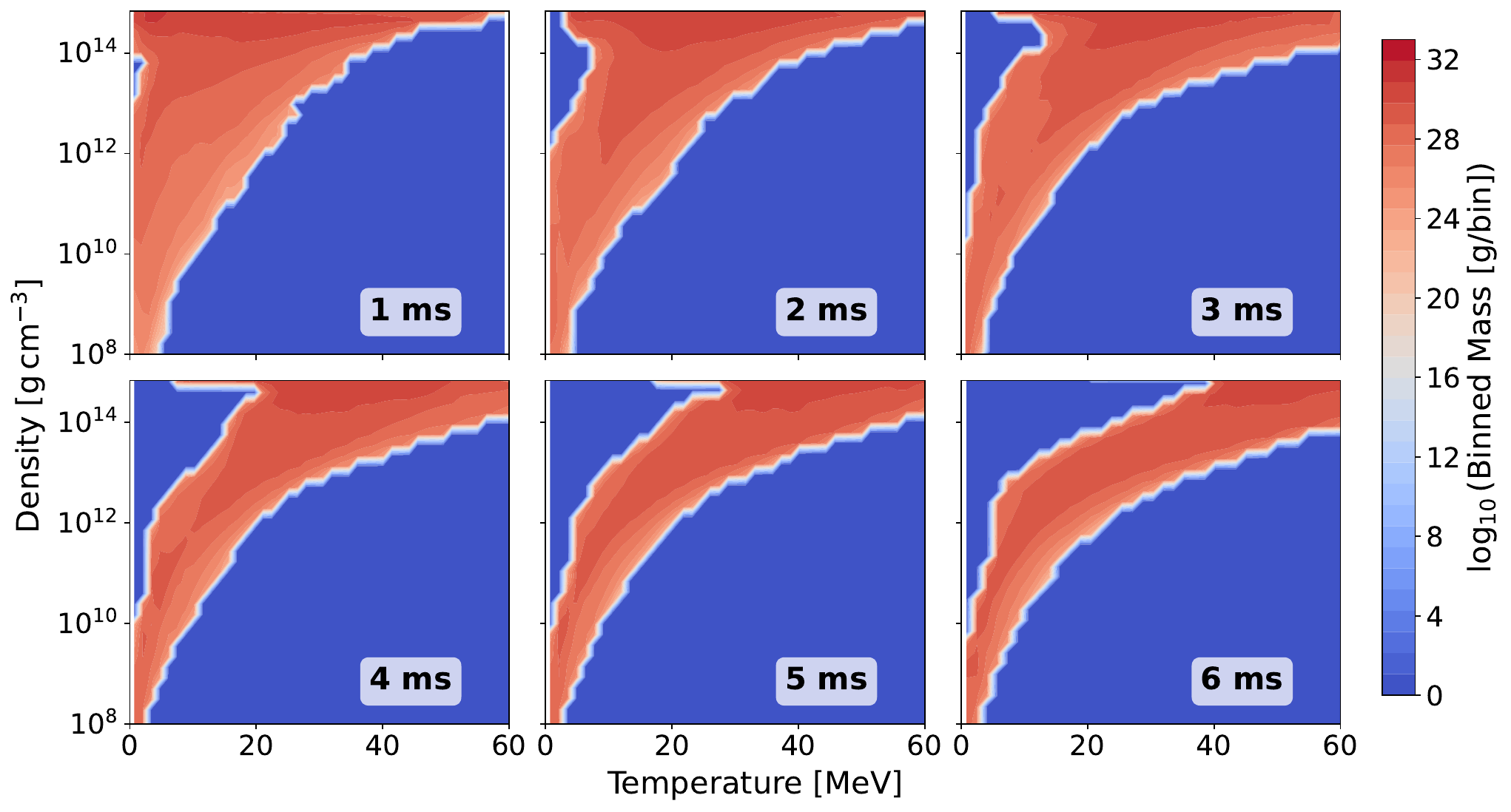}
 \caption{Mass binned in density and temperature, over the simulation domain. Regions outside the red-orange colored area mark the end of the phase space sampled by the simulation. Each image corresponds to a different time post-merger.}
\label{fig:mass_dom}
\end{figure*}

\begin{figure*}[t]
    \centering
    \includegraphics[width=\textwidth]{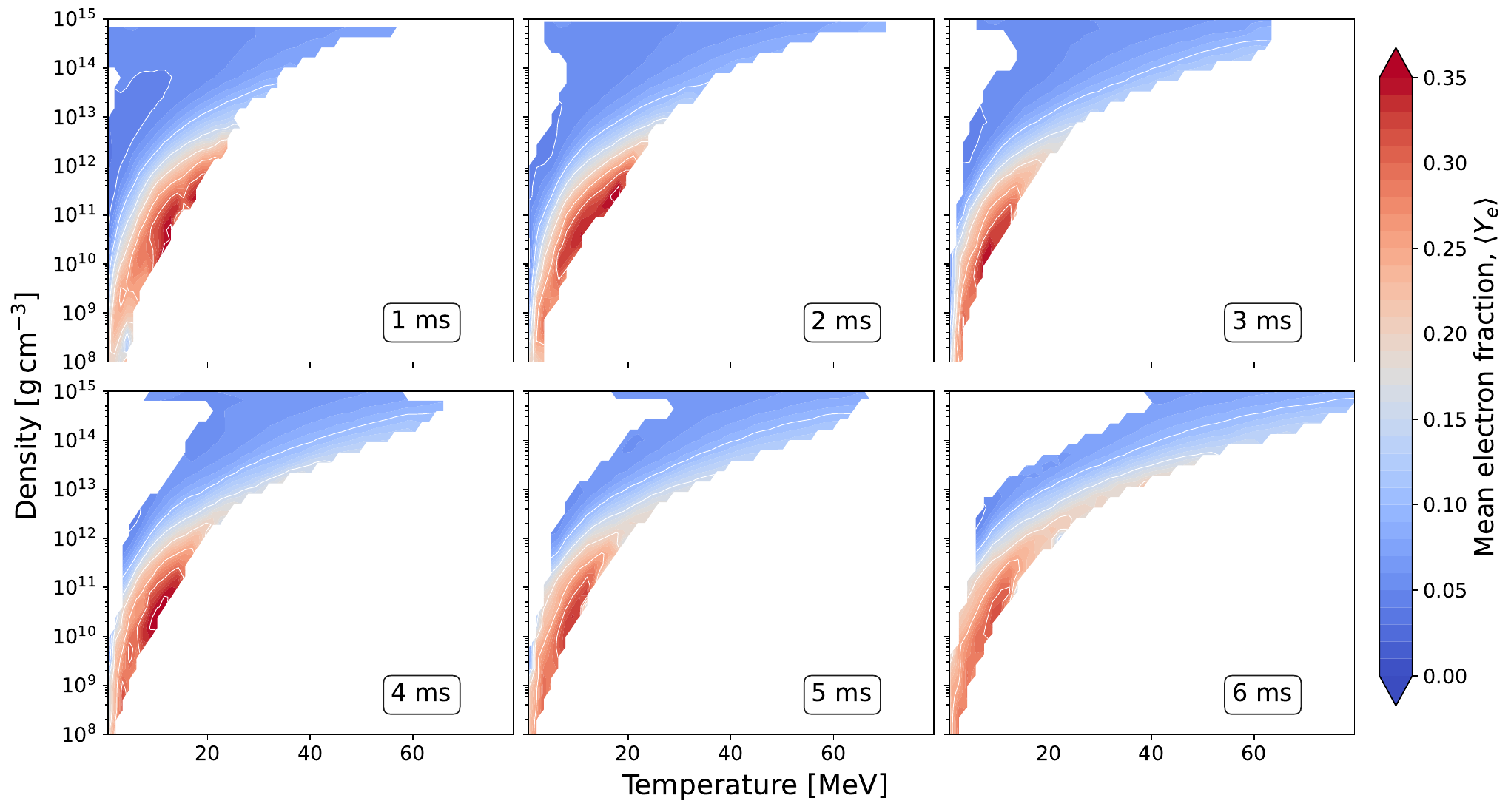}
    \caption{Distribution of $Y_e$ in the Density-Temperature plane of the simulation domain , at $(1-6)\,{\rm ms}$ post merger. We show the mass-weighted average value of $Y_e$ within each bin in $(\rho,T)$. White contour lines denote iso-contours of $\langle Y_e \rangle$ at values $0.05, 0.10, 0.15, 0.20, 0.25, 0.30,$ and $0.35$.}
    \label{fig:ye_dom}
\end{figure*}

\begin{figure*}[t]
    \centering
    \includegraphics[width=\textwidth]{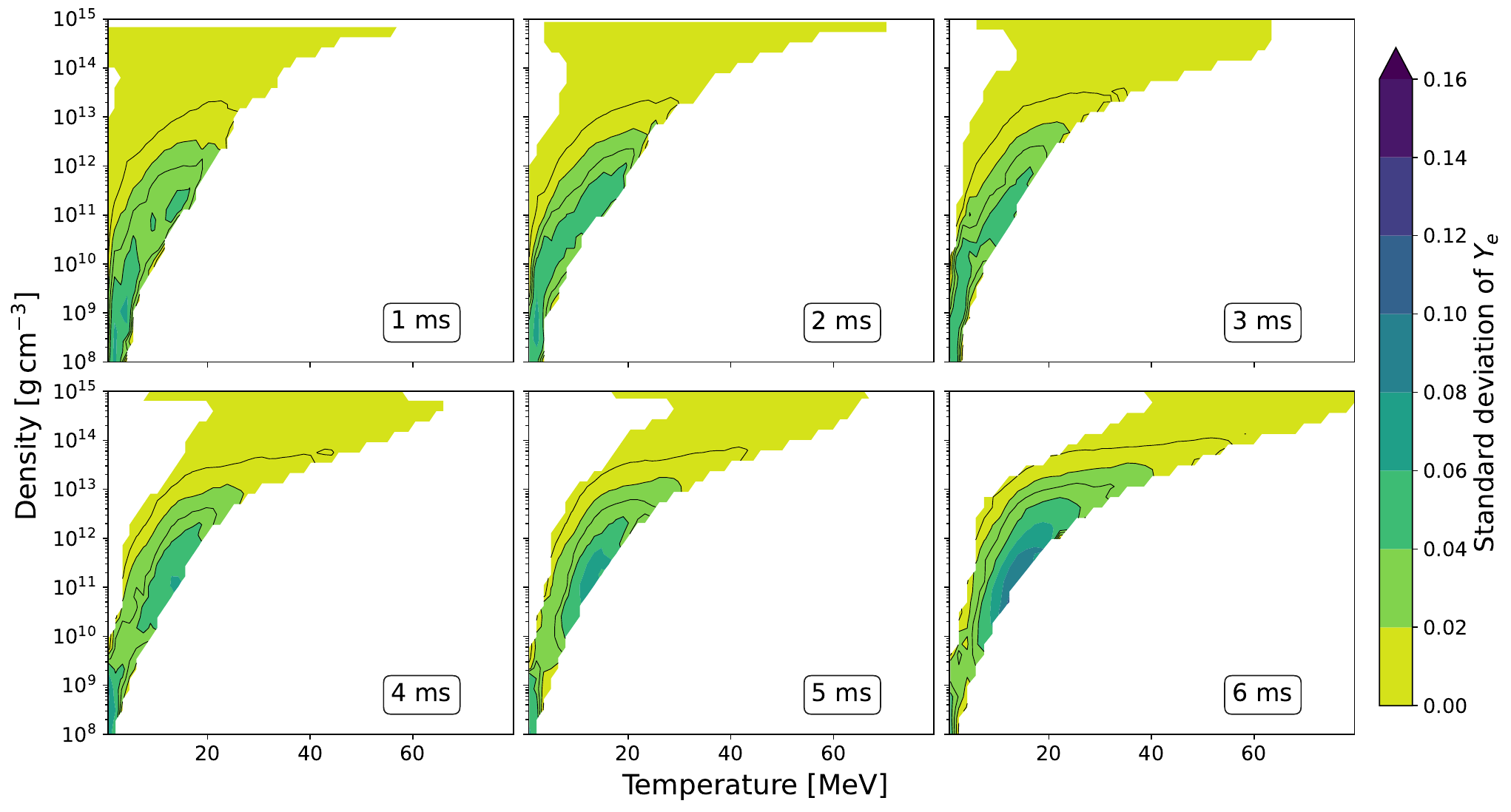}
    \caption{ Standard deviation of $Y_e$ within each $(\rho,T)$ bin, at $(1-6)\,{\rm ms}$ post merger.  Black contour lines denote the standard deviation at values $0.01, 0.02, 0.03,$ and $0.04$.}
    \label{fig:stddev_ye}
\end{figure*}

We transition to looking ar $\rho-T$ plots as a useful way to estimate the range of thermodynamic conditions encountered in mergers. To begin, we first analyze the total baryon mass binned in the $\rho-T$ plane as shown in Fig~\ref{fig:mass_dom}. This is done by constructing two-dimensional histograms in density and temperature space. We adopt logarithmically spaced bins in density and linearly spaced bins in temperature, using 50 bin edges in each dimension. This choice ensures uniform resolution across several orders of magnitude in density while maintaining adequate sampling of thermal variations.  Within each bin, we compute the total mass by summing the contributions from all fluid elements, weighted by the local rest-mass density, metric determinant, and Lorentz factor. As a result, the quantity shown represents the total mass contained in each $(\rho, T)$ bin. We see that within the first couple milliseconds, most of the mass is concentrated in $\rho \gtrsim 10^{13}\,\mathrm{g\,cm^{-3}}$, which characterizes the hot remnant core. As time evolves, we see heating of the core, likely due to shocks, and matter moving into the disk.
$Y_e$ parametrizes the composition of matter, $Y_e = n_p / (n_p + n_n)$, and serves as a useful guide to assess the neutron richness of the remnant at any point in our phase space. We plot $Y_e$ and its standard deviation in Fig~\ref{fig:ye_dom} and Fig~\ref{fig:stddev_ye}, respectively. We note that the standard deviation here is not a measure of simulation error, but rather of variations in the electron fraction of cells with similar density and temperature. There is an evident area of relatively higher electron fraction, $ Y_e \sim 0.2$ - $0.35$ at intermediate densities and temperatures, approximately $\rho \sim 10^{10}$ - $10^{12}\,\mathrm{g\,cm^{-3}}$ and $T \sim 10$ - $20$ MeV. This is the region of the disk and outflows where we would expect higher irradiation by neutrinos. We also note that matter at higher densities exhibits lower electron fractions and weak variation of $Y_e$ with temperature. This is in part because the equilibrium electron fraction in these dense regions is very low, and in part because neutrinos are trapped and only slowly diffuse out. In those regions, we do not reach `neutrinoless beta-equilibrium'. We instead reach first an equilibrium at constant total electron lepton number (including $e^\pm,\nu_e,\bar \nu_e$), with the total lepton number only evolving on the diffusion time scale. To further assess the consistency of thermodynamic conditions at fixed $\rho,T$, we calculate the standard deviation of the electron fraction within each bin, $\sigma_{Y_e}$. The largest values of $\sigma_{Y_e}$, reaching $\sim 0.1$ - $0.14$ arise for the same intermediate density, moderate-temperature region where the mean electron fraction is higher. This marks the region as one of higher interaction probability. Overall, these plots serve as a guide to understand which regions of phase space exist in our simulation, giving us a sense of where it is most important to know opacities. Further, we also need to know how representative the average value of $Y_e$ in Fig~\ref{fig:ye_dom} is through it's variation. The dense regions are well defined for a given $\rho, T$ as there isn't much variation in $Y_e$. In lower density regions, we have a mixture of cells with different $Y_e$ within each bin in our 2D plots.


\section{Relevant Reactions}
The properties of the post-merger remnant and its outflows are significantly influenced by the complex interplay of the four fundamental interactions. In the early stages post merger, electromagnetic and strong interactions proceed rapidly enough to approximately maintain thermodynamic and nuclear statistical equilibrium while the bulk dynamics of the system are governed by the balance between gravity and pressure gradients, the latter being in part due to strong interactions. Weak interactions, mediated by neutrino emission, absorption, and scattering, play a decisive role in regulating the composition and thermodynamic evolution of the outflowing material as well as the relative abundances of neutrino species. Specifically, we consider matter consisting of free nucleons ($n$, $p$), nuclei, photons, electrons and positrons ($e^{\pm}$), electron neutrinos and antineutrinos ($\nu_e$, $\bar{\nu}_e$), and heavy-lepton (anti) neutrinos, collectively denoted as $\nu_x$. In this case, we neglect the interaction with muons due to current unavailability of muonic interaction channels in Nulib and SpEC. The weak interactions considered in our analysis of simulation snapshots are the charged-current reactions with electrons and positrons, elastic and inelastic scattering (involving both charged and neutral current), $e^+e^-$ pair production/annihilation, and nucleon-nucleon Bremsstrahlung listed in Table~\ref{tab:reactions}. Not all of these reactions are included in the simulation ab-initio: the SpEC simulation ignores scattering on electrons and all pair processes for $(\nu_e,\bar\nu_e)$.

\begin{table}[t]
\centering
\caption{Neutrino reactions considered in this study. 
Here $\nu \in \{\nu_e,\bar{\nu}_e,\nu_x\}$ with $\nu_x \equiv \nu_\mu + \nu_\tau$, 
$N \in \{n,p\}$ denotes a nucleon, $A$ a nucleus, and $e^\pm$ electrons/positrons. 
Scattering on nucleons and nuclei is nearly elastic, while scattering on electrons is inelastic.}
\label{tab:reactions}
\begin{tabular}{ll}
\hline\hline
\textbf{Category} & \textbf{Reaction} \\
\hline
Charged Current 
& $\nu_e + n \leftrightarrow p + e^{-}$ \\
& $\bar{\nu}_e + p \leftrightarrow n + e^{+}$ \\
\hline
Scattering 
& $\nu + N \leftrightarrow \nu + N$ \\
& $\nu + A \leftrightarrow \nu + A$ \\
& $\nu + e^{\pm} \leftrightarrow \nu + e^{\pm}$ \\
\hline
Pair Processes 
& $e^{+} + e^{-} \leftrightarrow \nu + \bar{\nu}$ \\
& $N + N \leftrightarrow N + N + \nu + \bar{\nu}$ \\
\hline\hline
\end{tabular}
\end{table}

We present an overview of the role of weak interaction processes listed in Table~\ref{tab:reactions} and their expected contribution to the evolution of the remnant. Charged current reactions involving electron neutrinos and antineutrinos are the dominant mechanisms for cooling of the merger remnant and setting the electron fraction. In the absence of muons and for a fluid of no net charge, $Y_e = n_p / (n_p + n_n)$. It is worth mentioning that heavier and lighter $r-$process elements are produced due to low $Y_{e}$ ($\leq 0.25$) and high $Y_{e}$ outflows, respectively. In BNS merger simulations, the charged current reactions of heavy leptons are usually suppressed, albeit it has been recently shown that muon charged current reactions may be important in the densest regions of the remnants~\cite{Pajkos:2024iry,Ng:2024zve,Gieg:2024jxs,Gieg:2026beb}.

Under pair processes, three dominant mechanisms, namely electron-positron annihilation, plasmon decay and nucleon-nucleon Bremsstrahlung are discussed here for a crude qualitative understanding. In addition to charged current reactions, pair processes contribute to the absorption and emission of neutrinos. In pair annihilation/creation, neutrinos of all types ($\nu$ and $\overline{\nu}$) are involved. In the absence of charged current reactions of heavy leptons, pair processes are the dominant source of emission for $\nu_x$.

The scattering of neutrinos of all flavors on free nucleons, nuclei and leptons ($e^{\pm }$) may be either quasi-elastic or inelastic. Given the energy scale of the post-merger remnant $T=(1-50)$ MeV, the scattering of neutrinos on non-relativistic nucleons and nuclei is quasi-elastic and the scattering of neutrinos on relativistic electrons is inelastic. In the densest region of post-merger remnants, the diffusion of neutrinos is mostly limited by scattering. In the latter discussion, we will see that the contribution of charged current absorption and quasi-elastic scattering are of equal importance for $\nu_{e,a}$, while scattering is the dominant channel for neutrino-matter interactions for $\nu_x$. The inelastic scattering can provide a mechanism for the thermalization due to redistribution of energy for $\nu_x$.

In most of this manuscript, we look at the variation of gray opacity $\kappa$ everywhere in the remnant. Assuming thermal equilibrium, the opacity due to individual reactions is calculated using $\rho,T,Y_e$ from NuLib.
For an intuitive understanding of the physics of post-merger remnant, we first provide very qualitative estimates regarding the relative importance of different neutrino-matter interaction processes using approximate order-of-magnitude calculations. We note that these opacities are not what we actually use in the analysis of our simulations; they are only provided here as a way to gain intuition about the relative importance of different processes in various thermodynamic regimes. The absorption opacities $\kappa_{a}^{(\nu _{e}n)}$ and $\kappa _{a}^{(\overline{\nu }_{e}p)}$ of neutrons and protons, respectively, approximately scale as:~\cite{Bruenn:1985en,Burrows:2004vq,Foucart:2025cjr}

\begin{align}
\kappa_{a}^{(\nu_{e} n)}
&\thickapprox
\int \frac{2 d^{3}p_{n}}{n}
f_{n}(E)\left( 1 - f_{p}(E) \right)
\sigma^{(\nu_{e} n)}
\thickapprox
n_{n}\sigma^{(\nu_{e} n)} \\
\kappa_{a}^{(\overline{\nu}_{e} p)}
&\thickapprox
\int \frac{2 d^{3}p_{p}}{n}
f_{p}(E)\left( 1 - f_{n}(E) \right)
\sigma^{(\overline{\nu}_{e} p)}
\thickapprox
n_{p}\sigma^{(\overline{\nu}_{e} p)} .
\end{align}
with equilibrium Fermi-Dirac distribution function written as
\begin{equation}
f_{i}=\frac{1}{e^{\left( E_{i}-\mu _{i}\right) /k_{B}T}+1}
\end{equation}
and the cross-sections $\sigma ^{(\nu _{e}n)}$ and $\sigma ^{(\overline{\nu }
_{e}p)}$ (without recoil and weak-magnetism, $E_{\nu _{e}}>>\Delta
_{np}=m_{n}c^{2}-m_{p}c^{2}$), turn out to be
\begin{equation}
\sigma ^{(\nu _{e}n)}\approx \sigma ^{(\overline{\nu }_{e}p)}\thickapprox
1.38\sigma _{0}\left( \frac{E_{\nu _{e}}}{m_{e}c^{2}}\right) ^{2}
\end{equation}%
where $\sigma _{0}=1.705\times 10^{-44}$ $cm^{2}$. 
These rough estimates ignore all blocking factors. In the denser regions, the blocking factors of neutrinos and electrons is non-negligible and needs to be properly included. In most regions of the remnant, $n_{n}>n_{p}$ and hence, the absorption opacity $\kappa _{a}^{(\nu _{e}n)}>\kappa _{a}^{(\overline{\nu }_{e}p)}$.

In the calculation of  the neutrino emission rate $\eta $, given by a general form $\eta =\eta ^{\ast }\left( 1-f_{\nu }\right) $, the inclusion of the final state blocking factor $\left( 1-f_{\nu }\right) $ is quite subtle. Practically, in numerical simulations, it is convenient for $\eta $ and $\kappa $ to have no dependence on $f_{\nu }$ and to only depend on the properties of fluid.
This can be done through a simple mathematical transformation, but only for charged current reactions~\cite{Burrows:2004vq}.
By redefining the emission rate as $\eta ^{\ast }$ and the absorption opacity as 
$\kappa _{a}^{\ast }=\kappa _{a}/\left( 1-f_{\nu }^{(eq)}\right)$, the collision term of charged current reaction is
written as $\partial _{t}f_{\nu}=\eta _{\nu}-c\kappa _{\nu}f_{\nu }=\eta ^{\ast}-c\kappa _{a}^{\ast }f_{\nu}$ and the emissivity and absorption opacity are related by Kirchoff's law, $\eta ^{\ast}=c\kappa _{a}^{\ast }f_{\nu}^{(eq)}$. We note that while $\eta $ is dependent on $f_{\nu}$, the redefined $\eta ^{\ast }$ is independent of $f_{\nu}$. By calculating one of $\left( \eta ^{\ast},\kappa _{a}^{\ast }\right) $, the other can be obtained using Kirchoff's law. 

As already mentioned, neutrinos of all flavors ($\nu$ and $\overline{\nu }$) are involved in pair processes, which are the dominant source of neutrino emission for heavy-lepton neutrinos. The relative importance of different pair processes can be made explicit by comparing rough estimates of their emission rates. The total emissivity $Q_{e^- e^+}$ due to electron--positron annihilation for neutrinos in equilibrium with the fluid, is approximately given as,~\cite{Dicus:1972yr,Foucart:2022bth}
\begin{equation}
Q_{e^- e^+} \approx Q_0 \left( \frac{k_B T}{\mathrm{MeV}} \right)^9
\frac{
F_4(\eta_e)\,F_3(-\eta_e) + F_3(\eta_e)\,F_4(-\eta_e)
}{
2\,F_4(0)\,F_3(0)
}.
\end{equation}
where $Q_0 = 9.76 \times 10^{24}~\mathrm{erg\,cm^{-3}\,s^{-1}}$ (for $\nu_e \bar{\nu}_e$ emission), and $Q_0 = 4.17 \times 10^{24}~\mathrm{erg\,cm^{-3}\,s^{-1}}$ (for all other neutrino species). Comparing with $Q_{p e^-}$ and $Q_{n e^+}$, we see that the pair process producing $\nu_e \bar{\nu}_e$ dominates in hot and/or low-density regions of the fluid. In the former case, the equilibrium distribution of neutrinos is rapidly attained, while in the latter, the fluid is efficiently cooled by neutrino emission.

In addition the total neutrino emissivity per species, $Q_{nb}$ and $Q_{pl}$ due to nucleon-nucleon Bremsstrahlung~\cite{Brinkmann:1988vi,Hannestad:1997gc,Burrows:2004vq} and plasmon decay~\cite{Ruffert:1995fs},
respectively, are given approximately, for neutrinos in equilibrium with the fluid, by 
\begin{align}
Q_{nb} &\approx \left( 1.5 \times 10^{26}~\mathrm{erg\,cm^{-3}\,s^{-1}} \right)
\left( \frac{n_n}{10^{36}~\mathrm{cm^{-3}}} \right)^2
\left( \frac{k_B T}{\mathrm{MeV}} \right)^{5.5}, \\
Q_{pl} &\approx Q_{0,pl} \left( \frac{k_B T}{\mathrm{MeV}} \right)^9
\gamma^6 e^{-\gamma} (1+\gamma)
\left( 2 + \frac{\gamma^2}{1+\gamma} \right) B.
\end{align}
with $Q_{0,pl} = 6 \times 10^{23}~\mathrm{erg\,cm^{-3}\,s^{-1}}$ (for $\nu_e, \bar{\nu}_e$), and $Q_{0,pl} = 10^{21}~\mathrm{erg\,cm^{-3}\,s^{-1}}$ (for all other neutrino species) the blocking factor is
\begin{equation}
B = \langle 1 - f_\nu \rangle \, \langle 1 - f_{\bar{\nu}} \rangle,
\end{equation}
and
\begin{equation}
\gamma \approx 0.056 \sqrt{\frac{\pi^2 + 3\eta_e^2}{3}}.
\end{equation}

In comparison to electron--positron annihilation, Bremsstrahlung dominates in colder, denser regions. Since the required conditions for a dominant process to produce heavy-lepton neutrinos varies in the densest region of accretion disk, with $n_{n}\thicksim 10^{35-37}cm^{-3}$ and $T\thicksim 1$ MeV$-10$ MeV, we will need to consider both Bremsstrahlung and electron-positron pair processes. Plasmon decay is generally a subdominant reactions for the conditions encountered in merger remnants; we will not discuss it further here.
In the post-merger remnant, the emission of $\nu _{e}$ and $\overline{\nu }_{e}$ due to charged current is usually faster than the pair production of $\nu _{e}\overline{\nu }_{e}$. This situation is reversed in the case of $\nu_{x}\overline{\nu }_{x}$ production. In the simulation currently considered, emission of $\nu _{x}$ only occurs through pair processes. In $\nu \overline{\nu }$--pair production, the dependence of reaction rates on the distribution functions of $\nu$ and $\overline{\nu }$ is through blocking factors. In pair annihilation process, the reaction rate is directly proportional to product of $f_{\nu }f_{\overline{\nu }}$. Thus, an accurate treatment of pair process in simulation is difficult due to the nonlinear dependence of reaction rates on neutrino distribution functions. Existing simulations use equilibrium distribution of antineutrinos when calculating the annihilation rate of neutrinos, if pair processes are included at all. This is clearly a major approximation outside of the densest regions of the remnant.

Another important channel of interaction is via scattering. In the post-merger remnant, the scattering of neutrinos on nucleons and nuclei is quasi-elastic. To have a qualitative understanding about the relative importance of charge current and quasi elastic scattering processes, let us compare the scattering opacity $\kappa_s$ with that of $\kappa_a=\kappa_a^{(\nu_e n)}+\kappa_a^{(\overline{\nu}_e p)}$.
With the same approximation as used in obtaining $\sigma^{(\nu_e n)}$ and $\sigma^{(\overline{\nu}_e p)}$, we obtain
\begin{equation}
\sigma^{(s,X)} \approx C^{s,X}\sigma_0 \left( \frac{E_{\nu_e}}{1~\text{MeV}} \right)^2
\end{equation}
where $C^{s,X}=1.1$, $1.3$, $0.8$ for $X=p,n,\alpha$, respectively.
Combining the contributions due to protons and neutrons, the scattering opacity $\kappa_s$ is written as
\begin{equation}
\kappa_s \approx \left( 1.1 n_p + 1.3 n_n \right) \sigma_0 \left( \frac{E_{\nu_e}}{1~\text{MeV}} \right)^2
\end{equation}
Thus, absorption opacities $\kappa_a$ for charged current, and scattering opacities $\kappa_s$ are seen to be comparable. The latter will be dominant for heavy-lepton neutrinos $\nu_x$. In the remnant, usually $n_{\alpha} \ll n_n$ and the contribution of scattering opacity $\kappa_s$ due to $\alpha$ particle is sub-dominant (with a few exceptions in narrow regions of the parameter space covered by simulations). The same is true for scattering on heavier nuclei.

In the context of merger simulations, the inclusion of the inelastic scattering of neutrinos on electrons is quite complex and has not been attempted so far. We present a brief summary of the issues involved in capturing inelastic scattering here. A more detailed discussion can be found in~\cite{Foucart:2022bth}. The evolution of the distribution function of neutrinos due to inelastic scatterings can be written in the form,
\begin{equation}
\begin{split}
\left[ \frac{df_{\nu}}{d\tau} \right] &=
\left( 1 - f_{\nu} \right) \int \frac{d^{3}p'}{h^{3}} f_{\nu}' 
R^{\text{in}}\left( \nu, \nu', \cos \theta \right) \\
&\quad - f_{\nu} \int \frac{d^{3}p'}{h^{3}} \left( 1 - f_{\nu}' \right) 
R^{\text{out}}\left( \nu, \nu', \cos \theta \right)
\end{split}
\end{equation}
where $\theta$ is the scattering angle. The $f_{\nu}$ and $f_{\nu}'$ are distribution functions of neutrinos with energy-momentum $\left( \nu, \overrightarrow{p} \right)$ and $\left( \nu', \overrightarrow{p}' \right)$, respectively. Apart from the blocking factors, the collision term couples the distribution function of neutrinos at all possible values of four momenta $p$ and $p'$, which makes their inclusion in simulations expensive. Usually, one approximates the scattering kernel $R^{\text{in/out}}$ as a truncated expansion in
$\cos \theta$ written as
\begin{equation}
\begin{split}
R^{\text{in/out}}\left( \nu, \nu', \cos \theta \right) &=
\frac{1}{2} \Phi_0^{\text{in/out}}\left( \nu, \nu' \right) \\
&\quad + \frac{3}{2} \Phi_1^{\text{in/out}}\left( \nu, \nu' \right) \cos \theta.
\end{split}
\end{equation}
As an estimate, the effective opacity $\kappa_s$ due to the scattering of neutrinos on electrons is given by
\begin{equation}
\kappa_s \thicksim \sigma_0 \left[ \frac{E_{\nu}^{4} E_e}{(\text{MeV})^{5}} \right]
\left( 10^{30}~\text{cm}^{-3} \right)
\label{inela}
\end{equation}
where $E_{\nu}$ and $E_e$ are typical energies of neutrinos and electrons, respectively. At densities most sensitive to neutrino-matter interaction, the opacity due to inelastic scattering is subdominant in comparison to that of elastic scattering/charged current processes. However, opacities due to inelastic scattering and pair process may be comparable. This implies that, although the inelastic scattering of $\nu_e$ and $\overline{\nu}_e$ may not be important, it plays a crucial role in the thermalization of $\nu_x$.

\section{Interpolation}
In order to assess regions where neutrinos are more or less strongly coupled to the fluid, we require an estimate of the average absorption and scattering opacities, for their respective reactions. The following prescription is adopted for evaluation of neutrino opacities:
\begin{itemize}
\item Extracting neutrino opacities and emissivities from Nulib.
\item Interpolating to the local thermodynamic conditions of each fluid cell .
\item Computing energy-averaged mean opacities assuming equilibrium neutrino energy densities, when using gray opacities.
\item Averaging over all cells within a specific range of density and temperature through mass-weighting.
\end{itemize}

The absorption opacities ($\kappa_a$), emissivities ($\eta$) and scattering opacities ($\kappa_s$) are obtained from NuLib, an open-source neutrino interaction library. For elastic scattering and reactions involving a single neutrino, each quantity is extracted as an array of dimensions $(16, 3, 51, 65, 82)$ denoting neutrino energy bins, species, length of density grid, temperature grid and electron fraction grid, respectively. We consider three independent neutrino species, namely, electron neutrinos $\nu_e$,  electron anti-neutrinos $\bar{\nu}_e$ and heavy lepton (anti)neutrinos, $\nu_\mu + \nu_\tau$ collectively labeled as $\nu_x$. As SFHo assumes Nuclear Statistical Equilibrium and we do not include muons, it follows that the rest mass density, temperature and electron fraction encompass the thermodynamic state of matter completely~\cite{Arcones:2010yf}. For each neutrino species and energy, we obtain opacities at a given $\rho,T,Y_e$ through trilinear interpolation from the neutrino table.

Gray opacities are computed by weighting energy dependent opacities by the energy of neutrinos in equilibrium with the fluid in each energy bin. We emphasize that this weighting choice is an approximation from our general assumption of thermal equilibrium for the distribution. This is the method most commonly used in simulations relying on two-moment transport. We show results using the energy distribution from a Monte Carlo simulation in Sec.~\ref{sec:ooe}. Using the proper energy distribution mainly impacts the rate of pair annihilation and neutrino absorption outside the neutrinosphere. When using an equilibrium distribution, the resulting average opacity is then.
\begin{align}
\langle \kappa \rangle
=
\frac{\displaystyle \sum_{b}
\kappa(E_b)\,
\frac{\eta(E_b)}{\kappa_a(E_b)} 
}
{\displaystyle \sum_{E_b}
\frac{\eta(E_b)}{\kappa_a(E_b)}
},
\label{eq:kappa_avg}
\end{align}
with the sum being over all energy bins, and $E_b$ being the average energy of neutrinos in bin $b$. We discuss the calculation of rates for out-of-equilibrium neutrinos in Sec~\ref{sec:ooe}. Here, $\kappa$ without an index represents the opacity being calculated.

We also note that the above description does not apply to calculation of inelastic scattering rates. In thermal equilibrium, the electron distribution is described by the Fermi--Dirac function
\begin{equation}
f_e(E) = \frac{1}{\exp\!\left(\frac{E-\mu_e}{T}\right)+1},
\end{equation}
where $\mu_e$ is the electron chemical potential and $T$ is the temperature. The thermodynamic state of the fluid in merger simulations is typically described by the variables $(\rho, T, Y_e)$, where $\rho$ is the rest mass density and $Y_e$ is the electron fraction. For a given equation of state (EOS), these quantities uniquely determine the electron chemical potential $\mu_e(\rho,T,Y_e)$. It is convenient to introduce the dimensionless degeneracy parameter
\begin{equation}
I_\eta = \frac{\mu_e}{T},
\end{equation}
which characterizes the degree of electron degeneracy in the fluid. Because the electron occupation numbers entering neutrino interaction rates depend only on $T$ and the ratio $\mu_e/T$, the inelastic neutrino-electron scattering kernels are commonly tabulated as functions of $(T, I_\eta)$ rather than the full set of thermodynamic variables $(\rho, T, Y_e)$. In practice, $\mu_e$ is obtained by interpolating the EOS table at the local fluid state, allowing $I_\eta(\rho,T,Y_e)$ to be computed and henceforth used for evaluating the appropriate scattering kernels. This interpolation is performed using the same trilinear scheme as before.
The inelastic scattering kernels are obtained from NuLib. These kernels describe the probability of neutrino scattering between energy bins and are tabulated as functions of outgoing neutrino energy, incoming neutrino energy, the electron degeneracy parameter $I_\eta$, and temperature. In order to estimate the effects of this process within the simulation domain, we evaluate an effective inelastic scattering opacity following the handling of the rates as described in detail in Section~\ref{sec:inel} . This gives us an effective, energy-integrated interaction rate as a function of ($\rho$, T, $Y_e$). We note that this calculation is performed assuming an isotropic distribution of incoming neutrinos; i.e. we only use the zeroth order term in the expansion of the scattering kernel in $\theta$. This is likely a good first approximation in the context of merger remnants, as discussed in more detail in Section~\ref{sec:inel}.

\section{Opacity Results}

Absorption, scattering, and thermalization opacities characterize distinct aspects of neutrino--matter interactions in the post-merger remnant. The absorption opacity $\kappa_a$ quantifies processes that exchange energy and, for charged-current reactions, lepton number between neutrinos and the fluid, while the quasi-elastic scattering opacity $\kappa_s$ accounts for interactions that change neutrino direction with limited energy exchange (and no energy exchange within the assumptions of our numerical simulations). Inelastic scattering, which has so far been ignored in merger simulations, involves energy exchanges between neutrino and the fluid, but no change in lepton number. Thermalization is governed by a combination of these processes, often expressed through an effective opacity that determines how efficiently neutrinos approach equilibrium with the matter. It quantifies freeze-out from thermal equilibrium and is most important when assessing the last energy-changing interaction of neutrinos. These opacities play a central role in setting the neutrino decoupling region (or neutrinosphere), shaping the emergent neutrino spectra, and ultimately influencing the evolution of the electron fraction $Y_e$ and the composition of the ejecta.

\subsection{Absorption Opacity}

\begin{figure*}[t]
    \centering
    \includegraphics[width=\textwidth]{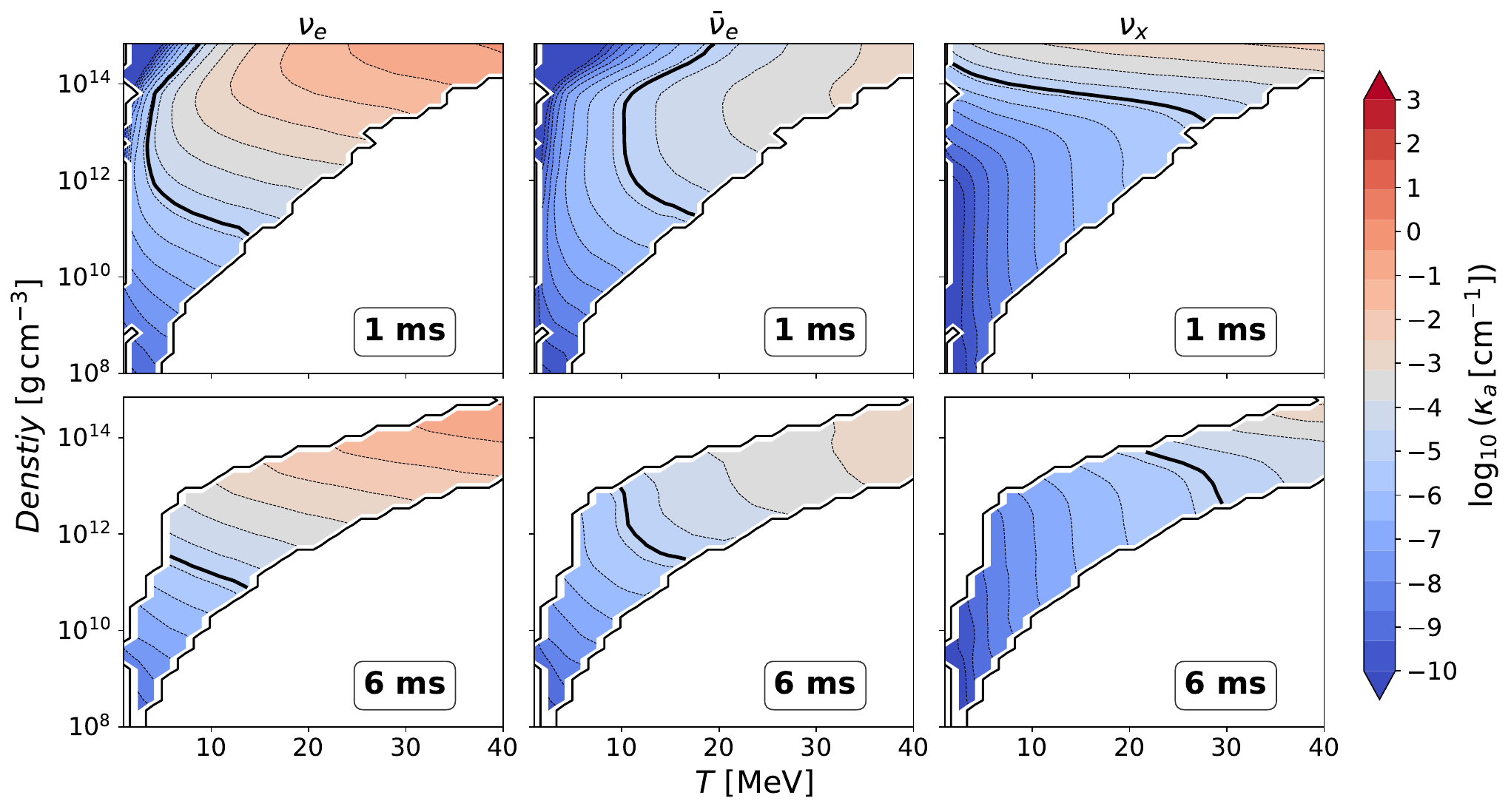}
    \caption{Mass-weighted absorption opacity $\log_{10}(\kappa_a)$ at $t = 1\,\mathrm{ms}$ (top) and $t = 6\,\mathrm{ms}$ (bottom) post-merger, shown for $\nu_e$ (left), $\bar{\nu}_e$ (centre), and $\nu_x$ (right). The remnant collapses to a black hole at $t \approx 6\,\mathrm{ms}$. In all plots, $\kappa_a$ is expressed in units of cm$^{-1}$ and the solid black line corresponds to $\kappa_a=10^{-5}\,{\rm cm}^{-1}$.}
    \label{fig:abs_dom}
\end{figure*}

In this section, we first focus on identifying where the absorption mean free path becomes comparable to the characteristic length scales of the remnant. We highlight the last interaction surface by defining it as an approximate characteristic mean free path of roughly $1\,\mathrm{km}$~\cite{Foucart:2025cjr}. This corresponds to $\kappa_a \sim 10^{-5}\mathrm{cm^{-1}}$. For visualization, we construct logarithmic density and linear temperature bins while calculating a fluid mass-weighted average absorption opacity in each bin. Contours of $\log_{10}(\kappa_a)$ are shown for $\nu_e$, $\bar{\nu}_e$, and $\nu_x$ at early and late times in Fig~\ref{fig:abs_dom}. We see that the location of last interaction is roughly between 5-15 MeV for $\nu_e$, between 10-15 MeV for $\bar{\nu}_e$ and between 1 - 25 MeV for $\nu_x$ depending on the density of the fluid. The evolution from 1 ms to 6 ms reflects the thermodynamic restructuring of the remnant rather than a qualitative change in interaction physics. While the density range at which the mean free path reaches $\sim 1\,\mathrm{km}$ is similar for $\nu_e$ and $\bar{\nu}_e$, around $\rho \sim 10^{11}\,\mathrm{g\,cm^{-3}}$, this transition occurs at significantly higher densities for $\nu_x$, around $\rho \sim 10^{13}\,\mathrm{g\,cm^{-3}}$.  This can be attributed to the lack of charged-current interactions on nucleons, leading to a longer mean free path, at least in the context of reactions currently considered in the simulation. We note that features in the high-density, low-temperature portion of the mapped thermodynamic domain ($\rho \gtrsim 10^{13}\mathrm{g\,cm^{-3}}$ and $T < 5\text{--}10\,\mathrm{MeV}$) are sensitive to the treatment of the transport method and associated dissipative heating. In these regions, even the equilibrium interaction rates used in simulations for charged-current reactions are inaccurate~\cite{Alford:2021ogv,Alford:2024xfb}, and these inaccuracies grow for pair processes or if the neutrino distribution departs from equilibrium.  Under such conditions, neutrinos may also contribute to an effective viscosity through their transport of energy and momentum and their role in restoring the fluid to its equilibrium composition. As shown in Fig.~\ref{fig:mass_dom}, these cold high-density regions disappear within a few milliseconds. The average time between absorption events $\tau=(\kappa_a c)^{-1}$ can, for charged-current reactions, be comparable to the evolution time scale. For as long as they exist in the merger remnant, these regions may have an out-of-equilibrium fluid composition and neutrino distribution functions.

\begin{figure*}[t]
    \centering
    \includegraphics[width=0.9\textwidth]{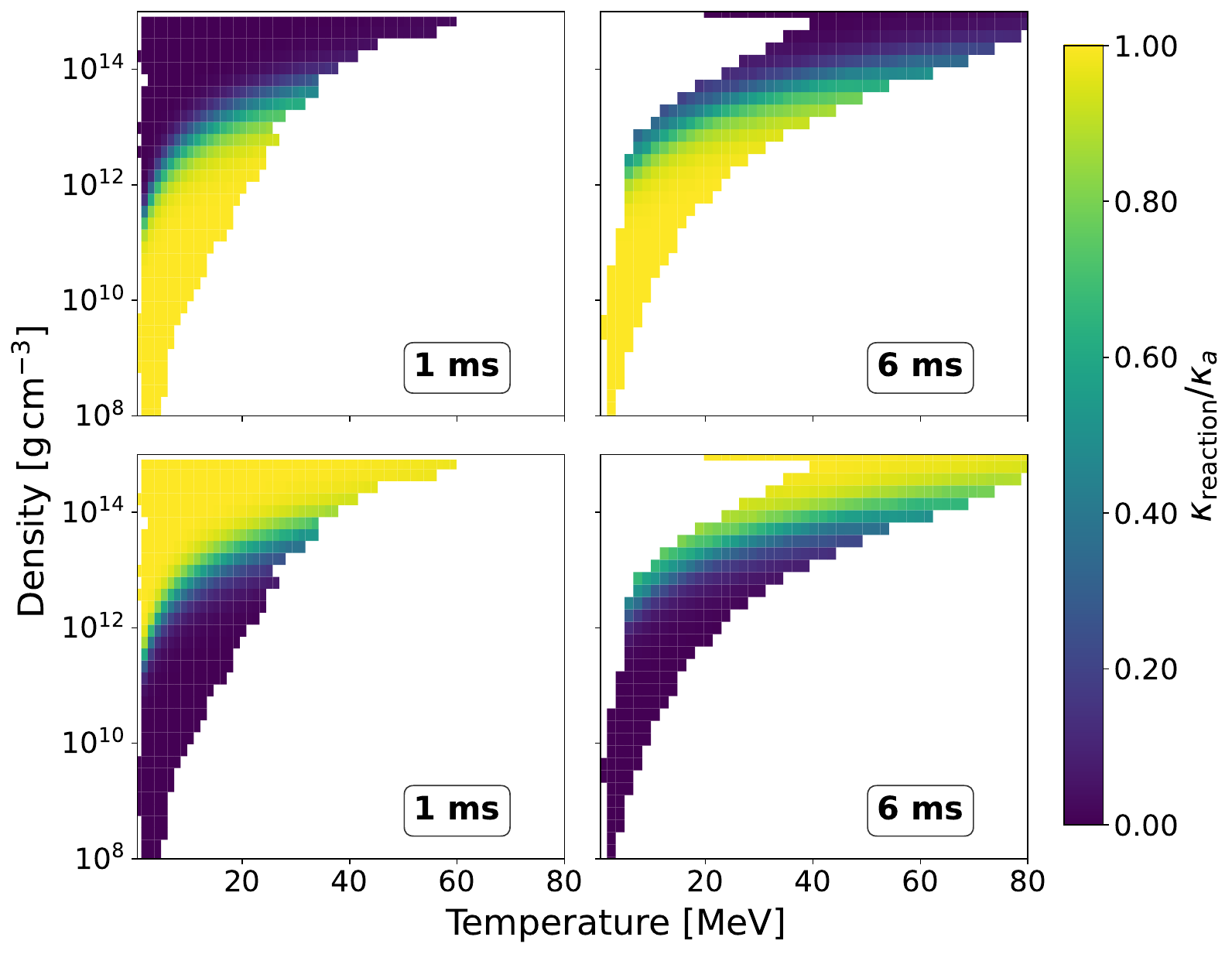}
\caption{{\it Top}: Fraction of the total absorption opacity due to $e^+e^-$ pair creation/annihilation for $\nu_x$ at 1\,ms and 6\,ms post-merger.
{\it Bottom}: Fraction of the total absorption opacity due to bremsstrahlung for $\nu_x$ at 1\,ms and 6\,ms post-merger.}
\label{fig:rel_abs}
\end{figure*}

Further, we determine the contribution of electron-positron pair annihilation and nucleon-nucleon Bremsstrahlung to the total absorption opacity for $\nu_x$. For $\nu_e$ and $\bar{\nu}_e$, these reactions are negligible everywhere when compared to charged-current reactions, when evaluated under the assumption that neutrinos remain in equilibrium with the fluid. This hierarchy, however, relies on the equilibrium assumption; in low-density regions where neutrinos decouple and the distribution function deviates from equilibrium, processes such as $\nu_e \bar{\nu}_e \rightarrow e^+ e^-$ can become important, particularly in polar regions. In cold high-density regions, we also need the full distribution function of neutrinos to estimate their contribution to the total absorption. We defer a more detailed discussion of these out-of-equilibrium effects to Sec.~\ref{sec:ooe}. For heavy-lepton neutrinos, $\nu_x$, the situation is qualitatively different. As charged-current reactions are absent, $e^+e^-$ annihilation and Bremsstrahlung play a significant role in setting the total absorption opacity. In Fig.~\ref{fig:rel_abs}, we show the fractional contribution of the different reactions to the total opacity, and find that $e^+e^-$ annihilation and Bremsstrahlung both contribute significantly to the total reaction rate for $\rho\sim 10^{12-14}\,{\rm g/cm^3}$, while dominating at low and high densities, respectively. If we compare this to the regions where neutrinos should be in thermal equilibrium with the fluid, discussed in more detail in Sec.~\ref{sec:therm}, we see that both reactions are important in the decoupling regions. This implies that out-of-equilibrium distributions may lead to significant corrections to the reaction rates close to the neutrinosphere for both Bremmstrahlung and $e^+e^-$ pair creation/annihilation, though the latter is likely more important beyond a few milliseconds post-merger.

\subsection{Scattering Opacity}

\begin{figure*}[t]
    \centering
    \includegraphics[width=\textwidth]{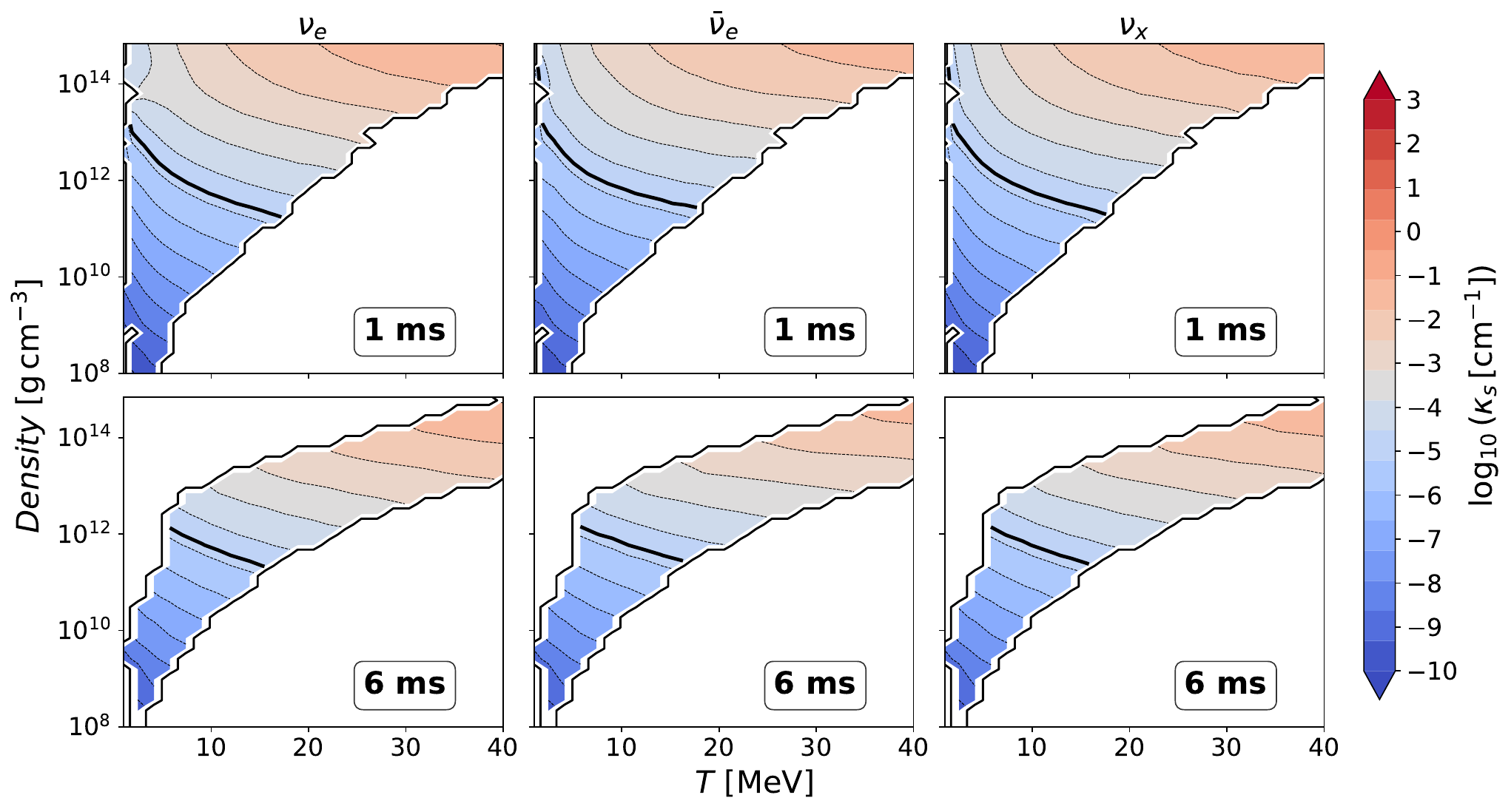}
    \caption{ Mass-weighted scattering opacity $\log_{10}(\kappa_s)$ at $t = 1\,\mathrm{ms}$ (top) and $t = 6\,\mathrm{ms}$ (bottom) post-merger, shown for $\nu_e$ (left), $\bar{\nu}_e$ (centre), and $\nu_x$ (right). The remnant collapses to a black hole at $t \approx 6\,\mathrm{ms}$. In all plots, $\kappa_s$ is expressed in units of cm$^{-1}$ and the solid black line corresponds to $\kappa_s=10^{-5}\,{\rm cm}^{-1}$.}
    \label{fig:scat_dom}
\end{figure*}

In addition to absorption opacity, it is important to determine the average scattering opacities as they have a combined effect on the transport properties of neutrinos. While absorption causes energy and lepton number changes, elastic scattering determines the angular distribution and diffusion of neutrinos through the fluid. To visualize the changes in these effects, we extract mass-weighted energy integrated scattering opacity using the same prescription as followed for average absorption opacity. We can infer from Fig~\ref{fig:scat_dom} that as opposed to absorption opacity, there is no particular hierarchy of species noted for last interaction surface by scattering. All of the last scattering regions are around $\rho \sim 10^{12}\,\mathrm{g\,cm^{-3}}$ and between 5-15 MeV. This is because all dense regions have large opacities here. We note however that this is partially an effect of assuming equilibrium neutrino spectra. In practice, escaping $\nu_x$ have higher average energies than escaping $\bar\nu_e$ and $\nu_e$ close to the surface of last scattering, an effect that will push the surface of last scattering to slightly lower densities for $\nu_x$. 

\begin{figure*}[t]
    \centering
    \includegraphics[width=0.9\textwidth]{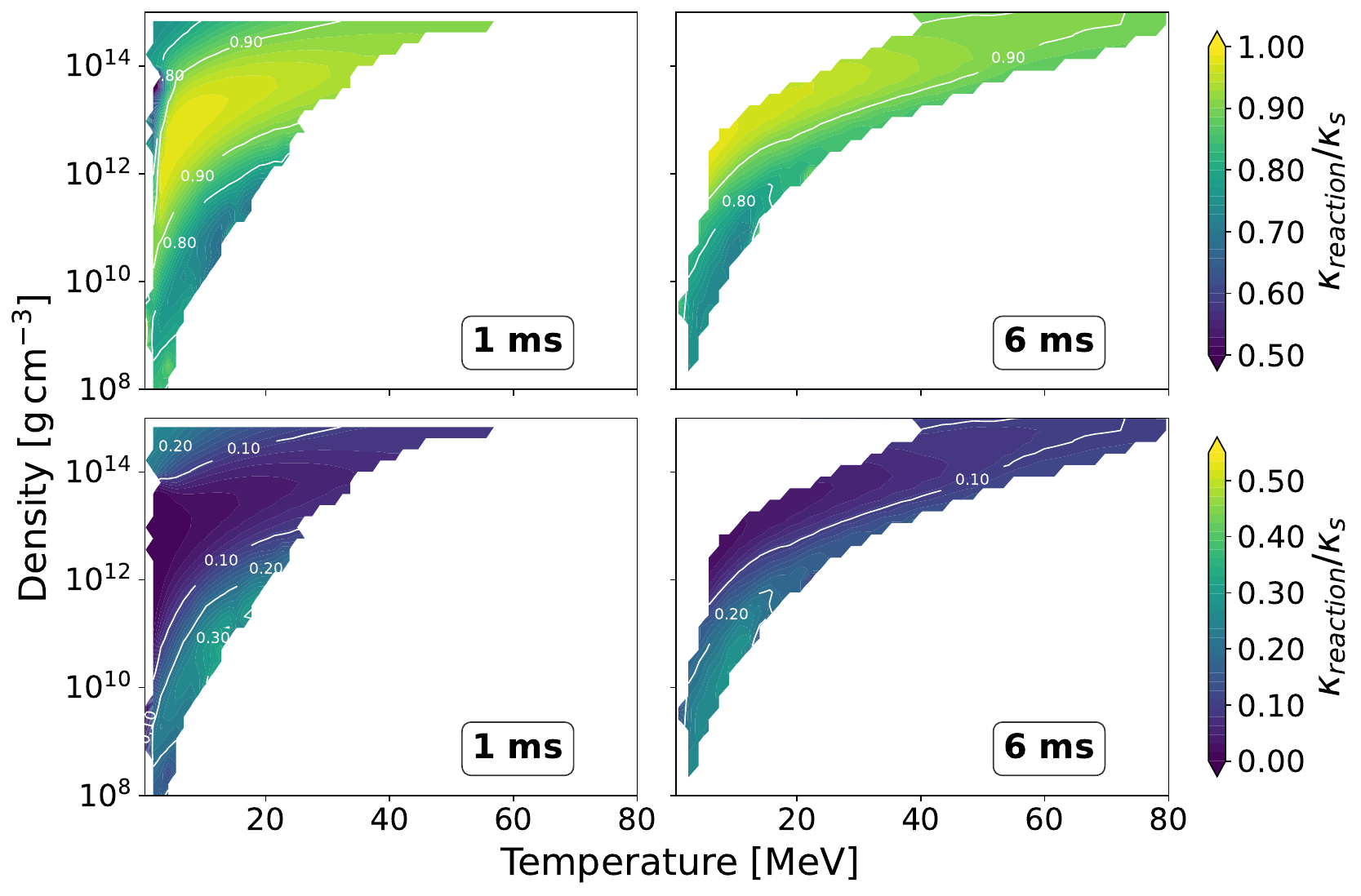}
\caption{ {\it Top}: Fractional contribution of scattering on neutrons to $\kappa_s$ for $\nu_e$ at $1\,\mathrm{ms}$ and $6\,\mathrm{ms}$ post merger.
{\it Bottom}: Same for scattering on protons.
Note the difference in the range of the color scale between the two sets of plots.}
\label{fig:rel_scat}
\end{figure*}

We now present the fraction of opacity due to different absorption and scattering reactions in Fig.~\ref{fig:rel_scat}. This will aid our assessment of the contribution from each reactions. We note that proton scattering is seen even in areas of high neutron richness. Specifically, in Fig~\ref{fig:rel_scat} at high densities and low temperatures, scattering on both neutrons and protons need to be considered. This might be attributed to the the fact that neutrino--nucleon scattering rates depend not only on the nucleon number densities but also on the underlying weak interaction couplings, which differ between neutrons and protons. 
Furthermore, nuclear many-body effects and nucleon correlations can modify effective scattering rates, altering the relative importance of neutron and proton contributions under these conditions. As a result, accurate modeling of neutrino transport in the inner remnant requires inclusion of scattering on both neutrons and protons. We also examined scattering on nuclei and $\alpha$ particles. In the latter case, we find subdominant contributions, at the $\sim 2\%$ level, while scattering on heavy nuclei becomes important at low temperatures and intermediate densities, $\rho \sim 10^{13},\mathrm{\,g\,cm^{-3}}$. Consequently, scattering is effectively dominated by free nucleons throughout most of the domain considered, and in the entirety of the simulation domain beyond the first few milliseconds of post-merger evolution. 

\subsection{Inelastic Scattering on Electrons}
\label{sec:inel}

Inelastic scattering of neutrinos on electrons is an important but comparatively underexplored component of neutrino transport in neutron star merger remnants. Unlike scattering on nucleons and nuclei, which is nearly elastic and primarily redistributes neutrino momentum, neutrino–electron scattering is intrinsically inelastic due to small electron mass compared to the typical neutrino energies and the nucleon mass. This enables efficient energy exchange between neutrinos and the surrounding fluid. Such interactions shape the neutrino spectra, regulate thermal decoupling, and influence energy deposition in the remnant and outflows. Despite its relevance, electron inelastic scattering hasn't been used in merger simulations, though some core collapse supernova (CCSNe) models include it~\cite{Mezzacappa:1993gn, Kotake:2018ypf}. This is because accurately modeling the energy-coupling kernel is computationally demanding. As modern simulations increasingly aim to capture detailed spectral transport and microphysical accuracy, it becomes essential to quantify the contribution of neutrino–electron scattering to the effective opacity. This can be especially important for heavy-lepton neutrinos, which lack strong charged-current absorption channels.

The rate of inelastic scattering for neutrinos of energy $\epsilon$ can be obtained from the NuLib tables using 
\beq
\kappa_{\rm inel} = \frac{1}{(hc)^3} \int d\epsilon' (\epsilon')^2 \int d\Omega' [1-f(\epsilon',\Omega')] R^{\rm out}(\epsilon,\epsilon',\mu')
\eeq
with $\epsilon'$ the energy of the neutrino post-scattering, $\Omega'$ its angular direction, and $\mu'=\cos\theta'$ with $\theta'$ the angle between the incoming and outgoing neutrinos. The kernel $R^{\rm out}$ is expanded into Legendre polynomials in $\mu'$. To estimate scattering rates from the limited information about the neutrino distribution function available in our Monte Carlo simulations, we leverage the fact that inelastic scattering is, in binary neutron star mergers, subdominant with respect to elastic scattering. Accordingly, we expect significant scatter rates only in regions where the neutrino distribution function is close to isotropic. In that case, we only use the first term in the expension of $R^{\rm out}$,
\beq
R^{\rm out}(\epsilon,\epsilon',\mu') = \frac{1}{2} \Phi^{\rm out}_0(\epsilon,\epsilon')
\eeq 
and estimate
\beq
\kappa_{\rm inel} = \frac{2\pi}{(hc)^3} \int d\epsilon' (\epsilon')^2 [1-\langle f(\epsilon')\rangle] \Phi_0^{\rm out} (\epsilon,\epsilon').
\eeq
with $\langle f(\epsilon')\rangle$ the angular-averaged value of the distribution function. We will use a value of $\langle f(\epsilon')\rangle$ obtained directly from a Monte Carlo simulation better designed for this purpose in Sec~\ref{sec:ooe}; here, as our simulation does not have Monte Carlo packets distributed in a way that allows for a measurement of $f_\nu$, we use the distribution function of neutrinos in equilibrium with the fluid (for which the assumption of an isotropic distribution is of course exact). The integral is evaluated by summation over the energy bins used in the Monte Carlo simulation, as NuLib provides $\Phi_0^{\rm out}$ with the same energy discretization. We note that as inelastic scattering is not included in the simulation itself, the calculation is not entirely self-consistent; but it provides a first estimate of the impact that inelastic scattering would have, should it be included.

\begin{figure*}[t]
    \centering
    \includegraphics[width=\textwidth]{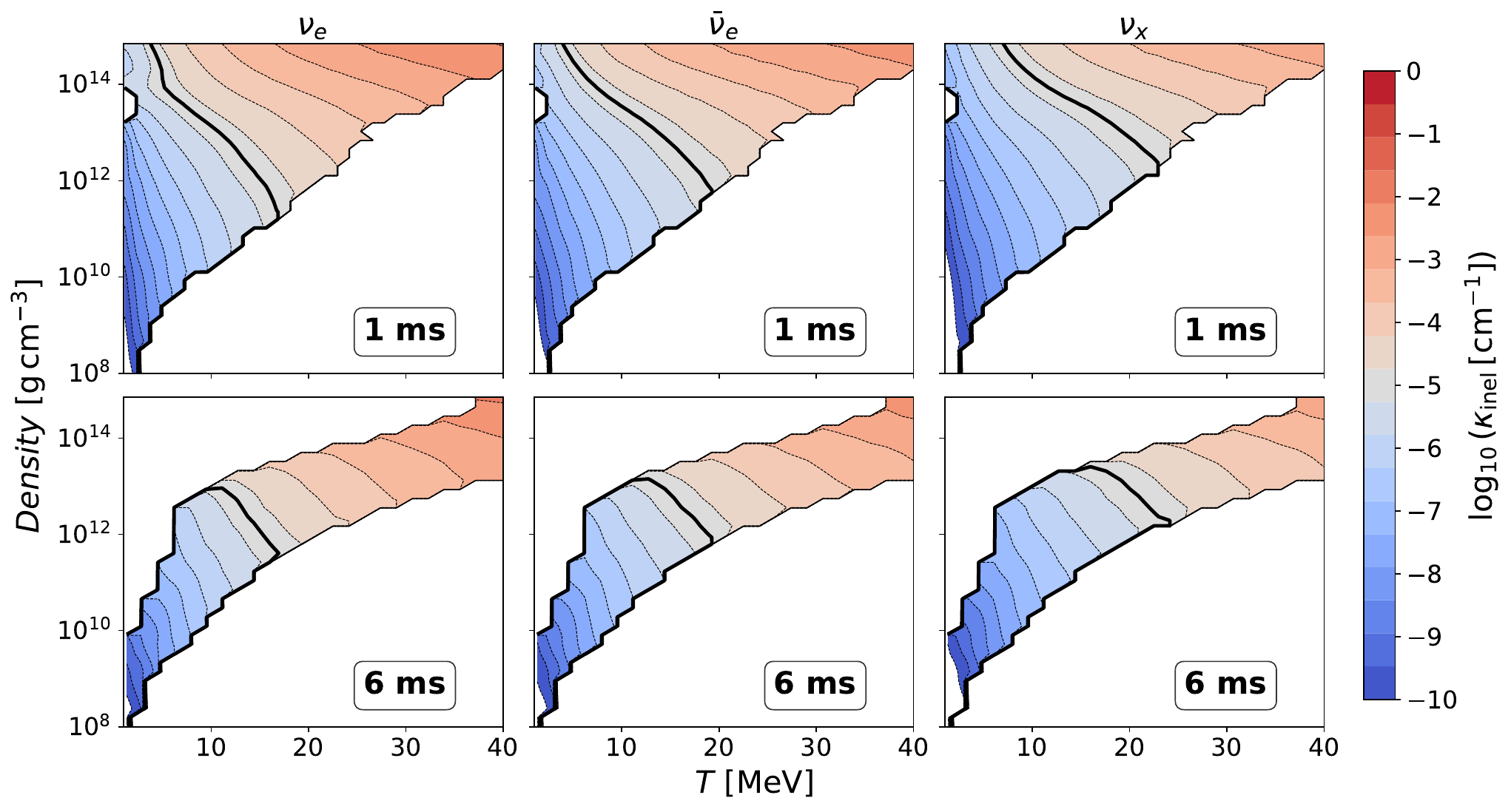}
 \caption{Average scattering opacity due to inelastic scattering of neutrinos on electrons, in log scale, at $1\,\mathrm{ms}$ and $6\,\mathrm{ms}$ post merger for $\nu_e$, $\bar{\nu}_e$ and $\nu_x$. In all plots, the solid black line corresponds to $\kappa_{inel}=10^{-5}\,{\rm cm}^{-1}$.}
\label{fig:inelas_scat}
\end{figure*}

Our results are shown in Fig.~\ref{fig:inelas_scat}. As inelastic scattering allows energy exchange between neutrinos and matter, it strengthens thermal equilibration. We can infer from Fig.~\ref{fig:inelas_scat} and a comparison with previous results for absorption opacities that inelastic scattering will be particularly important to the thermalization of heavy-lepton neutrinos in their decoupling regions. It can also impact the rate of thermalization of electron-type neutrinos in the cold high-density regions discussed in previous sections. We provide a more quantitative analysis of these results in the next section.

\subsection{Thermalization}
\label{sec:therm}

Neutrino transport is constituted by multiple channels of interaction that influence neutrino propagation in unique ways. So far, we have considered separately the opacity of the fluid to absorption, elastic scattering, and inelastic scattering interactions. However, it is also important to identify the surface at which freeze-out from thermal equilibrium occurs for a complete picture of neutrino-matter interactions. To this end, in the absence of inelastic scattering, we define the thermalization opacity, $\kappa_{\mathrm{th}}$, which is the geometric mean of the absorption opacity and the total opacity as:
\begin{align}
\kappa_{\mathrm{th}} = \sqrt{\kappa_a \left(\kappa_a + \kappa_s \right)} .
\end{align}
The region where $\kappa_{\mathrm{th}}\sim (1\,{\rm km})^{-1}$ approximately defines the thermalization surface, i.e. the location where neutrinos decouple thermally from the composition of the fluid. At this surface, neutrinos undergo their last effective energy-changing interactions, and their energy spectra are effectively set. Because only a subset of all reactions contributes to maintaining thermal equilibrium, thermalization surfaces lie deeper within the remnant than diffusion surfaces. 
With the inclusion of inelastic scattering, and under the simplifying assumption that inelastic scatterings on electrons are as efficient at redistributing energy than absorption/emission processes, the formula is modified to:
\begin{equation}
\kappa_{\mathrm{th}}
=
\sqrt{
\left(\kappa_a + \kappa_{\mathrm{inel}}\right)
\left(
\kappa_a + \kappa_{\mathrm{inel}} + \kappa_s
\right)}.
\label{eq:kappa_th_inel}
\end{equation}

\begin{figure*}[t]
    \centering
    \includegraphics[width=\textwidth]{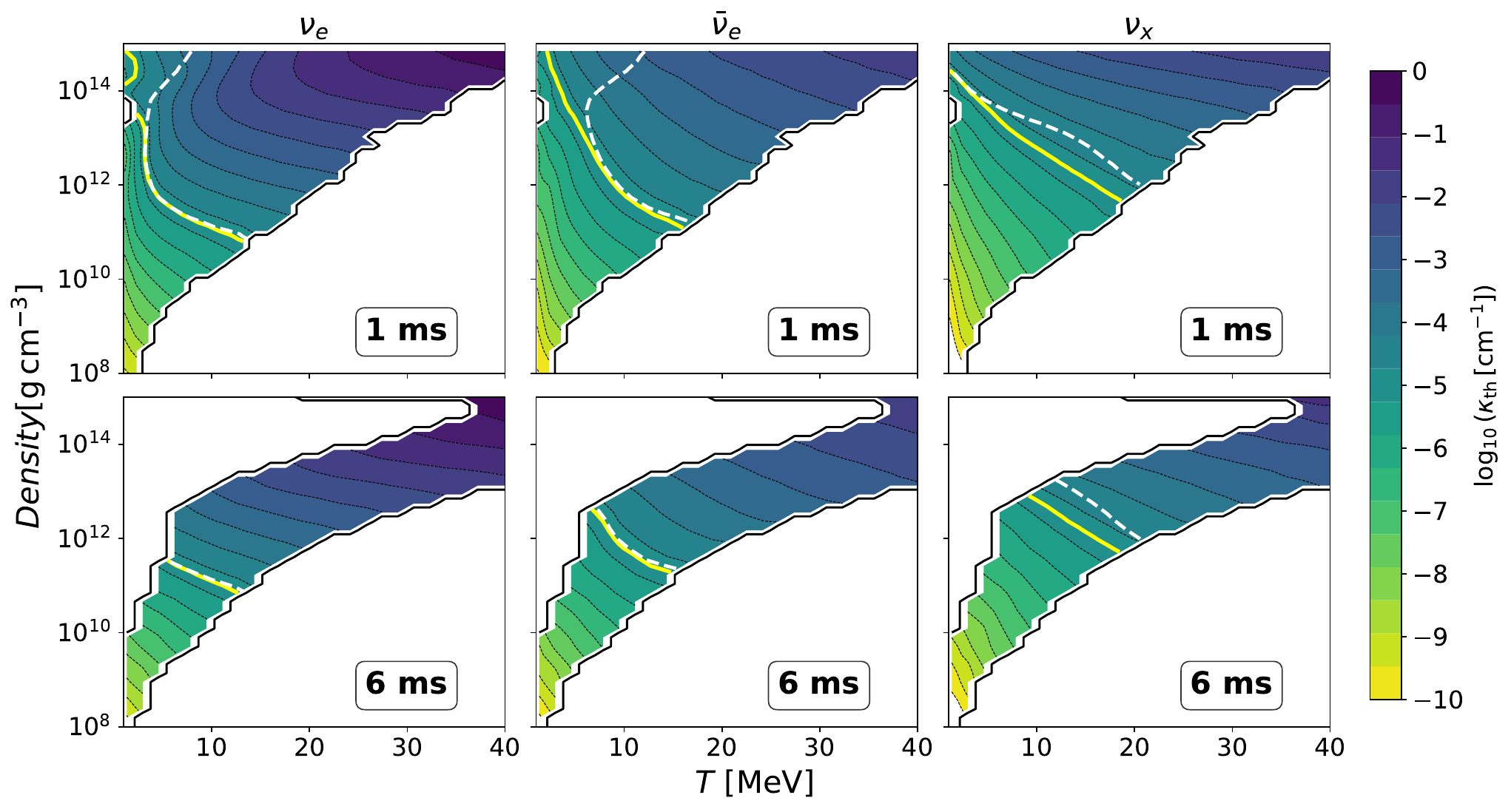}
\caption{Thermalization opacity (including inelastic scattering) shown in log scale at $1\,\mathrm{ms}$ (top) and $6\,\mathrm{ms}$ (bottom) post-merger for $\nu_e$ (left), $\bar{\nu}_e$ (middle) and $\nu_x$ (right). On each plot, we also show for reference the $\kappa=10^{-5}\,{\rm cm}^{-1}$ surface both with (solid yellow line) and without (dashed white line) inelastic scattering.}
\label{fig:therm_inelas}
\end{figure*}

Here, $\kappa_{\mathrm{th}}$ denotes the thermalization opacity. The absorption opacity is $\kappa_a$ and $\kappa_s$ represents the elastic scattering opacity (e.g., on nucleons and nuclei). $\kappa_{\mathrm{inel}}$ corresponds to the inelastic neutrino-electron scattering opacity. In Fig.~\ref{fig:therm_inelas}, we show the impact on the neutrinosphere due to $\kappa_{\mathrm{th}}$ evaluated according to Eq.~\ref{eq:kappa_th_inel}. We quantify the impact of inelastic neutrino-electron scattering on the thermalization opacity of heavy-lepton neutrinos ($\nu_x$) by examining the thermodynamic locations at which $\kappa_{\mathrm{th}} \approx 10^{-5}\,\mathrm{cm^{-1}}$, both with and without the inclusion of $\kappa_{\mathrm{inel}}$. At $1\,\mathrm{ms}$ post-merger, the addition of inelastic scattering systematically shifts the $\kappa_{\mathrm{th}}=10^{-5}\,{\rm cm}^{-1}$ contour toward lower densities. In the hot regime ($T \sim 18$--$20\,\mathrm{MeV}$), the density at which this threshold is reached decreases from $\rho \sim 1.4\times10^{12}\,\mathrm{g\,cm^{-3}}$ to $\rho \sim (5$--$9)\times10^{11}\,\mathrm{g\,cm^{-3}}$, corresponding to a factor of $\sim 2$--$3$ reduction. The effect is more pronounced at intermediate temperatures ($T \sim 10$--$12\,\mathrm{MeV}$), where the threshold density shifts from $\rho \sim 10^{13}\,\mathrm{g\,cm^{-3}}$ to a few $\times10^{12}\,\mathrm{g\,cm^{-3}}$, i.e. a factor of $\sim 4$--$5$ decrease. In contrast, the shift is modest in the cool, high-density regime ($T \lesssim 5\,\mathrm{MeV}$), where absorption and elastic scattering already dominate the coupling. 
A similar qualitative behavior is observed at $6\,\mathrm{ms}$. The inclusion of $\kappa_{\mathrm{inel}}$ again moves the thermalization contour to lower densities, indicating enhanced energy coupling in more optically thin regions. At the hottest end ($T \sim 19$--$21\,\mathrm{MeV}$), the density decreases by a factor of $\sim 1.5$--$2$, while at intermediate temperatures ($T \sim 12$--$16\,\mathrm{MeV}$) the shift is larger, reaching factors of $\sim 5$--$10$. Unlike the case of heavy-lepton neutrinos, inelastic neutrino-electron scattering does not substantially alter the thermodynamic location of the $\nu_e$ and $\bar{\nu}_e$ thermal decoupling layer. Including inelastic scattering shifts the contour toward slightly lower densities at comparable temperatures, typically at the level of tens of percent and at most a factor of order unity. The effect is modest, reflecting the fact that charged-current absorption remains an important energy-exchange channel for $\nu_e$ and  $\bar{\nu}_e$. However, in the cold, high density regions the inclusion of inelastic scattering enhances the thermalization opacity, giving approximately 4-5 orders of magntiude difference. We note that in those cold dense regions, the contour $\kappa_{\mathrm{th}} \approx 10^{-5}\,\mathrm{cm^{-1}}$ can no longer be interpreted as an approximate neutrinosphere, as these regions lie deep within the remnant and neutrinos cannot diffuse out of those regions before they disappear. Instead, we should see these regions as having thermalization time scales $\tau_{\rm th} \sim (\kappa_{\rm th} c)^{-1}$. We see that the inclusion of inelastic scattering brings $\tau_{\rm th}$ from values much higher than the dynamical time scale of the remnant to values comparable to that time scale ($\tau_{\rm th}\sim 1\,{\rm ms}$ corresponds to $\kappa_{\rm th}\sim 3\times 10^{-6}\,{\rm cm}^{-1}$).

\subsection{Out of equilibrium $f_\nu$}
\label{sec:ooe}

In a Monte Carlo simulation, we have direct access to a discretization of the distribution function of neutrinos -- though a very low resolution one. So low resolution in fact that, for the simulation discussed so far, a single packet can represent a change $\Delta f_\nu > 1000$ to the distribution function within a given energy bin and fluid cell. To obtain more accurate information about the impact of an out-of-equilibrium neutrino distribution on opacities, we performed a simulation of a neutron star merger with a modified weighting of the neutrino packets inspired by~\cite{Foucart:2025nub}; effectively, packets representing low-energy neutrinos are given lower weights (i.e. each packet represents fewer neutrinos), to limit $\Delta f_\nu \lesssim 0.1$ in the first few milliseconds post-merger. We consider a snapshot of the simulation $1.5\,{\rm ms}$ post-merger. Using this simulation, we re-evaluate all the energy-averaged opacities using the actual neutrino spectrum within the simulation. We also calculate the rates of inelastic scattering on electrons and $\nu\bar\nu \rightarrow e^+e^-$ pair annihilation i.e. the rates for which the assumption of an equilibrium distribution in our calculation of opacities is most problematic. For inelastic scattering, we proceed as in Sec.~\ref{sec:inel}, except that $\langle f_{\nu}(\epsilon')\rangle$ is estimated from the average energy of neutrinos within a grid cell and energy bin during a simulation time step. For pair annihilation, and under the same assumption as in Sec.~\ref{sec:inel}, we get
\beq
\kappa_{e^+e^-} =  \frac{2\pi \alpha_{\rm form}}{(hc)^3} \int d\epsilon' (\epsilon')^2 \langle \bar f(\epsilon')\rangle \Phi_0^{\rm a} (\epsilon,\epsilon')
\eeq
with $\bar f$ the distribution function of antineutrinos. The form factor $\alpha_{\rm form}$ is added here because even with modified packet weighting, resolving both the energy and angular distribution of neutrinos is beyond our current simulations. However, pair annihilation rates are the dominant channel for energy deposition in the polar regions, where neutrino distributions are very anisotropic, and these rates are sensitive to the relative orientation of neutrinos and antineutrinos. On average, in low-density regions where electron blocking factors are negligible, this correction is
\beq
\alpha_{\rm form} \approx \frac{T^{\mu\nu} \bar T_{\mu\nu}}{T^{\mu\nu}_{\rm iso} \bar T_{\mu\nu}^{\rm iso}}
\eeq 
with $T,\bar T$ the stress-energy tensor of neutrinos and antineutrinos, and $T_{\rm iso},\bar T_{\rm iso}$ the value they would take if the energy density of neutrinos in the fluid frame was the same, but the distribution was isotropic. We calculate a single value of $\alpha_{\rm form}$ for each grid cell and each neutrino species.
\begin{figure*}[t]
    \centering
    \includegraphics[width=\textwidth]{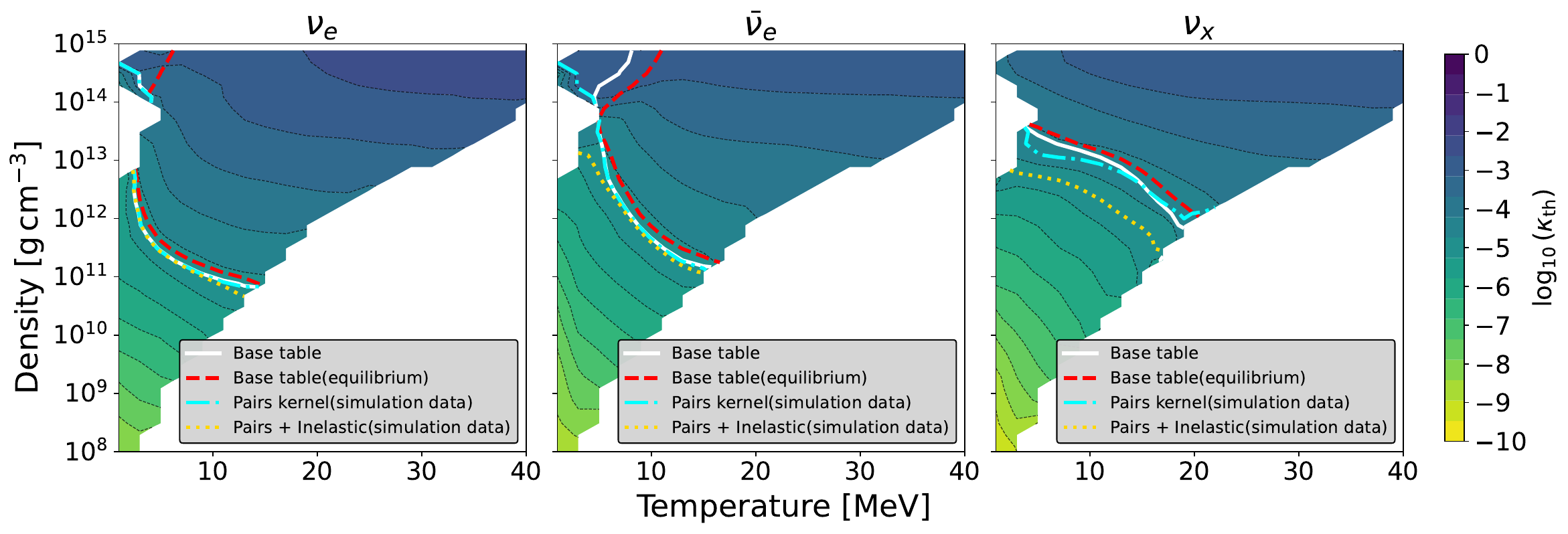}
    \caption{Thermalization opacity for $\nu_e$ (left), $\bar{\nu}_e$ (middle), and $\nu_x$ (right), shown over the density--temperature plane. The background colormap corresponds to the thermalization opacity computed using      the opacity prescription that includes both pair kernels and inelastic scattering. Overlaid contours indicate the approximate location of neutrinosphere for different opacity treatments: 1. Base table (no inelastic scattering, pairs only for $\nu_x$ using equilibrium distributions, with energy averaging weighted by the neutrino energy density in the simulation; 2. Same base table with gray treatment of table (energy averaging performed using equilibrium spectra); 3. the table including pair kernels, and 4. the table including both pair kernels and inelastic scattering, both using the energy distribution from the simulation.}
    \label{fig:comp_ooe}
\end{figure*}
While obviously approximate, these expressions capture the leading order corrections to the reaction rates: the dependence of pair annihilation on the energy spectrum of neutrinos and antineutrinos and the average effect of their angular distribution. 

We show results for the thermalization opacity in Fig.~\ref{fig:comp_ooe}, using four distinct methods to calculate opacities. We first consider the method used so far in this paper, weighting the opacities by the equilibrium energy density of neutrinos. In this setup, we ignore inelastic scattering and only include pair processes for $\nu_x$, as in our existing simulations. This is akin to the simplest method available to all moment schemes. We then consider the same set of reactions, but weighting opacities by the actual energy of neutrinos in the simulation. This provides an estimate of the effective opacities actually used by our Monte Carlo simulations. Finally, we include the additional effects of,first, a calculation of $e^+e^-$ pair processes using kernels (as described above), and then including both $e^+e^-$ pair processes and inelastic scattering using kernels.
The main effect of the correction in dense regions is on $\nu_x$: pair processes are more sensitive to neutrino energies than charged current reactions, and $\nu_x$ go through an extended region where absorption opacities are small but scattering opacities are high. In that region, neutrinos are out of thermal equilibrium with the fluid, and corrected pair annihilation rates are significantly higher than the original rates. As already discussed, inelastic scatterings also lead to thermalization of $\nu_x$ at lower densities/temperature. In regions where neutrinos are largely free-streaming, the main impact of our corrections is, unsurprisingly, to increase the rate of $\nu\bar\nu$ annihilation for all species. These are also the regions where the form factor $\alpha_{\rm form}$ plays the most important role, as in those regions neutrino distributions are very forward-peaked. Assuming an isotropic distribution of neutrinos overestimates the annihilation rate.
\begin{figure}
    \centering
    \includegraphics[width=0.48\textwidth]{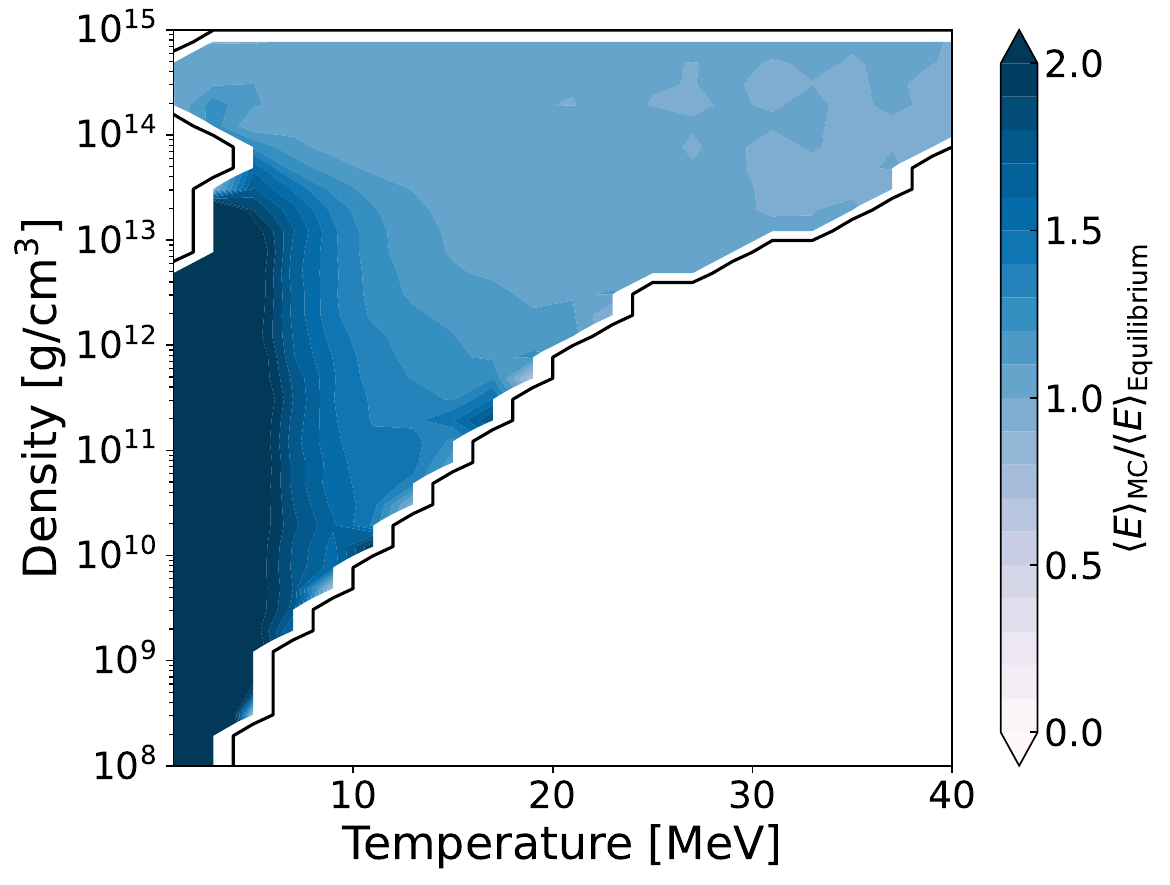}
    \caption{Ratio of average neutrino energies computed using the Monte Carlo energy spectrum over the equilibrium energy spectrum, shown over the density--temperature plane for $\nu_x$ (capped at a ratio of $2$).}
    \label{fig:ratio_en}
\end{figure}

\begin{figure*}
    \centering
    \includegraphics[width=0.48\textwidth]{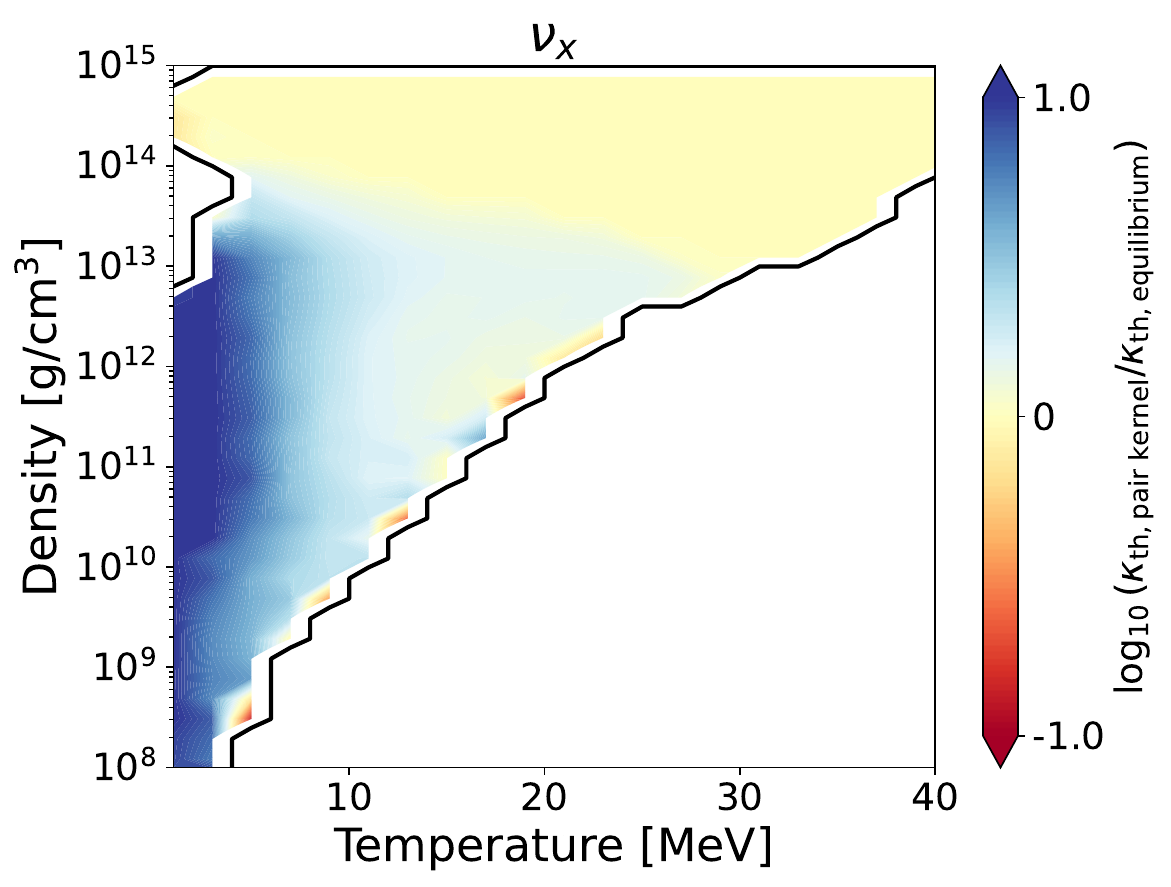}
    \hfill
    \includegraphics[width=0.48\textwidth]{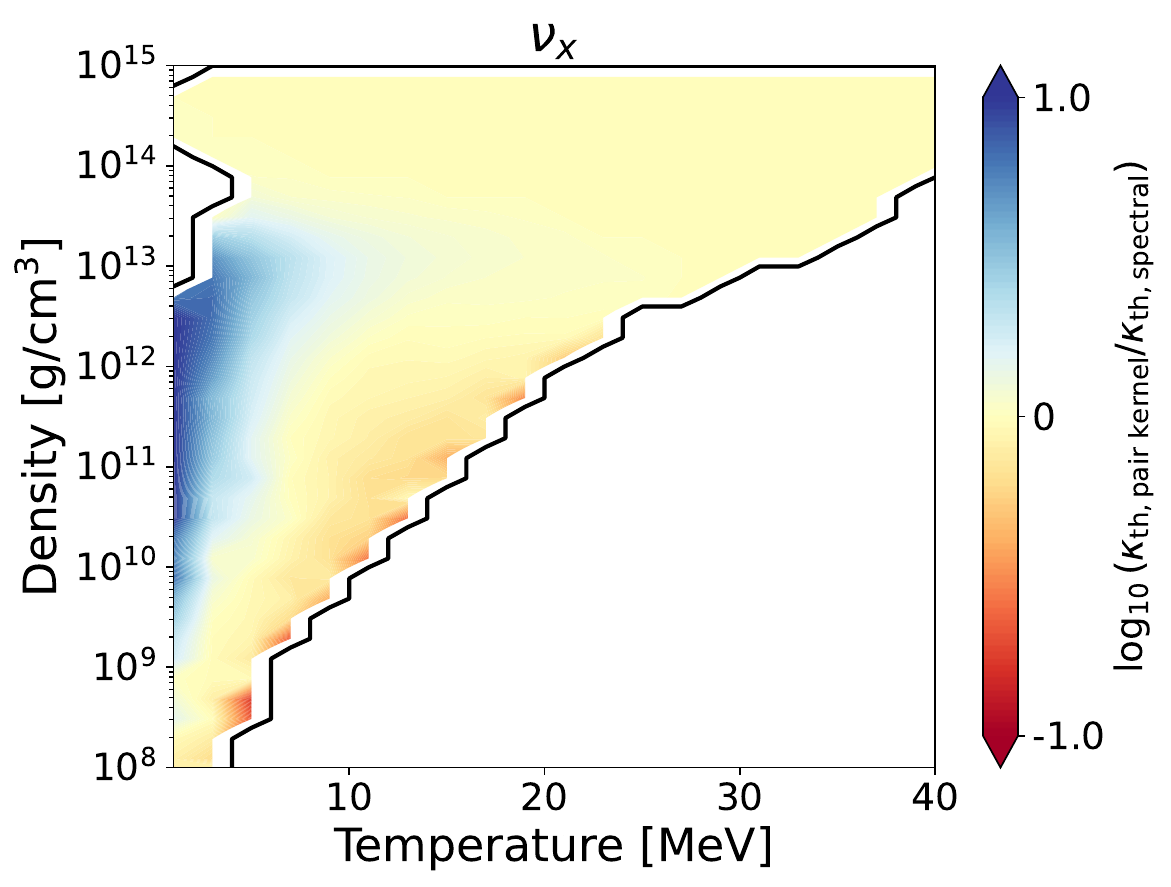}
    \caption{
    (left): Ratio of the thermalization opacity computed using kernels for $e^+e^-$ pair creation to the value computed assuming equilibrium distribution functions of neutrinos, for $\nu_x$.
    (right): Ratio of the thermalization opacity computed using kernels for $e^+e^-$ pair creation to the value computed using the energy spectrum from the simulations in averaging opacities, but an equilibrium distribution in the computation of the energy-dependent opacity (i.e. without using kernels).}
    \label{fig:ratio_pair}
\end{figure*}

The latter point is more visible in Fig.~\ref{fig:ratio_en} and  Fig.~\ref{fig:ratio_pair}. There, we present a comparison of the mean neutrino energy and the thermalization opacity across the density--temperature phase space for $\nu_x$.  Outside of the decoupling region, similar effects are observed for $\nu_e$ and $\bar{\nu}_e$. Fig.~\ref{fig:ratio_en} shows the ratio of the Monte Carlo mean energy to the equilibrium expectation, while the left panel of Fig.~\ref{fig:ratio_pair} shows the ratio of the thermalization opacity computed with the pair kernels and form factor to that computed without using kernels, and assuming an equilibrium distribution of neutrinos. For $\nu_x$, the inclusion of the pair kernels leads to significant enhancements of the thermalization opacities in nearly all regions where neutrinos are out of equilibrium with the fluid. This is, of course, particularly visible in low-temperature regions, where the energy of neutrinos in the simulation is much higher than the energy of neutrinos in thermal equilibrium with the fluid. The right panel of Fig~\ref{fig:ratio_pair} shows the ratio of the thermalization opacity computed with the pair kernels and form factor to its value without using kernels nor form factors, but accounting for the spectrum of neutrinos in the simulation (i.e. in the $\nu\bar\nu$ reaction, the absorption rate of $\nu$ is computed using the spectrum of $\nu$ in the simulation, but an equilibrium distribution for $\bar\nu$). This is again what Monte Carlo simulations do when they do not use kernels. We see that the effects there are more subtle. In some of the hotter, high-density regions, the reduction in absorption rate due to the form factor even reduces the interaction rates more than it is increased by the out-of-equilibrium energy spectrum of $\bar\nu$. In most regions however, the effect remains an increase in the interaction rate.

\subsection{Energy Dependent Analysis}
\begin{figure*}[t]
    \centering
    \includegraphics[width=0.98\textwidth,height=0.90\textheight,keepaspectratio]{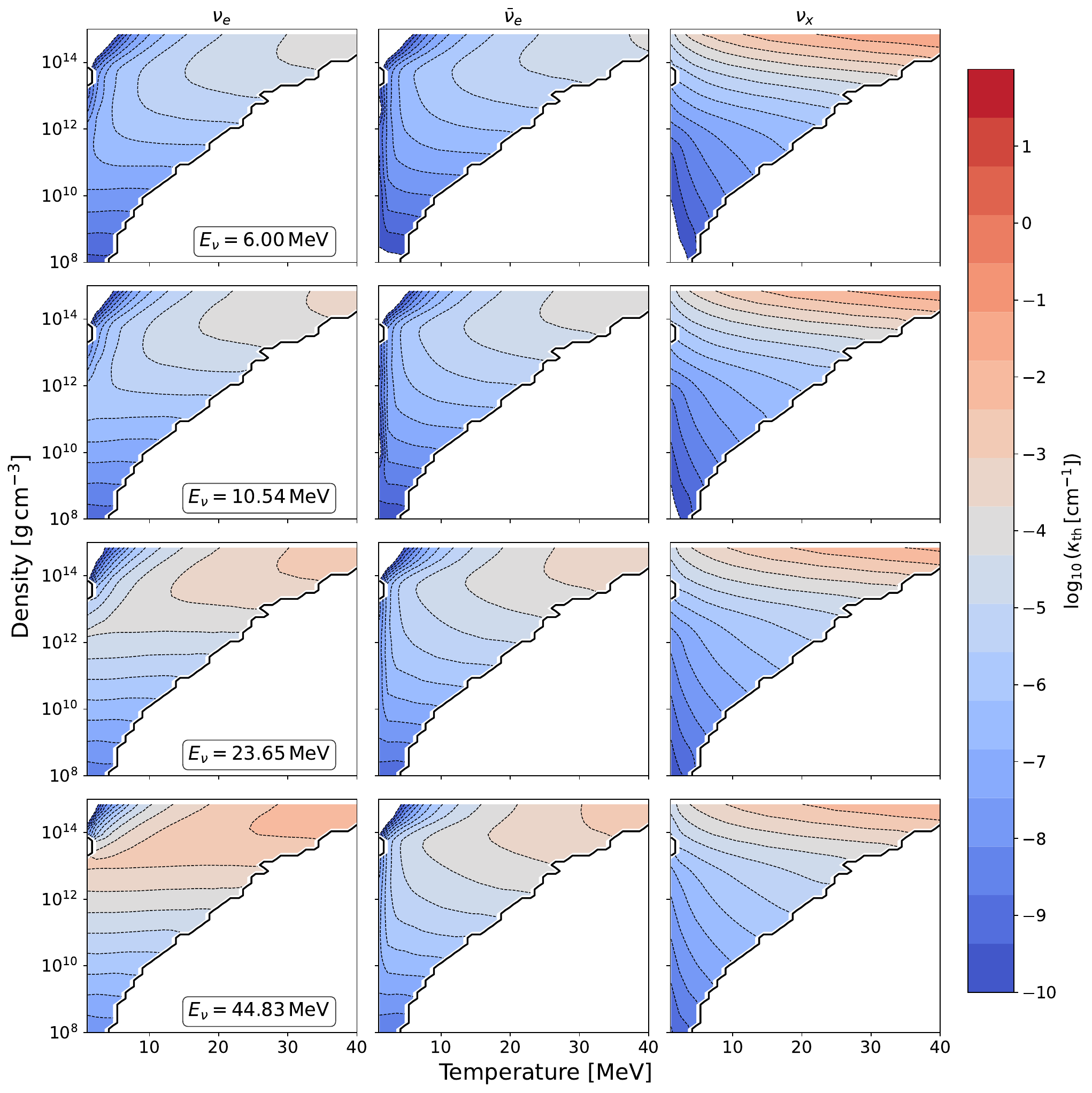}
    \caption{Energy-dependent contribution to the thermalization opacity at $1\,\mathrm{ms}$ post-merger. Rows correspond to increasing energy bins (1, 2, 4, 6 from top to bottom), while columns show electron neutrinos ($\nu_e$), electron antineutrinos ($\bar{\nu}_e$), and heavy-lepton neutrinos ($\nu_x$), respectively.}
    \label{fig:en_bin}
\end{figure*}

Figure~\ref{fig:en_bin} shows the energy-dependent thermalization opacity $\kappa_{\mathrm{th}}$ in the $\rho$--$T$ plane at $t=1\,\mathrm{ms}$ post-merger for $\nu_e$, $\bar{\nu}_e$, and $\nu_x$. The average energy of neutrinos emitted by neutron star mergers is $\sim 10-30\,{\rm MeV}$. We show here results for roughly $(6,10,23,45)\,{\rm MeV}$. Comparing the two lowest energy bins exhibits an enhancement in the dense, low-temperature region, indicating that higher-energy neutrinos decouple farther out than lower-energy neutrinos. We see that as we increase in energy, opacities rapidly increase, especially for $\nu_e$ and $\bar\nu_e$. We see that even at 10MeV (close to the average energy of escaping $\nu_e$), the mean free path of neutrinos inside the 'gray' neutrinosphere can be multiple kilometers. This highlights the limits of that energy-averaged analysis. We should not expect low-energy neutrinos to be in thermal equilibrium with the fluid at the estimated location of the neutrinosphere. On the other hand, higher energy neutrinos will only decouple well within the accretion disk ($\rho \sim 10^{11}\,{\rm g/cm^3}$ for $\nu_e,\bar\nu_e$). The highest energy neutrinos only decouple from the fluid in regions where the fluid temperatures are low enough that the equilibrium distribution of these neutrinos is heavily suppressed -- and accordingly, very few of those neutrinos escape the remnant. This is already visible for $45\,{\rm MeV}$ neutrinos here. Fig.~\ref{fig:en_bin} provides us with a visualization of the much broader range of densities and temperatures where most of the neutrinos emitted by merger remnants decouple thermally from the fluid. The heavy lepton neutrinos show the weakest dependence on energy variation, because $\kappa_s\gg \kappa_a$ for $\nu_x$ and scattering rates on nucleons are only weakly dependent on neutrino energies. 

\section{Discussion}

In this paper, we have investigated the relevance of different neutrino-matter interactions in binary neutron star merger remnants. This has been done by post-processing SpEC simulations of binary neutron stars with masses of $1.26\, M_\odot$ and $1.36\, M_\odot$. Those simulations used MC neutrino transport, and the SFHo equation of state. For that system, the remnant collapses to a black hole $6\,\mathrm{ms}$ after merger. We have quantified and examined opacities at three levels of refinement: energy-integrated (gray) opacities weighted by an equilibrium neutrino spectrum, energy-averaged opacities averaged over the simulated neutrino spectrum, and opacities at fixed neutrino energies. For inelastic scattering and $\nu + \bar{\nu} \leftrightarrow e^{+} + e^{-}$, we have also computed opacities using kernels and simulated distribution functions of both neutrinos involved in the reaction. Incorporating these approaches is especially relevant in neutron star mergers because the remnant contains regions spanning optically thick, semi-transparent, and free-streaming regimes simultaneously. The last two approaches are only possible thanks to SpEC's use of Monte Carlo transport to evolve neutrinos.

The energy-integrated analysis captures the global structure of neutrino decoupling and highlights the evolution of neutrinospheres as the merger remnant undergoes heating, neutrino emission, and contraction prior to black hole formation. The opacity maps provide a direct insight into the trapped, diffusive, and free-streaming regimes, identifying the regions where simulations are most sensitive to the underlying neutrino interaction rates. The results obtained within that approximation are found to be consistent with studies done previously using gray M1 transport~\cite{Endrizzi:2019trv}. This is corroborated by estimating the location of surface where the mean free path to absorption is $\lambda \sim 1\,{\rm km}$ (a typical lengthscale for the thermodynamical properties of the remnant to change), which is seen to be roughly between 5-15 MeV for $\nu_e$, between 10-15 MeV for $\bar{\nu}_e$ and between 1 - 25 MeV for $\nu_x$, depending on the density of the fluid. At late time, that surface is around density $\rho\sim 10^{11}\,{\rm g/cm^3}$ for $\nu_e$, $\rho\sim 10^{12}\,{\rm g/cm^3}$ for $\bar \nu_e$, and $\rho\sim 10^{13}\,{\rm g/cm^3}$ for $\nu_x$. Quasi-elastic scattering on the other hand, shows no specific hierarchy for any species. The region where the mean free path for scattering is $\sim 1\,{\rm km}$ is roughly between 5-15MeV and $\rho \sim 10^{11-12}\,\mathrm{g\,cm^{-3}}$. 

By analyzing the total opacity in terms of individual reaction channels, we identify the dominant interaction mechanisms for each neutrino species in different thermodynamical regimes. Charged-current processes are the primary contributors to absorption opacity for both $\nu_e$ and $\bar{\nu}_e$, reinforcing their central role in determining the electron fraction of the outflows and regulating cooling of the remnant. Production of electron-positron pairs and nucleon--nucleon Bremsstrahlung, while negligible for electron-flavor neutrinos, provide significant contributions to heavy-lepton neutrino opacity. Electron-positron pair production/annihilation dominates in moderately dense, high-temperature regions, while Bremsstrahlung becomes increasingly important at higher densities. We note that without the use of kernels, these absorption opacities are calculated from the emission rates of neutrinos by application of Kirchoff's law. This accurately represents the opacities as used in simulations today, not necessarily the correct reaction rates. For pair processes, this is in fact a better representation of which reaction dominates the emission of neutrinos, while absorption rates are simply chosen to get the correct equilibrium distribution function in simulations. The scattering opacity is dominated by neutrino interactions with nucleons throughout most of the remnant. In particular, scattering on both neutrons and protons contributes significantly to the total opacity in high-density, low-temperature regions where nucleon populations remain large but absorption processes are comparatively suppressed. These scattering interactions extend the diffusive regime of neutrinos beyond the thermal decoupling surface and play a key role in shaping the angular distribution and escape timescales of neutrinos. 

The method discussed so far to calculate opacities is essentially the treatment used for incorporating pair processes in the M1 scheme that do not attempt to correct for the expected energy spectrum of neutrinos. We note that under this approximation, the reaction rates themselves are, largely, inaccurate as soon as neutrinos are out of thermal equilibrium with the fluid. 

A key finding of this work comes from the inclusion of inelastic electron scattering in our opacity calculations. Inelastic scattering introduces energy exchange between neutrinos and the fluid and therefore directly impacts the thermalization of neutrinos. We find an overall increase in the thermalization opacity. This effect is particularly important for heavy-lepton neutrinos, whose interaction relies mostly on neutral-current interactions in the decoupling region. Thus, for $\nu_x$, inelastic scattering significantly extends the region over which neutrinos remain in energy equilibrium with the fluid. For $\nu_e$ and $\bar{\nu}_e$, inelastic scattering provides an additional channel for energy exchange in high density, low temperature regions where charged-current absorption becomes inefficient. While previously known to be important in CCSNe and hypermassive neutron stars, our work presents the first investigation of the impact on the decoupling layer when we include inelastic scattering. This will have significant impact on neutrino luminosities . The inclusion of this interaction in future simulations will be necessary to accurately capture energy exchanges between neutrinos and matter. 

We leverage the information available from MC simulations to assess the impact of relaxing the assumptions of equilibrium neutrino distribution. To this end, we first investigate changes introduced by using the simulated distribution function of neutrinos in the calculation of energy-averaged opacities, while still working with the interaction rates currently used in our simulations. These rates are calculated using the NuLib library, and assuming equilibrium distributions in blocking factors and whenever the distribution function of a second neutrino is required. This is thus a representation of the average opacities seen by neutrinos in our current MC simulations. As a second step, we then recalculate the reaction rates using kernels and the simulated energy distribution of neutrinos for $e^+e^-$ pair creation/annihilation and inelastic scattering on electrons. This is a view of what may become possible if kernels are used in simulations. 

In terms of effects on the decoupling regions, we find that for all species, there is a slight shift towards lower densities when average opacities are calculated using the simulated neutrino spectrum instead of the equilibrium spectrum. This can be interpreted as a consequence of the simulated neutrino spectrum having higher neutrino energy than the equilibrium spectrum. Further, this gives us an idea of how the choice of energy closure impacts M1 calculations. The full calculation of pair kernels and inelastic scattering from simulated distribution functions and kernels shows minor effects for $\nu_e$ and $\bar{\nu}_e$. For $\nu_x$, the inclusion of inelastic scattering remains the most important effect. In low density--high temperature regions, the equilibration timescale is affected by both the inclusion of pair kernels as well as from the simulated neutrino spectrum. Outside of the decoupling region, for all neutrinos, we see major differences from the inclusion of simulated energy spectra over the equilibrium spectra. For $\nu + \bar{\nu} \leftrightarrow e^{+} + e^{-}$, the effect of including pair annihilation rates calculated using kernels and the simulated distribution function is also significant but can lead to either an increase or decrease in reaction rate. This is partly due to the rates being reduced when neutrinos are propagating in nearly parallel directions. We note that the latter effect is only approximately accounted for in our analysis. We can gauge the sensitivity of the reaction rates to the true neutrino distribution function from these results. 

Finally, we note that the usage of energy--averaged opacities are an approximation and it is important to quantify the contribution from discrete energy bins. By visualizing the opacity of relatively low energy neutrinos ($6-10\,{\rm MeV}$), we see that they are out of thermal equilibrium with the fluid well inside of the decoupling region estimated from energy-averaged opacities. This demonstrates that interaction rates for lower energy neutrinos need to be accurately computed at densities higher than that of the decoupling regions.

Our study represents an assessment of which neutrino-matter interactions matter in the thermodynamic phase space sampled by merger simulations and how their contribution varies under several different conditions. While not truly complete, a series of such studies are imperative in our requirement for consistent inclusion of neutrino-matter interactions under a broad range of nuclear conditions and to ultimately gauge the composition of the outflows which determines the nucleosynthesis yields and electromagnetic signatures of binary neutron star mergers. Even subtle changes to the locations of the decoupling regions compel us to strive towards more consistent and exact reaction rates.

\begin{acknowledgments}
S.R. and F.F. acknowledge support from the Department of Energy, Office of Science, Office of Nuclear Physics, under contract number DE-SC0020435. L.K. acknowledges support from the National Science Foundation under Grants No.~PHY-2407742; No.~PHY-2207342; and No.~OAC-2513338 at Cornell. M.S. acknowledges support from NSF grants PHY-2309211, PHY-2309231, and OAC-2513339, and NASA award 80NSSC26K0340 at Caltech. L.K and M.S. also thank the Sherman Fairchild Foundation for their support.
\end{acknowledgments}

\appendix

\bibliography{References/paper_SamanthaNotes}

\begin{thebibliography}{75}%
\makeatletter
\providecommand \@ifxundefined [1]{%
 \@ifx{#1\undefined}
}%
\providecommand \@ifnum [1]{%
 \ifnum #1\expandafter \@firstoftwo
 \else \expandafter \@secondoftwo
 \fi
}%
\providecommand \@ifx [1]{%
 \ifx #1\expandafter \@firstoftwo
 \else \expandafter \@secondoftwo
 \fi
}%
\providecommand \natexlab [1]{#1}%
\providecommand \enquote  [1]{``#1''}%
\providecommand \bibnamefont  [1]{#1}%
\providecommand \bibfnamefont [1]{#1}%
\providecommand \citenamefont [1]{#1}%
\providecommand \href@noop [0]{\@secondoftwo}%
\providecommand \href [0]{\begingroup \@sanitize@url \@href}%
\providecommand \@href[1]{\@@startlink{#1}\@@href}%
\providecommand \@@href[1]{\endgroup#1\@@endlink}%
\providecommand \@sanitize@url [0]{\catcode `\\12\catcode `\$12\catcode
  `\&12\catcode `\#12\catcode `\^12\catcode `\_12\catcode `\%12\relax}%
\providecommand \@@startlink[1]{}%
\providecommand \@@endlink[0]{}%
\providecommand \url  [0]{\begingroup\@sanitize@url \@url }%
\providecommand \@url [1]{\endgroup\@href {#1}{\urlprefix }}%
\providecommand \urlprefix  [0]{URL }%
\providecommand \Eprint [0]{\href }%
\providecommand \doibase [0]{https://doi.org/}%
\providecommand \selectlanguage [0]{\@gobble}%
\providecommand \bibinfo  [0]{\@secondoftwo}%
\providecommand \bibfield  [0]{\@secondoftwo}%
\providecommand \translation [1]{[#1]}%
\providecommand \BibitemOpen [0]{}%
\providecommand \bibitemStop [0]{}%
\providecommand \bibitemNoStop [0]{.\EOS\space}%
\providecommand \EOS [0]{\spacefactor3000\relax}%
\providecommand \BibitemShut  [1]{\csname bibitem#1\endcsname}%
\let\auto@bib@innerbib\@empty
%</preamble>
\bibitem [{\citenamefont {Rosswog}(2015)}]{Rosswog:2015nja}%
  \BibitemOpen
  \bibfield  {author} {\bibinfo {author} {\bibfnamefont {S.}~\bibnamefont
  {Rosswog}},\ }\bibfield  {title} {\bibinfo {title} {{The multi-messenger
  picture of compact binary mergers}},\ }\href
  {https://doi.org/10.1142/S0218271815300128} {\bibfield  {journal} {\bibinfo
  {journal} {Int. J. Mod. Phys. D}\ }\textbf {\bibinfo {volume} {24}},\
  \bibinfo {pages} {1530012} (\bibinfo {year} {2015})},\ \Eprint
  {https://arxiv.org/abs/1501.02081} {arXiv:1501.02081 [astro-ph.HE]}
  \BibitemShut {NoStop}%
\bibitem [{\citenamefont {Baiotti}\ and\ \citenamefont
  {Rezzolla}(2017)}]{Baiotti:2016qnr}%
  \BibitemOpen
  \bibfield  {author} {\bibinfo {author} {\bibfnamefont {L.}~\bibnamefont
  {Baiotti}}\ and\ \bibinfo {author} {\bibfnamefont {L.}~\bibnamefont
  {Rezzolla}},\ }\bibfield  {title} {\bibinfo {title} {{Binary neutron star
  mergers: a review of Einstein's richest laboratory}},\ }\href
  {https://doi.org/10.1088/1361-6633/aa67bb} {\bibfield  {journal} {\bibinfo
  {journal} {Rept. Prog. Phys.}\ }\textbf {\bibinfo {volume} {80}},\ \bibinfo
  {pages} {096901} (\bibinfo {year} {2017})},\ \Eprint
  {https://arxiv.org/abs/1607.03540} {arXiv:1607.03540 [gr-qc]} \BibitemShut
  {NoStop}%
\bibitem [{\citenamefont {Barack}\ \emph {et~al.}(2019)\citenamefont {Barack}
  \emph {et~al.}}]{Barack:2018yly}%
  \BibitemOpen
  \bibfield  {author} {\bibinfo {author} {\bibfnamefont {L.}~\bibnamefont
  {Barack}} \emph {et~al.},\ }\bibfield  {title} {\bibinfo {title} {{Black
  holes, gravitational waves and fundamental physics: a roadmap}},\ }\href
  {https://doi.org/10.1088/1361-6382/ab0587} {\bibfield  {journal} {\bibinfo
  {journal} {Class. Quant. Grav.}\ }\textbf {\bibinfo {volume} {36}},\ \bibinfo
  {pages} {143001} (\bibinfo {year} {2019})},\ \Eprint
  {https://arxiv.org/abs/1806.05195} {arXiv:1806.05195 [gr-qc]} \BibitemShut
  {NoStop}%
\bibitem [{\citenamefont {Bernuzzi}(2020)}]{Bernuzzi:2020tgt}%
  \BibitemOpen
  \bibfield  {author} {\bibinfo {author} {\bibfnamefont {S.}~\bibnamefont
  {Bernuzzi}},\ }\bibfield  {title} {\bibinfo {title} {{Correction: Neutron
  star merger remnants [doi: 10.1007/s10714-020-02752-5]}},\ }\href
  {https://doi.org/10.1007/s10714-024-03291-z} {\bibfield  {journal} {\bibinfo
  {journal} {Gen. Rel. Grav.}\ }\textbf {\bibinfo {volume} {52}},\ \bibinfo
  {pages} {108} (\bibinfo {year} {2020})},\ \Eprint
  {https://arxiv.org/abs/2004.06419} {arXiv:2004.06419 [astro-ph.HE]}
  \BibitemShut {NoStop}%
\bibitem [{\citenamefont {Dhani}\ \emph {et~al.}(2025)\citenamefont {Dhani},
  \citenamefont {Camilletti}, \citenamefont {Radice}, \citenamefont {Kashyap},
  \citenamefont {Sathyaprakash}, \citenamefont {Logoteta},\ and\ \citenamefont
  {Perego}}]{Dhani:2025axt}%
  \BibitemOpen
  \bibfield  {author} {\bibinfo {author} {\bibfnamefont {A.}~\bibnamefont
  {Dhani}}, \bibinfo {author} {\bibfnamefont {A.}~\bibnamefont {Camilletti}},
  \bibinfo {author} {\bibfnamefont {D.}~\bibnamefont {Radice}}, \bibinfo
  {author} {\bibfnamefont {R.}~\bibnamefont {Kashyap}}, \bibinfo {author}
  {\bibfnamefont {B.}~\bibnamefont {Sathyaprakash}}, \bibinfo {author}
  {\bibfnamefont {D.}~\bibnamefont {Logoteta}},\ and\ \bibinfo {author}
  {\bibfnamefont {A.}~\bibnamefont {Perego}},\ }\bibfield  {title} {\bibinfo
  {title} {{Remnant properties of binary neutron star mergers undergoing prompt
  collapse}},\ }\href@noop {} {\bibfield  {journal} {\bibinfo  {journal}
  {arXiv}\ } (\bibinfo {year} {2025})},\ \Eprint
  {https://arxiv.org/abs/2507.19431} {arXiv:2507.19431 [gr-qc]} \BibitemShut
  {NoStop}%
\bibitem [{\citenamefont {Hotokezaka}\ \emph {et~al.}(2011)\citenamefont
  {Hotokezaka}, \citenamefont {Kyutoku}, \citenamefont {Okawa}, \citenamefont
  {Shibata},\ and\ \citenamefont {Kiuchi}}]{Hotokezaka:2011dh}%
  \BibitemOpen
  \bibfield  {author} {\bibinfo {author} {\bibfnamefont {K.}~\bibnamefont
  {Hotokezaka}}, \bibinfo {author} {\bibfnamefont {K.}~\bibnamefont {Kyutoku}},
  \bibinfo {author} {\bibfnamefont {H.}~\bibnamefont {Okawa}}, \bibinfo
  {author} {\bibfnamefont {M.}~\bibnamefont {Shibata}},\ and\ \bibinfo {author}
  {\bibfnamefont {K.}~\bibnamefont {Kiuchi}},\ }\bibfield  {title} {\bibinfo
  {title} {{Binary Neutron Star Mergers: Dependence on the Nuclear Equation of
  State}},\ }\href {https://doi.org/10.1103/PhysRevD.83.124008} {\bibfield
  {journal} {\bibinfo  {journal} {Phys. Rev. D}\ }\textbf {\bibinfo {volume}
  {83}},\ \bibinfo {pages} {124008} (\bibinfo {year} {2011})},\ \Eprint
  {https://arxiv.org/abs/1105.4370} {arXiv:1105.4370 [astro-ph.HE]}
  \BibitemShut {NoStop}%
\bibitem [{\citenamefont {Bauswein}\ \emph {et~al.}(2013)\citenamefont
  {Bauswein}, \citenamefont {Baumgarte},\ and\ \citenamefont
  {Janka}}]{Bauswein:2013jpa}%
  \BibitemOpen
  \bibfield  {author} {\bibinfo {author} {\bibfnamefont {A.}~\bibnamefont
  {Bauswein}}, \bibinfo {author} {\bibfnamefont {T.~W.}\ \bibnamefont
  {Baumgarte}},\ and\ \bibinfo {author} {\bibfnamefont {H.~T.}\ \bibnamefont
  {Janka}},\ }\bibfield  {title} {\bibinfo {title} {{Prompt merger collapse and
  the maximum mass of neutron stars}},\ }\href
  {https://doi.org/10.1103/PhysRevLett.111.131101} {\bibfield  {journal}
  {\bibinfo  {journal} {Phys. Rev. Lett.}\ }\textbf {\bibinfo {volume} {111}},\
  \bibinfo {pages} {131101} (\bibinfo {year} {2013})},\ \Eprint
  {https://arxiv.org/abs/1307.5191} {arXiv:1307.5191 [astro-ph.SR]}
  \BibitemShut {NoStop}%
\bibitem [{\citenamefont {Hotokezaka}\ \emph {et~al.}(2013)\citenamefont
  {Hotokezaka}, \citenamefont {Kiuchi}, \citenamefont {Kyutoku}, \citenamefont
  {Okawa}, \citenamefont {Sekiguchi}, \citenamefont {Shibata},\ and\
  \citenamefont {Taniguchi}}]{hotokezaka2013mass}%
  \BibitemOpen
  \bibfield  {author} {\bibinfo {author} {\bibfnamefont {K.}~\bibnamefont
  {Hotokezaka}}, \bibinfo {author} {\bibfnamefont {K.}~\bibnamefont {Kiuchi}},
  \bibinfo {author} {\bibfnamefont {K.}~\bibnamefont {Kyutoku}}, \bibinfo
  {author} {\bibfnamefont {H.}~\bibnamefont {Okawa}}, \bibinfo {author}
  {\bibfnamefont {Y.-i.}\ \bibnamefont {Sekiguchi}}, \bibinfo {author}
  {\bibfnamefont {M.}~\bibnamefont {Shibata}},\ and\ \bibinfo {author}
  {\bibfnamefont {K.}~\bibnamefont {Taniguchi}},\ }\bibfield  {title} {\bibinfo
  {title} {Mass ejection from the merger of binary neutron stars},\ }\href@noop
  {} {\bibfield  {journal} {\bibinfo  {journal} {Physical Review D—Particles,
  Fields, Gravitation, and Cosmology}\ }\textbf {\bibinfo {volume} {87}},\
  \bibinfo {pages} {024001} (\bibinfo {year} {2013})}\BibitemShut {NoStop}%
\bibitem [{\citenamefont {Symbalisty}\ and\ \citenamefont
  {Schramm}(1982)}]{Symbalisty:1982ni}%
  \BibitemOpen
  \bibfield  {author} {\bibinfo {author} {\bibfnamefont {E.}~\bibnamefont
  {Symbalisty}}\ and\ \bibinfo {author} {\bibfnamefont {D.~N.}\ \bibnamefont
  {Schramm}},\ }\bibfield  {title} {\bibinfo {title} {Neutron star collisions
  and the r-process},\ }\href {https://www.osti.gov/biblio/6636544} {\bibfield
  {journal} {\bibinfo  {journal} {Astrophys. Lett.; (United States)}\ }\textbf
  {\bibinfo {volume} {22:4}} (\bibinfo {year} {1982})}\BibitemShut {NoStop}%
\bibitem [{\citenamefont {Thielemann}\ \emph {et~al.}(2017)\citenamefont
  {Thielemann}, \citenamefont {Eichler}, \citenamefont {Panov},\ and\
  \citenamefont {Wehmeyer}}]{Thielemann:2017acv}%
  \BibitemOpen
  \bibfield  {author} {\bibinfo {author} {\bibfnamefont {F.~K.}\ \bibnamefont
  {Thielemann}}, \bibinfo {author} {\bibfnamefont {M.}~\bibnamefont {Eichler}},
  \bibinfo {author} {\bibfnamefont {I.~V.}\ \bibnamefont {Panov}},\ and\
  \bibinfo {author} {\bibfnamefont {B.}~\bibnamefont {Wehmeyer}},\ }\bibfield
  {title} {\bibinfo {title} {{Neutron Star Mergers and Nucleosynthesis of Heavy
  Elements}},\ }\href {https://doi.org/10.1146/annurev-nucl-101916-123246}
  {\bibfield  {journal} {\bibinfo  {journal} {Ann. Rev. Nucl. Part. Sci.}\
  }\textbf {\bibinfo {volume} {67}},\ \bibinfo {pages} {253} (\bibinfo {year}
  {2017})},\ \Eprint {https://arxiv.org/abs/1710.02142} {arXiv:1710.02142
  [astro-ph.HE]} \BibitemShut {NoStop}%
\bibitem [{\citenamefont {Li}\ and\ \citenamefont
  {Paczynski}(1998)}]{Li:1998bw}%
  \BibitemOpen
  \bibfield  {author} {\bibinfo {author} {\bibfnamefont {L.-X.}\ \bibnamefont
  {Li}}\ and\ \bibinfo {author} {\bibfnamefont {B.}~\bibnamefont {Paczynski}},\
  }\bibfield  {title} {\bibinfo {title} {{Transient events from neutron star
  mergers}},\ }\href {https://doi.org/10.1086/311680} {\bibfield  {journal}
  {\bibinfo  {journal} {Astrophys. J. Lett.}\ }\textbf {\bibinfo {volume}
  {507}},\ \bibinfo {pages} {L59} (\bibinfo {year} {1998})},\ \Eprint
  {https://arxiv.org/abs/astro-ph/9807272} {arXiv:astro-ph/9807272}
  \BibitemShut {NoStop}%
\bibitem [{\citenamefont {Metzger}\ \emph {et~al.}(2010)\citenamefont
  {Metzger}, \citenamefont {Mart{\'\i}nez-Pinedo}, \citenamefont {Darbha},
  \citenamefont {Quataert}, \citenamefont {Arcones}, \citenamefont {Kasen},
  \citenamefont {Thomas}, \citenamefont {Nugent}, \citenamefont {Panov},\ and\
  \citenamefont {Zinner}}]{Metzger:2010sy}%
  \BibitemOpen
  \bibfield  {author} {\bibinfo {author} {\bibfnamefont {B.~D.}\ \bibnamefont
  {Metzger}}, \bibinfo {author} {\bibfnamefont {G.}~\bibnamefont
  {Mart{\'\i}nez-Pinedo}}, \bibinfo {author} {\bibfnamefont {S.}~\bibnamefont
  {Darbha}}, \bibinfo {author} {\bibfnamefont {E.}~\bibnamefont {Quataert}},
  \bibinfo {author} {\bibfnamefont {A.}~\bibnamefont {Arcones}}, \bibinfo
  {author} {\bibfnamefont {D.}~\bibnamefont {Kasen}}, \bibinfo {author}
  {\bibfnamefont {R.}~\bibnamefont {Thomas}}, \bibinfo {author} {\bibfnamefont
  {P.}~\bibnamefont {Nugent}}, \bibinfo {author} {\bibfnamefont {I.~V.}\
  \bibnamefont {Panov}},\ and\ \bibinfo {author} {\bibfnamefont {N.~T.}\
  \bibnamefont {Zinner}},\ }\bibfield  {title} {\bibinfo {title}
  {Electromagnetic counterparts of compact object mergers powered by the
  radioactive decay of r-process nuclei},\ }\href
  {https://doi.org/10.1111/j.1365-2966.2010.16864.x} {\bibfield  {journal}
  {\bibinfo  {journal} {Mon. Not. Roy. Astron. Soc.}\ }\textbf {\bibinfo
  {volume} {406}},\ \bibinfo {pages} {2650} (\bibinfo {year}
  {2010})}\BibitemShut {NoStop}%
\bibitem [{\citenamefont {Tanaka}(2016)}]{Tanaka:2016sbx}%
  \BibitemOpen
  \bibfield  {author} {\bibinfo {author} {\bibfnamefont {M.}~\bibnamefont
  {Tanaka}},\ }\bibfield  {title} {\bibinfo {title} {{Kilonova/Macronova
  Emission from Compact Binary Mergers}},\ }\href
  {https://doi.org/10.1155/2016/6341974} {\bibfield  {journal} {\bibinfo
  {journal} {Adv. Astron.}\ }\textbf {\bibinfo {volume} {2016}},\ \bibinfo
  {pages} {6341974} (\bibinfo {year} {2016})},\ \Eprint
  {https://arxiv.org/abs/1605.07235} {arXiv:1605.07235 [astro-ph.HE]}
  \BibitemShut {NoStop}%
\bibitem [{\citenamefont {Metzger}(2017)}]{Metzger:2016pju}%
  \BibitemOpen
  \bibfield  {author} {\bibinfo {author} {\bibfnamefont {B.~D.}\ \bibnamefont
  {Metzger}},\ }\bibfield  {title} {\bibinfo {title} {{Kilonovae}},\ }\href
  {https://doi.org/10.1007/s41114-017-0006-z} {\bibfield  {journal} {\bibinfo
  {journal} {Living Rev. Rel.}\ }\textbf {\bibinfo {volume} {20}},\ \bibinfo
  {pages} {3} (\bibinfo {year} {2017})},\ \Eprint
  {https://arxiv.org/abs/1610.09381} {arXiv:1610.09381 [astro-ph.HE]}
  \BibitemShut {NoStop}%
\bibitem [{\citenamefont {Barnes}\ and\ \citenamefont
  {Kasen}(2013)}]{Barnes:2013wka}%
  \BibitemOpen
  \bibfield  {author} {\bibinfo {author} {\bibfnamefont {J.}~\bibnamefont
  {Barnes}}\ and\ \bibinfo {author} {\bibfnamefont {D.}~\bibnamefont {Kasen}},\
  }\bibfield  {title} {\bibinfo {title} {{Effect of a High Opacity on the Light
  Curves of Radioactively Powered Transients from Compact Object Mergers}},\
  }\href {https://doi.org/10.1088/0004-637X/775/1/18} {\bibfield  {journal}
  {\bibinfo  {journal} {Astrophys. J.}\ }\textbf {\bibinfo {volume} {775}},\
  \bibinfo {pages} {18} (\bibinfo {year} {2013})},\ \Eprint
  {https://arxiv.org/abs/1303.5787} {arXiv:1303.5787 [astro-ph.HE]}
  \BibitemShut {NoStop}%
\bibitem [{\citenamefont {Kullmann}\ \emph {et~al.}(2022)\citenamefont
  {Kullmann}, \citenamefont {Goriely}, \citenamefont {Just}, \citenamefont
  {Ardevol-Pulpillo}, \citenamefont {Bauswein},\ and\ \citenamefont
  {Janka}}]{Kullmann:2021gvo}%
  \BibitemOpen
  \bibfield  {author} {\bibinfo {author} {\bibfnamefont {I.}~\bibnamefont
  {Kullmann}}, \bibinfo {author} {\bibfnamefont {S.}~\bibnamefont {Goriely}},
  \bibinfo {author} {\bibfnamefont {O.}~\bibnamefont {Just}}, \bibinfo {author}
  {\bibfnamefont {R.}~\bibnamefont {Ardevol-Pulpillo}}, \bibinfo {author}
  {\bibfnamefont {A.}~\bibnamefont {Bauswein}},\ and\ \bibinfo {author}
  {\bibfnamefont {H.~T.}\ \bibnamefont {Janka}},\ }\bibfield  {title} {\bibinfo
  {title} {{Dynamical ejecta of neutron star mergers with nucleonic weak
  processes I: nucleosynthesis}},\ }\href
  {https://doi.org/10.1093/mnras/stab3393} {\bibfield  {journal} {\bibinfo
  {journal} {Mon. Not. Roy. Astron. Soc.}\ }\textbf {\bibinfo {volume} {510}},\
  \bibinfo {pages} {2804} (\bibinfo {year} {2022})},\ \Eprint
  {https://arxiv.org/abs/2109.02509} {arXiv:2109.02509 [astro-ph.HE]}
  \BibitemShut {NoStop}%
\bibitem [{\citenamefont {Just}\ \emph {et~al.}(2022)\citenamefont {Just},
  \citenamefont {Kullmann}, \citenamefont {Goriely}, \citenamefont {Bauswein},
  \citenamefont {Janka},\ and\ \citenamefont {Collins}}]{Just:2021vzy}%
  \BibitemOpen
  \bibfield  {author} {\bibinfo {author} {\bibfnamefont {O.}~\bibnamefont
  {Just}}, \bibinfo {author} {\bibfnamefont {I.}~\bibnamefont {Kullmann}},
  \bibinfo {author} {\bibfnamefont {S.}~\bibnamefont {Goriely}}, \bibinfo
  {author} {\bibfnamefont {A.}~\bibnamefont {Bauswein}}, \bibinfo {author}
  {\bibfnamefont {H.-T.}\ \bibnamefont {Janka}},\ and\ \bibinfo {author}
  {\bibfnamefont {C.~E.}\ \bibnamefont {Collins}},\ }\bibfield  {title}
  {\bibinfo {title} {{Dynamical ejecta of neutron star mergers with nucleonic
  weak processes {\textendash} II: kilonova emission}},\ }\href
  {https://doi.org/10.1093/mnras/stab3327} {\bibfield  {journal} {\bibinfo
  {journal} {Mon. Not. Roy. Astron. Soc.}\ }\textbf {\bibinfo {volume} {510}},\
  \bibinfo {pages} {2820} (\bibinfo {year} {2022})},\ \Eprint
  {https://arxiv.org/abs/2109.14617} {arXiv:2109.14617 [astro-ph.HE]}
  \BibitemShut {NoStop}%
\bibitem [{\citenamefont {Rosswog}\ and\ \citenamefont
  {Korobkin}(2024)}]{Rosswog:2022tus}%
  \BibitemOpen
  \bibfield  {author} {\bibinfo {author} {\bibfnamefont {S.}~\bibnamefont
  {Rosswog}}\ and\ \bibinfo {author} {\bibfnamefont {O.}~\bibnamefont
  {Korobkin}},\ }\bibfield  {title} {\bibinfo {title} {{Heavy Elements and
  Electromagnetic Transients from Neutron Star Mergers}},\ }\href
  {https://doi.org/10.1002/andp.202200306} {\bibfield  {journal} {\bibinfo
  {journal} {Annalen Phys.}\ }\textbf {\bibinfo {volume} {536}},\ \bibinfo
  {pages} {2200306} (\bibinfo {year} {2024})},\ \Eprint
  {https://arxiv.org/abs/2208.14026} {arXiv:2208.14026 [astro-ph.HE]}
  \BibitemShut {NoStop}%
\bibitem [{\citenamefont {Eichler}\ \emph {et~al.}(1989)\citenamefont
  {Eichler}, \citenamefont {Livio}, \citenamefont {Piran},\ and\ \citenamefont
  {Schramm}}]{Eichler:1989ve}%
  \BibitemOpen
  \bibfield  {author} {\bibinfo {author} {\bibfnamefont {D.}~\bibnamefont
  {Eichler}}, \bibinfo {author} {\bibfnamefont {M.}~\bibnamefont {Livio}},
  \bibinfo {author} {\bibfnamefont {T.}~\bibnamefont {Piran}},\ and\ \bibinfo
  {author} {\bibfnamefont {D.~N.}\ \bibnamefont {Schramm}},\ }\bibfield
  {title} {\bibinfo {title} {{Nucleosynthesis, Neutrino Bursts and Gamma-Rays
  from Coalescing Neutron Stars}},\ }\href {https://doi.org/10.1038/340126a0}
  {\bibfield  {journal} {\bibinfo  {journal} {Nature}\ }\textbf {\bibinfo
  {volume} {340}},\ \bibinfo {pages} {126} (\bibinfo {year}
  {1989})}\BibitemShut {NoStop}%
\bibitem [{\citenamefont {Janka}\ and\ \citenamefont
  {Ruffert}(1996)}]{Janka:1995cq}%
  \BibitemOpen
  \bibfield  {author} {\bibinfo {author} {\bibfnamefont {H.~T.}\ \bibnamefont
  {Janka}}\ and\ \bibinfo {author} {\bibfnamefont {M.}~\bibnamefont
  {Ruffert}},\ }\bibfield  {title} {\bibinfo {title} {{Can neutrinos from
  neutron star mergers power gamma-ray bursts?}},\ }\href@noop {} {\bibfield
  {journal} {\bibinfo  {journal} {Astron. Astrophys.}\ }\textbf {\bibinfo
  {volume} {307}},\ \bibinfo {pages} {L33} (\bibinfo {year} {1996})},\ \Eprint
  {https://arxiv.org/abs/astro-ph/9512144} {arXiv:astro-ph/9512144}
  \BibitemShut {NoStop}%
\bibitem [{\citenamefont {Nakar}(2007)}]{Nakar:2007yr}%
  \BibitemOpen
  \bibfield  {author} {\bibinfo {author} {\bibfnamefont {E.}~\bibnamefont
  {Nakar}},\ }\bibfield  {title} {\bibinfo {title} {{Short-Hard Gamma-Ray
  Bursts}},\ }\href {https://doi.org/10.1016/j.physrep.2007.02.005} {\bibfield
  {journal} {\bibinfo  {journal} {Phys. Rept.}\ }\textbf {\bibinfo {volume}
  {442}},\ \bibinfo {pages} {166} (\bibinfo {year} {2007})},\ \Eprint
  {https://arxiv.org/abs/astro-ph/0701748} {arXiv:astro-ph/0701748}
  \BibitemShut {NoStop}%
\bibitem [{\citenamefont {Lee}\ and\ \citenamefont
  {Ramirez-Ruiz}(2007)}]{Lee:2007js}%
  \BibitemOpen
  \bibfield  {author} {\bibinfo {author} {\bibfnamefont {W.~H.}\ \bibnamefont
  {Lee}}\ and\ \bibinfo {author} {\bibfnamefont {E.}~\bibnamefont
  {Ramirez-Ruiz}},\ }\bibfield  {title} {\bibinfo {title} {{The Progenitors of
  Short Gamma-Ray Bursts}},\ }\href {https://doi.org/10.1088/1367-2630/9/1/017}
  {\bibfield  {journal} {\bibinfo  {journal} {New J. Phys.}\ }\textbf {\bibinfo
  {volume} {9}},\ \bibinfo {pages} {17} (\bibinfo {year} {2007})},\ \Eprint
  {https://arxiv.org/abs/astro-ph/0701874} {arXiv:astro-ph/0701874}
  \BibitemShut {NoStop}%
\bibitem [{\citenamefont {Meszaros}\ and\ \citenamefont
  {Rees}(2014)}]{Meszaros:2014pca}%
  \BibitemOpen
  \bibfield  {author} {\bibinfo {author} {\bibfnamefont {P.}~\bibnamefont
  {Meszaros}}\ and\ \bibinfo {author} {\bibfnamefont {M.~J.}\ \bibnamefont
  {Rees}},\ }\bibfield  {title} {\bibinfo {title} {{Gamma-Ray Bursts}},\
  }\href@noop {} {\bibfield  {journal} {\bibinfo  {journal} {arXiv}\ }
  (\bibinfo {year} {2014})},\ \Eprint {https://arxiv.org/abs/1401.3012}
  {arXiv:1401.3012 [astro-ph.HE]} \BibitemShut {NoStop}%
\bibitem [{\citenamefont {Abbott}\ \emph
  {et~al.}(2017{\natexlab{a}})\citenamefont {Abbott} \emph
  {et~al.}}]{LIGOScientific:2017vwq}%
  \BibitemOpen
  \bibfield  {author} {\bibinfo {author} {\bibfnamefont {B.~P.}\ \bibnamefont
  {Abbott}} \emph {et~al.} (\bibinfo {collaboration} {LIGO Scientific,
  Virgo}),\ }\bibfield  {title} {\bibinfo {title} {{GW170817: Observation of
  Gravitational Waves from a Binary Neutron Star Inspiral}},\ }\href
  {https://doi.org/10.1103/PhysRevLett.119.161101} {\bibfield  {journal}
  {\bibinfo  {journal} {Phys. Rev. Lett.}\ }\textbf {\bibinfo {volume} {119}},\
  \bibinfo {pages} {161101} (\bibinfo {year} {2017}{\natexlab{a}})},\ \Eprint
  {https://arxiv.org/abs/1710.05832} {arXiv:1710.05832 [gr-qc]} \BibitemShut
  {NoStop}%
\bibitem [{\citenamefont {Grado}(2023)}]{Grado:2023pyb}%
  \BibitemOpen
  \bibfield  {author} {\bibinfo {author} {\bibfnamefont {A.}~\bibnamefont
  {Grado}},\ }\bibfield  {title} {\bibinfo {title} {{Einstein Telescope, the
  future generation of ground based gravitational wave detectors}},\ }\href
  {https://doi.org/10.1088/1742-6596/2429/1/012041} {\bibfield  {journal}
  {\bibinfo  {journal} {J. Phys. Conf. Ser.}\ }\textbf {\bibinfo {volume}
  {2429}},\ \bibinfo {pages} {012041} (\bibinfo {year} {2023})}\BibitemShut
  {NoStop}%
\bibitem [{\citenamefont {Abbott}\ \emph
  {et~al.}(2017{\natexlab{b}})\citenamefont {Abbott} \emph
  {et~al.}}]{LIGOScientific:2017zic}%
  \BibitemOpen
  \bibfield  {author} {\bibinfo {author} {\bibfnamefont {B.~P.}\ \bibnamefont
  {Abbott}} \emph {et~al.} (\bibinfo {collaboration} {LIGO Scientific, Virgo,
  Fermi-GBM, INTEGRAL}),\ }\bibfield  {title} {\bibinfo {title} {{Gravitational
  Waves and Gamma-rays from a Binary Neutron Star Merger: GW170817 and GRB
  170817A}},\ }\href {https://doi.org/10.3847/2041-8213/aa920c} {\bibfield
  {journal} {\bibinfo  {journal} {Astrophys. J. Lett.}\ }\textbf {\bibinfo
  {volume} {848}},\ \bibinfo {pages} {L13} (\bibinfo {year}
  {2017}{\natexlab{b}})},\ \Eprint {https://arxiv.org/abs/1710.05834}
  {arXiv:1710.05834 [astro-ph.HE]} \BibitemShut {NoStop}%
\bibitem [{\citenamefont {Abbott}\ \emph
  {et~al.}(2017{\natexlab{c}})\citenamefont {Abbott} \emph
  {et~al.}}]{LIGOScientific:2017ync}%
  \BibitemOpen
  \bibfield  {author} {\bibinfo {author} {\bibfnamefont {B.~P.}\ \bibnamefont
  {Abbott}} \emph {et~al.} (\bibinfo {collaboration} {LIGO Scientific, Virgo,
  Fermi GBM, INTEGRAL, IceCube, AstroSat Cadmium Zinc Telluride Imager Team,
  IPN, Insight-Hxmt, ANTARES, Swift, AGILE Team, 1M2H Team, Dark Energy Camera
  GW-EM, DES, DLT40, GRAWITA, Fermi-LAT, ATCA, ASKAP, Las Cumbres Observatory
  Group, OzGrav, DWF (Deeper Wider Faster Program), AST3, CAASTRO, VINROUGE,
  MASTER, J-GEM, GROWTH, JAGWAR, CaltechNRAO, TTU-NRAO, NuSTAR, Pan-STARRS,
  MAXI Team, TZAC Consortium, KU, Nordic Optical Telescope, ePESSTO, GROND,
  Texas Tech University, SALT Group, TOROS, BOOTES, MWA, CALET, IKI-GW
  Follow-up, H.E.S.S., LOFAR, LWA, HAWC, Pierre Auger, ALMA, Euro VLBI Team, Pi
  of Sky, Chandra Team at McGill University, DFN, ATLAS Telescopes, High Time
  Resolution Universe Survey, RIMAS, RATIR, SKA South Africa/MeerKAT}),\
  }\bibfield  {title} {\bibinfo {title} {{Multi-messenger Observations of a
  Binary Neutron Star Merger}},\ }\href
  {https://doi.org/10.3847/2041-8213/aa91c9} {\bibfield  {journal} {\bibinfo
  {journal} {Astrophys. J. Lett.}\ }\textbf {\bibinfo {volume} {848}},\
  \bibinfo {pages} {L12} (\bibinfo {year} {2017}{\natexlab{c}})},\ \Eprint
  {https://arxiv.org/abs/1710.05833} {arXiv:1710.05833 [astro-ph.HE]}
  \BibitemShut {NoStop}%
\bibitem [{\citenamefont {Kasen}\ \emph {et~al.}(2017)\citenamefont {Kasen},
  \citenamefont {Metzger}, \citenamefont {Barnes}, \citenamefont {Quataert},\
  and\ \citenamefont {Ramirez-Ruiz}}]{Kasen:2017sxr}%
  \BibitemOpen
  \bibfield  {author} {\bibinfo {author} {\bibfnamefont {D.}~\bibnamefont
  {Kasen}}, \bibinfo {author} {\bibfnamefont {B.}~\bibnamefont {Metzger}},
  \bibinfo {author} {\bibfnamefont {J.}~\bibnamefont {Barnes}}, \bibinfo
  {author} {\bibfnamefont {E.}~\bibnamefont {Quataert}},\ and\ \bibinfo
  {author} {\bibfnamefont {E.}~\bibnamefont {Ramirez-Ruiz}},\ }\bibfield
  {title} {\bibinfo {title} {{Origin of the heavy elements in binary
  neutron-star mergers from a gravitational wave event}},\ }\href
  {https://doi.org/10.1038/nature24453} {\bibfield  {journal} {\bibinfo
  {journal} {Nature}\ }\textbf {\bibinfo {volume} {551}},\ \bibinfo {pages}
  {80} (\bibinfo {year} {2017})},\ \Eprint {https://arxiv.org/abs/1710.05463}
  {arXiv:1710.05463 [astro-ph.HE]} \BibitemShut {NoStop}%
\bibitem [{\citenamefont {Villar}\ \emph {et~al.}(2017)\citenamefont {Villar}
  \emph {et~al.}}]{Villar:2017wcc}%
  \BibitemOpen
  \bibfield  {author} {\bibinfo {author} {\bibfnamefont {V.~A.}\ \bibnamefont
  {Villar}} \emph {et~al.},\ }\bibfield  {title} {\bibinfo {title} {{The
  Combined Ultraviolet, Optical, and Near-Infrared Light Curves of the Kilonova
  Associated with the Binary Neutron Star Merger GW170817: Unified Data Set,
  Analytic Models, and Physical Implications}},\ }\href
  {https://doi.org/10.3847/2041-8213/aa9c84} {\bibfield  {journal} {\bibinfo
  {journal} {Astrophys. J. Lett.}\ }\textbf {\bibinfo {volume} {851}},\
  \bibinfo {pages} {L21} (\bibinfo {year} {2017})},\ \Eprint
  {https://arxiv.org/abs/1710.11576} {arXiv:1710.11576 [astro-ph.HE]}
  \BibitemShut {NoStop}%
\bibitem [{\citenamefont {Perego}\ \emph {et~al.}(2017)\citenamefont {Perego},
  \citenamefont {Radice},\ and\ \citenamefont {Bernuzzi}}]{Perego:2017wtu}%
  \BibitemOpen
  \bibfield  {author} {\bibinfo {author} {\bibfnamefont {A.}~\bibnamefont
  {Perego}}, \bibinfo {author} {\bibfnamefont {D.}~\bibnamefont {Radice}},\
  and\ \bibinfo {author} {\bibfnamefont {S.}~\bibnamefont {Bernuzzi}},\
  }\bibfield  {title} {\bibinfo {title} {{AT 2017gfo: An Anisotropic and
  Three-component Kilonova Counterpart of GW170817}},\ }\href
  {https://doi.org/10.3847/2041-8213/aa9ab9} {\bibfield  {journal} {\bibinfo
  {journal} {Astrophys. J. Lett.}\ }\textbf {\bibinfo {volume} {850}},\
  \bibinfo {pages} {L37} (\bibinfo {year} {2017})},\ \Eprint
  {https://arxiv.org/abs/1711.03982} {arXiv:1711.03982 [astro-ph.HE]}
  \BibitemShut {NoStop}%
\bibitem [{\citenamefont {Wollaeger}\ \emph {et~al.}(2018)\citenamefont
  {Wollaeger}, \citenamefont {Korobkin}, \citenamefont {Fontes}, \citenamefont
  {Rosswog}, \citenamefont {Even}, \citenamefont {Fryer}, \citenamefont
  {Sollerman}, \citenamefont {Hungerford}, \citenamefont {van Rossum},\ and\
  \citenamefont {Wollaber}}]{Wollaeger:2017ahm}%
  \BibitemOpen
  \bibfield  {author} {\bibinfo {author} {\bibfnamefont {R.~T.}\ \bibnamefont
  {Wollaeger}}, \bibinfo {author} {\bibfnamefont {O.}~\bibnamefont {Korobkin}},
  \bibinfo {author} {\bibfnamefont {C.~J.}\ \bibnamefont {Fontes}}, \bibinfo
  {author} {\bibfnamefont {S.~K.}\ \bibnamefont {Rosswog}}, \bibinfo {author}
  {\bibfnamefont {W.~P.}\ \bibnamefont {Even}}, \bibinfo {author}
  {\bibfnamefont {C.~L.}\ \bibnamefont {Fryer}}, \bibinfo {author}
  {\bibfnamefont {J.}~\bibnamefont {Sollerman}}, \bibinfo {author}
  {\bibfnamefont {A.~L.}\ \bibnamefont {Hungerford}}, \bibinfo {author}
  {\bibfnamefont {D.~R.}\ \bibnamefont {van Rossum}},\ and\ \bibinfo {author}
  {\bibfnamefont {A.~B.}\ \bibnamefont {Wollaber}},\ }\bibfield  {title}
  {\bibinfo {title} {{Impact of ejecta morphology and composition on the
  electromagnetic signatures of neutron star mergers}},\ }\href
  {https://doi.org/10.1093/mnras/sty1018} {\bibfield  {journal} {\bibinfo
  {journal} {Mon. Not. Roy. Astron. Soc.}\ }\textbf {\bibinfo {volume} {478}},\
  \bibinfo {pages} {3298} (\bibinfo {year} {2018})},\ \Eprint
  {https://arxiv.org/abs/1705.07084} {arXiv:1705.07084 [astro-ph.HE]}
  \BibitemShut {NoStop}%
\bibitem [{\citenamefont {Kawaguchi}\ \emph {et~al.}(2018)\citenamefont
  {Kawaguchi}, \citenamefont {Shibata},\ and\ \citenamefont
  {Tanaka}}]{Kawaguchi:2018ptg}%
  \BibitemOpen
  \bibfield  {author} {\bibinfo {author} {\bibfnamefont {K.}~\bibnamefont
  {Kawaguchi}}, \bibinfo {author} {\bibfnamefont {M.}~\bibnamefont {Shibata}},\
  and\ \bibinfo {author} {\bibfnamefont {M.}~\bibnamefont {Tanaka}},\
  }\bibfield  {title} {\bibinfo {title} {{Radiative transfer simulation for the
  optical and near-infrared electromagnetic counterparts to GW170817}},\ }\href
  {https://doi.org/10.3847/2041-8213/aade02} {\bibfield  {journal} {\bibinfo
  {journal} {Astrophys. J. Lett.}\ }\textbf {\bibinfo {volume} {865}},\
  \bibinfo {pages} {L21} (\bibinfo {year} {2018})},\ \Eprint
  {https://arxiv.org/abs/1806.04088} {arXiv:1806.04088 [astro-ph.HE]}
  \BibitemShut {NoStop}%
\bibitem [{\citenamefont {Miller}\ \emph {et~al.}(2019)\citenamefont {Miller},
  \citenamefont {Ryan}, \citenamefont {Dolence}, \citenamefont {Burrows},
  \citenamefont {Fontes}, \citenamefont {Fryer}, \citenamefont {Korobkin},
  \citenamefont {Lippuner}, \citenamefont {Mumpower},\ and\ \citenamefont
  {Wollaeger}}]{Miller:2019dpt}%
  \BibitemOpen
  \bibfield  {author} {\bibinfo {author} {\bibfnamefont {J.~M.}\ \bibnamefont
  {Miller}}, \bibinfo {author} {\bibfnamefont {B.~R.}\ \bibnamefont {Ryan}},
  \bibinfo {author} {\bibfnamefont {J.~C.}\ \bibnamefont {Dolence}}, \bibinfo
  {author} {\bibfnamefont {A.}~\bibnamefont {Burrows}}, \bibinfo {author}
  {\bibfnamefont {C.~J.}\ \bibnamefont {Fontes}}, \bibinfo {author}
  {\bibfnamefont {C.~L.}\ \bibnamefont {Fryer}}, \bibinfo {author}
  {\bibfnamefont {O.}~\bibnamefont {Korobkin}}, \bibinfo {author}
  {\bibfnamefont {J.}~\bibnamefont {Lippuner}}, \bibinfo {author}
  {\bibfnamefont {M.~R.}\ \bibnamefont {Mumpower}},\ and\ \bibinfo {author}
  {\bibfnamefont {R.~T.}\ \bibnamefont {Wollaeger}},\ }\bibfield  {title}
  {\bibinfo {title} {{Full Transport Model of GW170817-Like Disk Produces a
  Blue Kilonova}},\ }\href {https://doi.org/10.1103/PhysRevD.100.023008}
  {\bibfield  {journal} {\bibinfo  {journal} {Phys. Rev. D}\ }\textbf {\bibinfo
  {volume} {100}},\ \bibinfo {pages} {023008} (\bibinfo {year} {2019})},\
  \Eprint {https://arxiv.org/abs/1905.07477} {arXiv:1905.07477 [astro-ph.HE]}
  \BibitemShut {NoStop}%
\bibitem [{\citenamefont {Ruffert}\ \emph {et~al.}(1997)\citenamefont
  {Ruffert}, \citenamefont {Janka}, \citenamefont {Takahashi},\ and\
  \citenamefont {Schaefer}}]{Ruffert:1996by}%
  \BibitemOpen
  \bibfield  {author} {\bibinfo {author} {\bibfnamefont {M.}~\bibnamefont
  {Ruffert}}, \bibinfo {author} {\bibfnamefont {H.~T.}\ \bibnamefont {Janka}},
  \bibinfo {author} {\bibfnamefont {K.}~\bibnamefont {Takahashi}},\ and\
  \bibinfo {author} {\bibfnamefont {G.}~\bibnamefont {Schaefer}},\ }\bibfield
  {title} {\bibinfo {title} {{Coalescing neutron stars: A Step towards physical
  models. 2. Neutrino emission, neutron tori, and gamma-ray bursts}},\
  }\href@noop {} {\bibfield  {journal} {\bibinfo  {journal} {Astron.
  Astrophys.}\ }\textbf {\bibinfo {volume} {319}},\ \bibinfo {pages} {122}
  (\bibinfo {year} {1997})},\ \Eprint {https://arxiv.org/abs/astro-ph/9606181}
  {arXiv:astro-ph/9606181} \BibitemShut {NoStop}%
\bibitem [{\citenamefont {Rosswog}\ and\ \citenamefont
  {Liebendoerfer}(2003)}]{Rosswog:2003rv}%
  \BibitemOpen
  \bibfield  {author} {\bibinfo {author} {\bibfnamefont {S.}~\bibnamefont
  {Rosswog}}\ and\ \bibinfo {author} {\bibfnamefont {M.}~\bibnamefont
  {Liebendoerfer}},\ }\bibfield  {title} {\bibinfo {title} {{High resolution
  calculations of merging neutron stars. 2: Neutrino emission}},\ }\href
  {https://doi.org/10.1046/j.1365-8711.2003.06579.x} {\bibfield  {journal}
  {\bibinfo  {journal} {Mon. Not. Roy. Astron. Soc.}\ }\textbf {\bibinfo
  {volume} {342}},\ \bibinfo {pages} {673} (\bibinfo {year} {2003})},\ \Eprint
  {https://arxiv.org/abs/astro-ph/0302301} {arXiv:astro-ph/0302301}
  \BibitemShut {NoStop}%
\bibitem [{\citenamefont {Deaton}\ \emph {et~al.}(2013)\citenamefont {Deaton},
  \citenamefont {Duez}, \citenamefont {Foucart}, \citenamefont {O'Connor},
  \citenamefont {Ott}, \citenamefont {Kidder}, \citenamefont {Muhlberger},
  \citenamefont {Scheel},\ and\ \citenamefont {Szilagyi}}]{Deaton:2013sla}%
  \BibitemOpen
  \bibfield  {author} {\bibinfo {author} {\bibfnamefont {M.~B.}\ \bibnamefont
  {Deaton}}, \bibinfo {author} {\bibfnamefont {M.~D.}\ \bibnamefont {Duez}},
  \bibinfo {author} {\bibfnamefont {F.}~\bibnamefont {Foucart}}, \bibinfo
  {author} {\bibfnamefont {E.}~\bibnamefont {O'Connor}}, \bibinfo {author}
  {\bibfnamefont {C.~D.}\ \bibnamefont {Ott}}, \bibinfo {author} {\bibfnamefont
  {L.~E.}\ \bibnamefont {Kidder}}, \bibinfo {author} {\bibfnamefont {C.~D.}\
  \bibnamefont {Muhlberger}}, \bibinfo {author} {\bibfnamefont {M.~A.}\
  \bibnamefont {Scheel}},\ and\ \bibinfo {author} {\bibfnamefont
  {B.}~\bibnamefont {Szilagyi}},\ }\bibfield  {title} {\bibinfo {title} {{Black
  Hole-Neutron Star Mergers with a Hot Nuclear Equation of State: Outflow and
  Neutrino-Cooled Disk for a Low-Mass, High-Spin Case}},\ }\href
  {https://doi.org/10.1088/0004-637X/776/1/47} {\bibfield  {journal} {\bibinfo
  {journal} {Astrophys. J.}\ }\textbf {\bibinfo {volume} {776}},\ \bibinfo
  {pages} {47} (\bibinfo {year} {2013})},\ \Eprint
  {https://arxiv.org/abs/1304.3384} {arXiv:1304.3384 [astro-ph.HE]}
  \BibitemShut {NoStop}%
\bibitem [{\citenamefont {Neilsen}\ \emph {et~al.}(2014)\citenamefont
  {Neilsen}, \citenamefont {Liebling}, \citenamefont {Anderson}, \citenamefont
  {Lehner}, \citenamefont {O'Connor},\ and\ \citenamefont
  {Palenzuela}}]{Neilsen:2014hha}%
  \BibitemOpen
  \bibfield  {author} {\bibinfo {author} {\bibfnamefont {D.}~\bibnamefont
  {Neilsen}}, \bibinfo {author} {\bibfnamefont {S.~L.}\ \bibnamefont
  {Liebling}}, \bibinfo {author} {\bibfnamefont {M.}~\bibnamefont {Anderson}},
  \bibinfo {author} {\bibfnamefont {L.}~\bibnamefont {Lehner}}, \bibinfo
  {author} {\bibfnamefont {E.}~\bibnamefont {O'Connor}},\ and\ \bibinfo
  {author} {\bibfnamefont {C.}~\bibnamefont {Palenzuela}},\ }\bibfield  {title}
  {\bibinfo {title} {{Magnetized Neutron Stars With Realistic Equations of
  State and Neutrino Cooling}},\ }\href
  {https://doi.org/10.1103/PhysRevD.89.104029} {\bibfield  {journal} {\bibinfo
  {journal} {Phys. Rev. D}\ }\textbf {\bibinfo {volume} {89}},\ \bibinfo
  {pages} {104029} (\bibinfo {year} {2014})},\ \Eprint
  {https://arxiv.org/abs/1403.3680} {arXiv:1403.3680 [gr-qc]} \BibitemShut
  {NoStop}%
\bibitem [{\citenamefont {Wanajo}\ \emph {et~al.}(2014)\citenamefont {Wanajo},
  \citenamefont {Sekiguchi}, \citenamefont {Nishimura}, \citenamefont {Kiuchi},
  \citenamefont {Kyutoku},\ and\ \citenamefont {Shibata}}]{Wanajo:2014wha}%
  \BibitemOpen
  \bibfield  {author} {\bibinfo {author} {\bibfnamefont {S.}~\bibnamefont
  {Wanajo}}, \bibinfo {author} {\bibfnamefont {Y.}~\bibnamefont {Sekiguchi}},
  \bibinfo {author} {\bibfnamefont {N.}~\bibnamefont {Nishimura}}, \bibinfo
  {author} {\bibfnamefont {K.}~\bibnamefont {Kiuchi}}, \bibinfo {author}
  {\bibfnamefont {K.}~\bibnamefont {Kyutoku}},\ and\ \bibinfo {author}
  {\bibfnamefont {M.}~\bibnamefont {Shibata}},\ }\bibfield  {title} {\bibinfo
  {title} {{Production of all the r-process nuclides in the dynamical ejecta of
  neutron star mergers}},\ }\href {https://doi.org/10.1088/2041-8205/789/2/L39}
  {\bibfield  {journal} {\bibinfo  {journal} {Astrophys. J. Lett.}\ }\textbf
  {\bibinfo {volume} {789}},\ \bibinfo {pages} {L39} (\bibinfo {year}
  {2014})},\ \Eprint {https://arxiv.org/abs/1402.7317} {arXiv:1402.7317
  [astro-ph.SR]} \BibitemShut {NoStop}%
\bibitem [{\citenamefont {Sekiguchi}\ \emph {et~al.}(2015)\citenamefont
  {Sekiguchi}, \citenamefont {Kiuchi}, \citenamefont {Kyutoku},\ and\
  \citenamefont {Shibata}}]{Sekiguchi:2015dma}%
  \BibitemOpen
  \bibfield  {author} {\bibinfo {author} {\bibfnamefont {Y.}~\bibnamefont
  {Sekiguchi}}, \bibinfo {author} {\bibfnamefont {K.}~\bibnamefont {Kiuchi}},
  \bibinfo {author} {\bibfnamefont {K.}~\bibnamefont {Kyutoku}},\ and\ \bibinfo
  {author} {\bibfnamefont {M.}~\bibnamefont {Shibata}},\ }\bibfield  {title}
  {\bibinfo {title} {{Dynamical mass ejection from binary neutron star mergers:
  Radiation-hydrodynamics study in general relativity}},\ }\href
  {https://doi.org/10.1103/PhysRevD.91.064059} {\bibfield  {journal} {\bibinfo
  {journal} {Phys. Rev. D}\ }\textbf {\bibinfo {volume} {91}},\ \bibinfo
  {pages} {064059} (\bibinfo {year} {2015})},\ \Eprint
  {https://arxiv.org/abs/1502.06660} {arXiv:1502.06660 [astro-ph.HE]}
  \BibitemShut {NoStop}%
\bibitem [{\citenamefont {Foucart}\ \emph {et~al.}(2015)\citenamefont
  {Foucart}, \citenamefont {O'Connor}, \citenamefont {Roberts}, \citenamefont
  {Duez}, \citenamefont {Haas}, \citenamefont {Kidder}, \citenamefont {Ott},
  \citenamefont {Pfeiffer}, \citenamefont {Scheel},\ and\ \citenamefont
  {Szilagyi}}]{Foucart:2015vpa}%
  \BibitemOpen
  \bibfield  {author} {\bibinfo {author} {\bibfnamefont {F.}~\bibnamefont
  {Foucart}}, \bibinfo {author} {\bibfnamefont {E.}~\bibnamefont {O'Connor}},
  \bibinfo {author} {\bibfnamefont {L.}~\bibnamefont {Roberts}}, \bibinfo
  {author} {\bibfnamefont {M.~D.}\ \bibnamefont {Duez}}, \bibinfo {author}
  {\bibfnamefont {R.}~\bibnamefont {Haas}}, \bibinfo {author} {\bibfnamefont
  {L.~E.}\ \bibnamefont {Kidder}}, \bibinfo {author} {\bibfnamefont {C.~D.}\
  \bibnamefont {Ott}}, \bibinfo {author} {\bibfnamefont {H.~P.}\ \bibnamefont
  {Pfeiffer}}, \bibinfo {author} {\bibfnamefont {M.~A.}\ \bibnamefont
  {Scheel}},\ and\ \bibinfo {author} {\bibfnamefont {B.}~\bibnamefont
  {Szilagyi}},\ }\bibfield  {title} {\bibinfo {title} {{Post-merger evolution
  of a neutron star-black hole binary with neutrino transport}},\ }\href
  {https://doi.org/10.1103/PhysRevD.91.124021} {\bibfield  {journal} {\bibinfo
  {journal} {Phys. Rev. D}\ }\textbf {\bibinfo {volume} {91}},\ \bibinfo
  {pages} {124021} (\bibinfo {year} {2015})},\ \Eprint
  {https://arxiv.org/abs/1502.04146} {arXiv:1502.04146 [astro-ph.HE]}
  \BibitemShut {NoStop}%
\bibitem [{\citenamefont {Radice}\ \emph {et~al.}(2022)\citenamefont {Radice},
  \citenamefont {Bernuzzi}, \citenamefont {Perego},\ and\ \citenamefont
  {Haas}}]{Radice:2021jtw}%
  \BibitemOpen
  \bibfield  {author} {\bibinfo {author} {\bibfnamefont {D.}~\bibnamefont
  {Radice}}, \bibinfo {author} {\bibfnamefont {S.}~\bibnamefont {Bernuzzi}},
  \bibinfo {author} {\bibfnamefont {A.}~\bibnamefont {Perego}},\ and\ \bibinfo
  {author} {\bibfnamefont {R.}~\bibnamefont {Haas}},\ }\bibfield  {title}
  {\bibinfo {title} {{A new moment-based general-relativistic
  neutrino-radiation transport code: Methods and first applications to neutron
  star mergers}},\ }\href {https://doi.org/10.1093/mnras/stac589} {\bibfield
  {journal} {\bibinfo  {journal} {Mon. Not. Roy. Astron. Soc.}\ }\textbf
  {\bibinfo {volume} {512}},\ \bibinfo {pages} {1499} (\bibinfo {year}
  {2022})},\ \Eprint {https://arxiv.org/abs/2111.14858} {arXiv:2111.14858
  [astro-ph.HE]} \BibitemShut {NoStop}%
\bibitem [{\citenamefont {Foucart}\ \emph {et~al.}(2016)\citenamefont
  {Foucart}, \citenamefont {O'Connor}, \citenamefont {Roberts}, \citenamefont
  {Kidder}, \citenamefont {Pfeiffer},\ and\ \citenamefont
  {Scheel}}]{Foucart:2016rxm}%
  \BibitemOpen
  \bibfield  {author} {\bibinfo {author} {\bibfnamefont {F.}~\bibnamefont
  {Foucart}}, \bibinfo {author} {\bibfnamefont {E.}~\bibnamefont {O'Connor}},
  \bibinfo {author} {\bibfnamefont {L.}~\bibnamefont {Roberts}}, \bibinfo
  {author} {\bibfnamefont {L.~E.}\ \bibnamefont {Kidder}}, \bibinfo {author}
  {\bibfnamefont {H.~P.}\ \bibnamefont {Pfeiffer}},\ and\ \bibinfo {author}
  {\bibfnamefont {M.~A.}\ \bibnamefont {Scheel}},\ }\bibfield  {title}
  {\bibinfo {title} {{Impact of an improved neutrino energy estimate on
  outflows in neutron star merger simulations}},\ }\href
  {https://doi.org/10.1103/PhysRevD.94.123016} {\bibfield  {journal} {\bibinfo
  {journal} {Phys. Rev. D}\ }\textbf {\bibinfo {volume} {94}},\ \bibinfo
  {pages} {123016} (\bibinfo {year} {2016})},\ \Eprint
  {https://arxiv.org/abs/1607.07450} {arXiv:1607.07450 [astro-ph.HE]}
  \BibitemShut {NoStop}%
\bibitem [{\citenamefont {Foucart}\ \emph {et~al.}(2021)\citenamefont
  {Foucart}, \citenamefont {Duez}, \citenamefont {Hebert}, \citenamefont
  {Kidder}, \citenamefont {Kovarik}, \citenamefont {Pfeiffer},\ and\
  \citenamefont {Scheel}}]{Foucart:2021mcb}%
  \BibitemOpen
  \bibfield  {author} {\bibinfo {author} {\bibfnamefont {F.}~\bibnamefont
  {Foucart}}, \bibinfo {author} {\bibfnamefont {M.~D.}\ \bibnamefont {Duez}},
  \bibinfo {author} {\bibfnamefont {F.}~\bibnamefont {Hebert}}, \bibinfo
  {author} {\bibfnamefont {L.~E.}\ \bibnamefont {Kidder}}, \bibinfo {author}
  {\bibfnamefont {P.}~\bibnamefont {Kovarik}}, \bibinfo {author} {\bibfnamefont
  {H.~P.}\ \bibnamefont {Pfeiffer}},\ and\ \bibinfo {author} {\bibfnamefont
  {M.~A.}\ \bibnamefont {Scheel}},\ }\bibfield  {title} {\bibinfo {title}
  {{Implementation of Monte Carlo Transport in the General Relativistic SpEC
  Code}},\ }\href {https://doi.org/10.3847/1538-4357/ac1737} {\bibfield
  {journal} {\bibinfo  {journal} {Astrophys. J.}\ }\textbf {\bibinfo {volume}
  {920}},\ \bibinfo {pages} {82} (\bibinfo {year} {2021})},\ \Eprint
  {https://arxiv.org/abs/2103.16588} {arXiv:2103.16588 [astro-ph.HE]}
  \BibitemShut {NoStop}%
\bibitem [{\citenamefont {Foucart}\ \emph {et~al.}(2023)\citenamefont
  {Foucart}, \citenamefont {Duez}, \citenamefont {Haas}, \citenamefont
  {Kidder}, \citenamefont {Pfeiffer}, \citenamefont {Scheel},\ and\
  \citenamefont {Spira-Savett}}]{Foucart:2022kon}%
  \BibitemOpen
  \bibfield  {author} {\bibinfo {author} {\bibfnamefont {F.}~\bibnamefont
  {Foucart}}, \bibinfo {author} {\bibfnamefont {M.~D.}\ \bibnamefont {Duez}},
  \bibinfo {author} {\bibfnamefont {R.}~\bibnamefont {Haas}}, \bibinfo {author}
  {\bibfnamefont {L.~E.}\ \bibnamefont {Kidder}}, \bibinfo {author}
  {\bibfnamefont {H.~P.}\ \bibnamefont {Pfeiffer}}, \bibinfo {author}
  {\bibfnamefont {M.~A.}\ \bibnamefont {Scheel}},\ and\ \bibinfo {author}
  {\bibfnamefont {E.}~\bibnamefont {Spira-Savett}},\ }\bibfield  {title}
  {\bibinfo {title} {{General relativistic simulations of collapsing binary
  neutron star mergers with Monte~Carlo neutrino transport}},\ }\href
  {https://doi.org/10.1103/PhysRevD.107.103055} {\bibfield  {journal} {\bibinfo
   {journal} {Phys. Rev. D}\ }\textbf {\bibinfo {volume} {107}},\ \bibinfo
  {pages} {103055} (\bibinfo {year} {2023})},\ \Eprint
  {https://arxiv.org/abs/2210.05670} {arXiv:2210.05670 [astro-ph.HE]}
  \BibitemShut {NoStop}%
\bibitem [{\citenamefont {Izquierdo}\ \emph {et~al.}(2024)\citenamefont
  {Izquierdo}, \citenamefont {Abalos},\ and\ \citenamefont
  {Palenzuela}}]{Izquierdo:2023fub}%
  \BibitemOpen
  \bibfield  {author} {\bibinfo {author} {\bibfnamefont {M.~R.}\ \bibnamefont
  {Izquierdo}}, \bibinfo {author} {\bibfnamefont {J.~F.}\ \bibnamefont
  {Abalos}},\ and\ \bibinfo {author} {\bibfnamefont {C.}~\bibnamefont
  {Palenzuela}},\ }\bibfield  {title} {\bibinfo {title} {{Guided moments
  formalism: A new efficient full-neutrino treatment for astrophysical
  simulations}},\ }\href {https://doi.org/10.1103/PhysRevD.109.043044}
  {\bibfield  {journal} {\bibinfo  {journal} {Phys. Rev. D}\ }\textbf {\bibinfo
  {volume} {109}},\ \bibinfo {pages} {043044} (\bibinfo {year} {2024})},\
  \Eprint {https://arxiv.org/abs/2312.09275} {arXiv:2312.09275 [astro-ph.HE]}
  \BibitemShut {NoStop}%
\bibitem [{\citenamefont {Shapiro}\ and\ \citenamefont
  {Teukolsky}(1983)}]{Shapiro:1983du}%
  \BibitemOpen
  \bibfield  {author} {\bibinfo {author} {\bibfnamefont {S.~L.}\ \bibnamefont
  {Shapiro}}\ and\ \bibinfo {author} {\bibfnamefont {S.~A.}\ \bibnamefont
  {Teukolsky}},\ }\href@noop {} {\emph {\bibinfo {title} {{Black Holes, White
  Dwarfs, and Neutron Stars: The Physics of Compact Objects}}}}\ (\bibinfo
  {publisher} {Wiley-Interscience},\ \bibinfo {address} {New York},\ \bibinfo
  {year} {1983})\BibitemShut {NoStop}%
\bibitem [{\citenamefont {Janka}(2012)}]{Janka:2012wk}%
  \BibitemOpen
  \bibfield  {author} {\bibinfo {author} {\bibfnamefont {H.-T.}\ \bibnamefont
  {Janka}},\ }\bibfield  {title} {\bibinfo {title} {{Explosion Mechanisms of
  Core-Collapse Supernovae}},\ }\href
  {https://doi.org/10.1146/annurev-nucl-102711-094901} {\bibfield  {journal}
  {\bibinfo  {journal} {Ann. Rev. Nucl. Part. Sci.}\ }\textbf {\bibinfo
  {volume} {62}},\ \bibinfo {pages} {407} (\bibinfo {year} {2012})},\ \Eprint
  {https://arxiv.org/abs/1206.2503} {arXiv:1206.2503 [astro-ph.SR]}
  \BibitemShut {NoStop}%
\bibitem [{\citenamefont {Thompson}\ and\ \citenamefont
  {Burrows}(2001)}]{Thompson:2000tu}%
  \BibitemOpen
  \bibfield  {author} {\bibinfo {author} {\bibfnamefont {T.~A.}\ \bibnamefont
  {Thompson}}\ and\ \bibinfo {author} {\bibfnamefont {A.}~\bibnamefont
  {Burrows}},\ }\bibfield  {title} {\bibinfo {title} {{Neutrino processes in
  supernovae and the physics of protoneutron star winds}},\ }\href
  {https://doi.org/10.1016/S0375-9474(01)00730-8} {\bibfield  {journal}
  {\bibinfo  {journal} {Nucl. Phys. A}\ }\textbf {\bibinfo {volume} {688}},\
  \bibinfo {pages} {377} (\bibinfo {year} {2001})},\ \Eprint
  {https://arxiv.org/abs/astro-ph/0009449} {arXiv:astro-ph/0009449}
  \BibitemShut {NoStop}%
\bibitem [{\citenamefont {Buras}\ \emph {et~al.}(2006)\citenamefont {Buras},
  \citenamefont {Janka}, \citenamefont {Rampp},\ and\ \citenamefont
  {Kifonidis}}]{Buras:2005tb}%
  \BibitemOpen
  \bibfield  {author} {\bibinfo {author} {\bibfnamefont {R.}~\bibnamefont
  {Buras}}, \bibinfo {author} {\bibfnamefont {H.-T.}\ \bibnamefont {Janka}},
  \bibinfo {author} {\bibfnamefont {M.}~\bibnamefont {Rampp}},\ and\ \bibinfo
  {author} {\bibfnamefont {K.}~\bibnamefont {Kifonidis}},\ }\bibfield  {title}
  {\bibinfo {title} {{Two-dimensional hydrodynamic core-collapse supernova
  simulations with spectral neutrino transport. 2. models for different
  progenitor stars}},\ }\href {https://doi.org/10.1051/0004-6361:20054654}
  {\bibfield  {journal} {\bibinfo  {journal} {Astron. Astrophys.}\ }\textbf
  {\bibinfo {volume} {457}},\ \bibinfo {pages} {281} (\bibinfo {year}
  {2006})},\ \Eprint {https://arxiv.org/abs/astro-ph/0512189}
  {arXiv:astro-ph/0512189} \BibitemShut {NoStop}%
\bibitem [{\citenamefont {Langanke}\ \emph {et~al.}(2008)\citenamefont
  {Langanke}, \citenamefont {Martinez-Pinedo}, \citenamefont {Muller},
  \citenamefont {Janka}, \citenamefont {Marek}, \citenamefont {Hix},
  \citenamefont {Juodagalvis},\ and\ \citenamefont
  {Sampaio}}]{Langanke:2007ua}%
  \BibitemOpen
  \bibfield  {author} {\bibinfo {author} {\bibfnamefont {K.}~\bibnamefont
  {Langanke}}, \bibinfo {author} {\bibfnamefont {G.}~\bibnamefont
  {Martinez-Pinedo}}, \bibinfo {author} {\bibfnamefont {B.}~\bibnamefont
  {Muller}}, \bibinfo {author} {\bibfnamefont {H.~T.}\ \bibnamefont {Janka}},
  \bibinfo {author} {\bibfnamefont {A.}~\bibnamefont {Marek}}, \bibinfo
  {author} {\bibfnamefont {W.~R.}\ \bibnamefont {Hix}}, \bibinfo {author}
  {\bibfnamefont {A.}~\bibnamefont {Juodagalvis}},\ and\ \bibinfo {author}
  {\bibfnamefont {J.~M.}\ \bibnamefont {Sampaio}},\ }\bibfield  {title}
  {\bibinfo {title} {{Effects of Inelastic Neutrino-Nucleus Scattering on
  Supernova Dynamics and Radiated Neutrino Spectra}},\ }\href
  {https://doi.org/10.1103/PhysRevLett.100.011101} {\bibfield  {journal}
  {\bibinfo  {journal} {Phys. Rev. Lett.}\ }\textbf {\bibinfo {volume} {100}},\
  \bibinfo {pages} {011101} (\bibinfo {year} {2008})},\ \Eprint
  {https://arxiv.org/abs/0706.1687} {arXiv:0706.1687 [astro-ph]} \BibitemShut
  {NoStop}%
\bibitem [{\citenamefont {Rrapaj}\ \emph {et~al.}(2015)\citenamefont {Rrapaj},
  \citenamefont {Holt}, \citenamefont {Bartl}, \citenamefont {Reddy},\ and\
  \citenamefont {Schwenk}}]{Rrapaj:2014yba}%
  \BibitemOpen
  \bibfield  {author} {\bibinfo {author} {\bibfnamefont {E.}~\bibnamefont
  {Rrapaj}}, \bibinfo {author} {\bibfnamefont {J.~W.}\ \bibnamefont {Holt}},
  \bibinfo {author} {\bibfnamefont {A.}~\bibnamefont {Bartl}}, \bibinfo
  {author} {\bibfnamefont {S.}~\bibnamefont {Reddy}},\ and\ \bibinfo {author}
  {\bibfnamefont {A.}~\bibnamefont {Schwenk}},\ }\bibfield  {title} {\bibinfo
  {title} {{Charged-current reactions in the supernova neutrino-sphere}},\
  }\href {https://doi.org/10.1103/PhysRevC.91.035806} {\bibfield  {journal}
  {\bibinfo  {journal} {Phys. Rev. C}\ }\textbf {\bibinfo {volume} {91}},\
  \bibinfo {pages} {035806} (\bibinfo {year} {2015})},\ \Eprint
  {https://arxiv.org/abs/1408.3368} {arXiv:1408.3368 [nucl-th]} \BibitemShut
  {NoStop}%
\bibitem [{\citenamefont {Endrizzi}\ \emph {et~al.}(2020)\citenamefont
  {Endrizzi}, \citenamefont {Perego}, \citenamefont {Fabbri}, \citenamefont
  {Branca}, \citenamefont {Radice}, \citenamefont {Bernuzzi}, \citenamefont
  {Giacomazzo}, \citenamefont {Pederiva},\ and\ \citenamefont
  {Lovato}}]{Endrizzi:2019trv}%
  \BibitemOpen
  \bibfield  {author} {\bibinfo {author} {\bibfnamefont {A.}~\bibnamefont
  {Endrizzi}}, \bibinfo {author} {\bibfnamefont {A.}~\bibnamefont {Perego}},
  \bibinfo {author} {\bibfnamefont {F.~M.}\ \bibnamefont {Fabbri}}, \bibinfo
  {author} {\bibfnamefont {L.}~\bibnamefont {Branca}}, \bibinfo {author}
  {\bibfnamefont {D.}~\bibnamefont {Radice}}, \bibinfo {author} {\bibfnamefont
  {S.}~\bibnamefont {Bernuzzi}}, \bibinfo {author} {\bibfnamefont
  {B.}~\bibnamefont {Giacomazzo}}, \bibinfo {author} {\bibfnamefont
  {F.}~\bibnamefont {Pederiva}},\ and\ \bibinfo {author} {\bibfnamefont
  {A.}~\bibnamefont {Lovato}},\ }\bibfield  {title} {\bibinfo {title}
  {{Thermodynamics conditions of matter in the neutrino decoupling region
  during neutron star mergers}},\ }\href
  {https://doi.org/10.1140/epja/s10050-019-00018-6} {\bibfield  {journal}
  {\bibinfo  {journal} {Eur. Phys. J. A}\ }\textbf {\bibinfo {volume} {56}},\
  \bibinfo {pages} {15} (\bibinfo {year} {2020})},\ \Eprint
  {https://arxiv.org/abs/1908.04952} {arXiv:1908.04952 [astro-ph.HE]}
  \BibitemShut {NoStop}%
\bibitem [{\citenamefont {Cheong}\ \emph {et~al.}(2025)\citenamefont {Cheong},
  \citenamefont {Foucart}, \citenamefont {Ng}, \citenamefont {Offermans},
  \citenamefont {Duez}, \citenamefont {Muhammed},\ and\ \citenamefont
  {Chawhan}}]{Cheong:2024cnb}%
  \BibitemOpen
  \bibfield  {author} {\bibinfo {author} {\bibfnamefont {P.~C.-K.}\
  \bibnamefont {Cheong}}, \bibinfo {author} {\bibfnamefont {F.}~\bibnamefont
  {Foucart}}, \bibinfo {author} {\bibfnamefont {H.~H.-Y.}\ \bibnamefont {Ng}},
  \bibinfo {author} {\bibfnamefont {A.}~\bibnamefont {Offermans}}, \bibinfo
  {author} {\bibfnamefont {M.~D.}\ \bibnamefont {Duez}}, \bibinfo {author}
  {\bibfnamefont {N.}~\bibnamefont {Muhammed}},\ and\ \bibinfo {author}
  {\bibfnamefont {P.}~\bibnamefont {Chawhan}},\ }\bibfield  {title} {\bibinfo
  {title} {{Influence of neutrino-electron scattering and neutrino-pair
  annihilation on hypermassive neutron star}},\ }\href
  {https://doi.org/10.1103/PhysRevD.111.043036} {\bibfield  {journal} {\bibinfo
   {journal} {Phys. Rev. D}\ }\textbf {\bibinfo {volume} {111}},\ \bibinfo
  {pages} {043036} (\bibinfo {year} {2025})},\ \Eprint
  {https://arxiv.org/abs/2410.20681} {arXiv:2410.20681 [astro-ph.HE]}
  \BibitemShut {NoStop}%
\bibitem [{\citenamefont {Duez}\ \emph {et~al.}(2008)\citenamefont {Duez},
  \citenamefont {Foucart}, \citenamefont {Kidder}, \citenamefont {Pfeiffer},
  \citenamefont {Scheel},\ and\ \citenamefont {Teukolsky}}]{Duez:2008rb}%
  \BibitemOpen
  \bibfield  {author} {\bibinfo {author} {\bibfnamefont {M.~D.}\ \bibnamefont
  {Duez}}, \bibinfo {author} {\bibfnamefont {F.}~\bibnamefont {Foucart}},
  \bibinfo {author} {\bibfnamefont {L.~E.}\ \bibnamefont {Kidder}}, \bibinfo
  {author} {\bibfnamefont {H.~P.}\ \bibnamefont {Pfeiffer}}, \bibinfo {author}
  {\bibfnamefont {M.~A.}\ \bibnamefont {Scheel}},\ and\ \bibinfo {author}
  {\bibfnamefont {S.~A.}\ \bibnamefont {Teukolsky}},\ }\bibfield  {title}
  {\bibinfo {title} {{Evolving black hole-neutron star binaries in general
  relativity using pseudospectral and finite difference methods}},\ }\href
  {https://doi.org/10.1103/PhysRevD.78.104015} {\bibfield  {journal} {\bibinfo
  {journal} {Phys. Rev. D}\ }\textbf {\bibinfo {volume} {78}},\ \bibinfo
  {pages} {104015} (\bibinfo {year} {2008})},\ \Eprint
  {https://arxiv.org/abs/0809.0002} {arXiv:0809.0002 [gr-qc]} \BibitemShut
  {NoStop}%
\bibitem [{\citenamefont {Foucart}\ \emph {et~al.}(2013)\citenamefont
  {Foucart}, \citenamefont {Deaton}, \citenamefont {Duez}, \citenamefont
  {Kidder}, \citenamefont {MacDonald}, \citenamefont {Ott}, \citenamefont
  {Pfeiffer}, \citenamefont {Scheel}, \citenamefont {Szilagyi},\ and\
  \citenamefont {Teukolsky}}]{Foucart:2012vn}%
  \BibitemOpen
  \bibfield  {author} {\bibinfo {author} {\bibfnamefont {F.}~\bibnamefont
  {Foucart}}, \bibinfo {author} {\bibfnamefont {M.~B.}\ \bibnamefont {Deaton}},
  \bibinfo {author} {\bibfnamefont {M.~D.}\ \bibnamefont {Duez}}, \bibinfo
  {author} {\bibfnamefont {L.~E.}\ \bibnamefont {Kidder}}, \bibinfo {author}
  {\bibfnamefont {I.}~\bibnamefont {MacDonald}}, \bibinfo {author}
  {\bibfnamefont {C.~D.}\ \bibnamefont {Ott}}, \bibinfo {author} {\bibfnamefont
  {H.~P.}\ \bibnamefont {Pfeiffer}}, \bibinfo {author} {\bibfnamefont {M.~A.}\
  \bibnamefont {Scheel}}, \bibinfo {author} {\bibfnamefont {B.}~\bibnamefont
  {Szilagyi}},\ and\ \bibinfo {author} {\bibfnamefont {S.~A.}\ \bibnamefont
  {Teukolsky}},\ }\bibfield  {title} {\bibinfo {title} {{Black hole-neutron
  star mergers at realistic mass ratios: Equation of state and spin orientation
  effects}},\ }\href {https://doi.org/10.1103/PhysRevD.87.084006} {\bibfield
  {journal} {\bibinfo  {journal} {Phys. Rev. D}\ }\textbf {\bibinfo {volume}
  {87}},\ \bibinfo {pages} {084006} (\bibinfo {year} {2013})},\ \Eprint
  {https://arxiv.org/abs/1212.4810} {arXiv:1212.4810 [gr-qc]} \BibitemShut
  {NoStop}%
\bibitem [{\citenamefont {O'Connor}(2015)}]{OConnor:2014sgn}%
  \BibitemOpen
  \bibfield  {author} {\bibinfo {author} {\bibfnamefont {E.}~\bibnamefont
  {O'Connor}},\ }\bibfield  {title} {\bibinfo {title} {{An Open-Source Neutrino
  Radiation Hydrodynamics Code for Core-Collapse Supernovae}},\ }\href
  {https://doi.org/10.1088/0067-0049/219/2/24} {\bibfield  {journal} {\bibinfo
  {journal} {Astrophys. J. Suppl.}\ }\textbf {\bibinfo {volume} {219}},\
  \bibinfo {pages} {24} (\bibinfo {year} {2015})},\ \Eprint
  {https://arxiv.org/abs/1411.7058} {arXiv:1411.7058 [astro-ph.HE]}
  \BibitemShut {NoStop}%
\bibitem [{\citenamefont {Foucart}(2025{\natexlab{a}})}]{Foucart:2025nub}%
  \BibitemOpen
  \bibfield  {author} {\bibinfo {author} {\bibfnamefont {F.}~\bibnamefont
  {Foucart}},\ }\bibfield  {title} {\bibinfo {title} {{Assessing the difficulty
  of capturing the distribution function of neutrinos in neutron star merger
  simulations}},\ }\href {https://doi.org/10.1103/gkk5-6wmh} {\bibfield
  {journal} {\bibinfo  {journal} {Phys. Rev. D}\ }\textbf {\bibinfo {volume}
  {112}},\ \bibinfo {pages} {023046} (\bibinfo {year} {2025}{\natexlab{a}})},\
  \Eprint {https://arxiv.org/abs/2504.21822} {arXiv:2504.21822 [astro-ph.HE]}
  \BibitemShut {NoStop}%
\bibitem [{\citenamefont {Alford}\ \emph {et~al.}(2026)\citenamefont {Alford},
  \citenamefont {Brodie}, \citenamefont {Foucart},\ and\ \citenamefont
  {Haber}}]{Alford:2026kwd}%
  \BibitemOpen
  \bibfield  {author} {\bibinfo {author} {\bibfnamefont {M.~G.}\ \bibnamefont
  {Alford}}, \bibinfo {author} {\bibfnamefont {L.}~\bibnamefont {Brodie}},
  \bibinfo {author} {\bibfnamefont {F.}~\bibnamefont {Foucart}},\ and\ \bibinfo
  {author} {\bibfnamefont {A.}~\bibnamefont {Haber}},\ }\bibfield  {title}
  {\bibinfo {title} {{Thermalization of Neutrinos in a Neutron Star Merger
  Simulation}},\ }\href@noop {} {\bibfield  {journal} {\bibinfo  {journal}
  {arXiv}\ } (\bibinfo {year} {2026})},\ \Eprint
  {https://arxiv.org/abs/2603.06788} {arXiv:2603.06788 [astro-ph.HE]}
  \BibitemShut {NoStop}%
\bibitem [{\citenamefont {Pajkos}\ and\ \citenamefont
  {Most}(2025)}]{Pajkos:2024iry}%
  \BibitemOpen
  \bibfield  {author} {\bibinfo {author} {\bibfnamefont {M.~A.}\ \bibnamefont
  {Pajkos}}\ and\ \bibinfo {author} {\bibfnamefont {E.~R.}\ \bibnamefont
  {Most}},\ }\bibfield  {title} {\bibinfo {title} {{Influence of muons, pions,
  and trapped neutrinos on neutron star mergers}},\ }\href
  {https://doi.org/10.1103/PhysRevD.111.043013} {\bibfield  {journal} {\bibinfo
   {journal} {Phys. Rev. D}\ }\textbf {\bibinfo {volume} {111}},\ \bibinfo
  {pages} {043013} (\bibinfo {year} {2025})},\ \Eprint
  {https://arxiv.org/abs/2409.09147} {arXiv:2409.09147 [astro-ph.HE]}
  \BibitemShut {NoStop}%
\bibitem [{\citenamefont {Ng}\ \emph {et~al.}(2025)\citenamefont {Ng},
  \citenamefont {Musolino}, \citenamefont {Tootle},\ and\ \citenamefont
  {Rezzolla}}]{Ng:2024zve}%
  \BibitemOpen
  \bibfield  {author} {\bibinfo {author} {\bibfnamefont {H.~H.-Y.}\
  \bibnamefont {Ng}}, \bibinfo {author} {\bibfnamefont {C.}~\bibnamefont
  {Musolino}}, \bibinfo {author} {\bibfnamefont {S.~D.}\ \bibnamefont
  {Tootle}},\ and\ \bibinfo {author} {\bibfnamefont {L.}~\bibnamefont
  {Rezzolla}},\ }\bibfield  {title} {\bibinfo {title} {{Accurate Muonic
  Interactions in Neutron Star Mergers and Impact on Heavy-element
  Nucleosynthesis}},\ }\href {https://doi.org/10.3847/2041-8213/add324}
  {\bibfield  {journal} {\bibinfo  {journal} {Astrophys. J. Lett.}\ }\textbf
  {\bibinfo {volume} {985}},\ \bibinfo {pages} {L36} (\bibinfo {year}
  {2025})},\ \Eprint {https://arxiv.org/abs/2411.19178} {arXiv:2411.19178
  [astro-ph.HE]} \BibitemShut {NoStop}%
\bibitem [{\citenamefont {Gieg}\ \emph {et~al.}(2025)\citenamefont {Gieg},
  \citenamefont {Schianchi}, \citenamefont {Ujevic},\ and\ \citenamefont
  {Dietrich}}]{Gieg:2024jxs}%
  \BibitemOpen
  \bibfield  {author} {\bibinfo {author} {\bibfnamefont {H.}~\bibnamefont
  {Gieg}}, \bibinfo {author} {\bibfnamefont {F.}~\bibnamefont {Schianchi}},
  \bibinfo {author} {\bibfnamefont {M.}~\bibnamefont {Ujevic}},\ and\ \bibinfo
  {author} {\bibfnamefont {T.}~\bibnamefont {Dietrich}},\ }\bibfield  {title}
  {\bibinfo {title} {{Role of muons in binary neutron star mergers: First
  simulations}},\ }\href {https://doi.org/10.1103/52ph-53yw} {\bibfield
  {journal} {\bibinfo  {journal} {Phys. Rev. D}\ }\textbf {\bibinfo {volume}
  {112}},\ \bibinfo {pages} {023036} (\bibinfo {year} {2025})},\ \Eprint
  {https://arxiv.org/abs/2409.04420} {arXiv:2409.04420 [gr-qc]} \BibitemShut
  {NoStop}%
\bibitem [{\citenamefont {Gieg}\ \emph {et~al.}(2026)\citenamefont {Gieg},
  \citenamefont {Jaeger}, \citenamefont {Ujevic},\ and\ \citenamefont
  {Dietrich}}]{Gieg:2026beb}%
  \BibitemOpen
  \bibfield  {author} {\bibinfo {author} {\bibfnamefont {H.}~\bibnamefont
  {Gieg}}, \bibinfo {author} {\bibfnamefont {R.}~\bibnamefont {Jaeger}},
  \bibinfo {author} {\bibfnamefont {M.}~\bibnamefont {Ujevic}},\ and\ \bibinfo
  {author} {\bibfnamefont {T.}~\bibnamefont {Dietrich}},\ }\bibfield  {title}
  {\bibinfo {title} {{Consistent Treatment of Muons in Binary Neutron Star
  Mergers}},\ }\href@noop {} {\bibfield  {journal} {\bibinfo  {journal}
  {arXiv}\ } (\bibinfo {year} {2026})},\ \Eprint
  {https://arxiv.org/abs/2604.14225} {arXiv:2604.14225 [astro-ph.HE]}
  \BibitemShut {NoStop}%
\bibitem [{\citenamefont {Bruenn}(1985)}]{Bruenn:1985en}%
  \BibitemOpen
  \bibfield  {author} {\bibinfo {author} {\bibfnamefont {S.~W.}\ \bibnamefont
  {Bruenn}},\ }\bibfield  {title} {\bibinfo {title} {Stellar core collapse:
  Numerical model and infall epoch},\ }\href {https://doi.org/10.1086/191056}
  {\bibfield  {journal} {\bibinfo  {journal} {Astrophys. J. Suppl. Ser.}\
  }\textbf {\bibinfo {volume} {58}},\ \bibinfo {pages} {771} (\bibinfo {year}
  {1985})}\BibitemShut {NoStop}%
\bibitem [{\citenamefont {Burrows}\ \emph {et~al.}(2006)\citenamefont
  {Burrows}, \citenamefont {Reddy},\ and\ \citenamefont
  {Thompson}}]{Burrows:2004vq}%
  \BibitemOpen
  \bibfield  {author} {\bibinfo {author} {\bibfnamefont {A.}~\bibnamefont
  {Burrows}}, \bibinfo {author} {\bibfnamefont {S.}~\bibnamefont {Reddy}},\
  and\ \bibinfo {author} {\bibfnamefont {T.~A.}\ \bibnamefont {Thompson}},\
  }\bibfield  {title} {\bibinfo {title} {{Neutrino opacities in nuclear
  matter}},\ }\href {https://doi.org/10.1016/j.nuclphysa.2004.06.012}
  {\bibfield  {journal} {\bibinfo  {journal} {Nucl. Phys. A}\ }\textbf
  {\bibinfo {volume} {777}},\ \bibinfo {pages} {356} (\bibinfo {year}
  {2006})},\ \Eprint {https://arxiv.org/abs/astro-ph/0404432}
  {arXiv:astro-ph/0404432} \BibitemShut {NoStop}%
\bibitem [{\citenamefont {Foucart}(2025{\natexlab{b}})}]{Foucart:2025cjr}%
  \BibitemOpen
  \bibfield  {author} {\bibinfo {author} {\bibfnamefont {F.}~\bibnamefont
  {Foucart}},\ }\bibfield  {title} {\bibinfo {title} {{Neutrinos in colliding
  neutron stars and black holes}},\ }\bibfield  {journal} {\bibinfo  {journal}
  {Encyclopedia of Particle Physics}\ }\href
  {https://doi.org/10.1016/B978-0-443-26598-3.00004-3}
  {10.1016/B978-0-443-26598-3.00004-3} (\bibinfo {year} {2025}{\natexlab{b}}),\
  \Eprint {https://arxiv.org/abs/2410.03646} {arXiv:2410.03646 [astro-ph.HE]}
  \BibitemShut {NoStop}%
\bibitem [{\citenamefont {Dicus}(1972)}]{Dicus:1972yr}%
  \BibitemOpen
  \bibfield  {author} {\bibinfo {author} {\bibfnamefont {D.~A.}\ \bibnamefont
  {Dicus}},\ }\bibfield  {title} {\bibinfo {title} {{Stellar energy-loss rates
  in a convergent theory of weak and electromagnetic interactions}},\ }\href
  {https://doi.org/10.1103/PhysRevD.6.941} {\bibfield  {journal} {\bibinfo
  {journal} {Phys. Rev. D}\ }\textbf {\bibinfo {volume} {6}},\ \bibinfo {pages}
  {941} (\bibinfo {year} {1972})}\BibitemShut {NoStop}%
\bibitem [{\citenamefont {Foucart}(2023)}]{Foucart:2022bth}%
  \BibitemOpen
  \bibfield  {author} {\bibinfo {author} {\bibfnamefont {F.}~\bibnamefont
  {Foucart}},\ }\bibfield  {title} {\bibinfo {title} {{Neutrino transport in
  general relativistic neutron star merger simulations}},\ }\href
  {https://doi.org/10.1007/s41115-023-00016-y} {\bibfield  {journal} {\bibinfo
  {journal} {Liv. Rev. Comput. Astrophys.}\ }\textbf {\bibinfo {volume} {9}},\
  \bibinfo {pages} {1} (\bibinfo {year} {2023})},\ \Eprint
  {https://arxiv.org/abs/2209.02538} {arXiv:2209.02538 [astro-ph.HE]}
  \BibitemShut {NoStop}%
\bibitem [{\citenamefont {Brinkmann}\ and\ \citenamefont
  {Turner}(1988)}]{Brinkmann:1988vi}%
  \BibitemOpen
  \bibfield  {author} {\bibinfo {author} {\bibfnamefont {R.~P.}\ \bibnamefont
  {Brinkmann}}\ and\ \bibinfo {author} {\bibfnamefont {M.~S.}\ \bibnamefont
  {Turner}},\ }\bibfield  {title} {\bibinfo {title} {{Numerical rates for
  nucleon-nucleon, axion bremsstrahlung}},\ }\href
  {https://doi.org/10.1103/PhysRevD.38.2338} {\bibfield  {journal} {\bibinfo
  {journal} {Phys. Rev. D}\ }\textbf {\bibinfo {volume} {38}},\ \bibinfo
  {pages} {2338} (\bibinfo {year} {1988})}\BibitemShut {NoStop}%
\bibitem [{\citenamefont {Hannestad}\ and\ \citenamefont
  {Raffelt}(1998)}]{Hannestad:1997gc}%
  \BibitemOpen
  \bibfield  {author} {\bibinfo {author} {\bibfnamefont {S.}~\bibnamefont
  {Hannestad}}\ and\ \bibinfo {author} {\bibfnamefont {G.}~\bibnamefont
  {Raffelt}},\ }\bibfield  {title} {\bibinfo {title} {{Supernova neutrino
  opacity from nucleon-nucleon Bremsstrahlung and related processes}},\ }\href
  {https://doi.org/10.1086/306303} {\bibfield  {journal} {\bibinfo  {journal}
  {Astrophys. J.}\ }\textbf {\bibinfo {volume} {507}},\ \bibinfo {pages} {339}
  (\bibinfo {year} {1998})},\ \Eprint {https://arxiv.org/abs/astro-ph/9711132}
  {arXiv:astro-ph/9711132} \BibitemShut {NoStop}%
\bibitem [{\citenamefont {Ruffert}\ \emph {et~al.}(1996)\citenamefont
  {Ruffert}, \citenamefont {Janka},\ and\ \citenamefont
  {Schaefer}}]{Ruffert:1995fs}%
  \BibitemOpen
  \bibfield  {author} {\bibinfo {author} {\bibfnamefont {M.~H.}\ \bibnamefont
  {Ruffert}}, \bibinfo {author} {\bibfnamefont {H.~T.}\ \bibnamefont {Janka}},\
  and\ \bibinfo {author} {\bibfnamefont {G.}~\bibnamefont {Schaefer}},\
  }\bibfield  {title} {\bibinfo {title} {{Coalescing neutron stars: A Step
  towards physical models. 1: Hydrodynamic evolution and gravitational wave
  emission}},\ }\href@noop {} {\bibfield  {journal} {\bibinfo  {journal}
  {Astron. Astrophys.}\ }\textbf {\bibinfo {volume} {311}},\ \bibinfo {pages}
  {532} (\bibinfo {year} {1996})},\ \Eprint
  {https://arxiv.org/abs/astro-ph/9509006} {arXiv:astro-ph/9509006}
  \BibitemShut {NoStop}%
\bibitem [{\citenamefont {Arcones}\ \emph {et~al.}(2010)\citenamefont
  {Arcones}, \citenamefont {Martinez-Pinedo}, \citenamefont {Roberts},\ and\
  \citenamefont {Woosley}}]{Arcones:2010yf}%
  \BibitemOpen
  \bibfield  {author} {\bibinfo {author} {\bibfnamefont {A.}~\bibnamefont
  {Arcones}}, \bibinfo {author} {\bibfnamefont {G.}~\bibnamefont
  {Martinez-Pinedo}}, \bibinfo {author} {\bibfnamefont {L.~F.}\ \bibnamefont
  {Roberts}},\ and\ \bibinfo {author} {\bibfnamefont {S.~E.}\ \bibnamefont
  {Woosley}},\ }\bibfield  {title} {\bibinfo {title} {Electron fraction
  constraints based on nuclear statistical equilibrium with beta equilibrium},\
  }\href {https://doi.org/10.1051/0004-6361/201014276} {\bibfield  {journal}
  {\bibinfo  {journal} {Astron. Astrophys.}\ }\textbf {\bibinfo {volume}
  {522}},\ \bibinfo {pages} {A25} (\bibinfo {year} {2010})},\ \Eprint
  {https://arxiv.org/abs/1008.3890} {arXiv:1008.3890 [astro-ph.SR]}
  \BibitemShut {NoStop}%
\bibitem [{\citenamefont {Alford}\ \emph {et~al.}(2021)\citenamefont {Alford},
  \citenamefont {Haber}, \citenamefont {Harris},\ and\ \citenamefont
  {Zhang}}]{Alford:2021ogv}%
  \BibitemOpen
  \bibfield  {author} {\bibinfo {author} {\bibfnamefont {M.~G.}\ \bibnamefont
  {Alford}}, \bibinfo {author} {\bibfnamefont {A.}~\bibnamefont {Haber}},
  \bibinfo {author} {\bibfnamefont {S.~P.}\ \bibnamefont {Harris}},\ and\
  \bibinfo {author} {\bibfnamefont {Z.}~\bibnamefont {Zhang}},\ }\bibfield
  {title} {\bibinfo {title} {{Beta Equilibrium Under Neutron Star Merger
  Conditions}},\ }\href {https://doi.org/10.3390/universe7110399} {\bibfield
  {journal} {\bibinfo  {journal} {Universe}\ }\textbf {\bibinfo {volume} {7}},\
  \bibinfo {pages} {399} (\bibinfo {year} {2021})},\ \Eprint
  {https://arxiv.org/abs/2108.03324} {arXiv:2108.03324 [nucl-th]} \BibitemShut
  {NoStop}%
\bibitem [{\citenamefont {Alford}\ \emph {et~al.}(2024)\citenamefont {Alford},
  \citenamefont {Haber},\ and\ \citenamefont {Zhang}}]{Alford:2024xfb}%
  \BibitemOpen
  \bibfield  {author} {\bibinfo {author} {\bibfnamefont {M.~G.}\ \bibnamefont
  {Alford}}, \bibinfo {author} {\bibfnamefont {A.}~\bibnamefont {Haber}},\ and\
  \bibinfo {author} {\bibfnamefont {Z.}~\bibnamefont {Zhang}},\ }\bibfield
  {title} {\bibinfo {title} {{Beyond modified Urca: The nucleon width
  approximation for flavor-changing processes~in dense matter}},\ }\href
  {https://doi.org/10.1103/PhysRevC.110.L052801} {\bibfield  {journal}
  {\bibinfo  {journal} {Phys. Rev. C}\ }\textbf {\bibinfo {volume} {110}},\
  \bibinfo {pages} {L052801} (\bibinfo {year} {2024})},\ \Eprint
  {https://arxiv.org/abs/2406.13717} {arXiv:2406.13717 [nucl-th]} \BibitemShut
  {NoStop}%
\bibitem [{\citenamefont {Mezzacappa}\ and\ \citenamefont
  {Bruenn}(1993)}]{Mezzacappa:1993gn}%
  \BibitemOpen
  \bibfield  {author} {\bibinfo {author} {\bibfnamefont {A.}~\bibnamefont
  {Mezzacappa}}\ and\ \bibinfo {author} {\bibfnamefont {S.~W.}\ \bibnamefont
  {Bruenn}},\ }\bibfield  {title} {\bibinfo {title} {Stellar core collapse: A
  boltzmann treatment of neutrino-electron scattering},\ }\href
  {https://doi.org/10.1086/172791} {\bibfield  {journal} {\bibinfo  {journal}
  {Astrophys. J.}\ }\textbf {\bibinfo {volume} {410}},\ \bibinfo {pages} {740}
  (\bibinfo {year} {1993})}\BibitemShut {NoStop}%
\bibitem [{\citenamefont {Kotake}\ \emph {et~al.}(2018)\citenamefont {Kotake},
  \citenamefont {Takiwaki}, \citenamefont {Fischer}, \citenamefont {Nakamura},\
  and\ \citenamefont {Mart{\'\i}nez-Pinedo}}]{Kotake:2018ypf}%
  \BibitemOpen
  \bibfield  {author} {\bibinfo {author} {\bibfnamefont {K.}~\bibnamefont
  {Kotake}}, \bibinfo {author} {\bibfnamefont {T.}~\bibnamefont {Takiwaki}},
  \bibinfo {author} {\bibfnamefont {T.}~\bibnamefont {Fischer}}, \bibinfo
  {author} {\bibfnamefont {K.}~\bibnamefont {Nakamura}},\ and\ \bibinfo
  {author} {\bibfnamefont {G.}~\bibnamefont {Mart{\'\i}nez-Pinedo}},\
  }\bibfield  {title} {\bibinfo {title} {{Impact of Neutrino Opacities on
  Core-Collapse Supernova Simulations}},\ }\href
  {https://doi.org/10.3847/1538-4357/aaa716} {\bibfield  {journal} {\bibinfo
  {journal} {Astrophys. J.}\ }\textbf {\bibinfo {volume} {853}},\ \bibinfo
  {pages} {170} (\bibinfo {year} {2018})},\ \Eprint
  {https://arxiv.org/abs/1801.02703} {arXiv:1801.02703 [astro-ph.HE]}
  \BibitemShut {NoStop}%
\end{thebibliography}%

\end{document}